\documentclass[12pt]{article}
\usepackage{graphicx}
\usepackage{longtable}
\makeatletter
 
  \@addtoreset{equation}{section}
\makeatother
\def\beqar {\begin{eqnarray}}
\def\eeqar {\end{eqnarray}}
\def\beq {\begin{equation}}
\def\eeq {\end{equation}}
\def\A{{\cal A}}
\def\B{{\cal B}}
\def\C{{\cal C}}
\def\F{{\cal F}}

\def\S{{\cal S}}

\def\P{{\cal P}}
\def\L{{\cal L}}
\def\R{{\cal R}}

\def\N{{\cal N}}

\def\al{\alpha}
\def\bt{\beta}
\def\del{\delta}

\def\ga{\gamma}
\def\Ga{\Gamma}

\def\ep{\epsilon}

\def\om{\omega}

\def\th{\theta}

\def\si{\sigma}

\def\d{\partial}

\def\Ad{{\dot A}}
\def\Bd{{\dot B}}

\def\bz{{\bar z}}
\def\bu{{\bar u}}

\def\hf{\frac{1}{2}}

\def\<{\langle}
\def\>{\rangle}

\def\Tr{{\rm Tr}}

\def\Path{{\rm P}}

\def\cp{{\bf CP}}

\begin{document}

\begin{titlepage}
\null\vspace{-62pt} \pagestyle{empty}
\begin{center}
%\rightline{} \rightline{CCNY-HEP-/05}
\vspace{1.0truein}

{\Large\bf Holonomies of gauge fields in twistor space 4: \\
\vspace{.35cm}
\hspace{-.3cm}
functional MHV rules and one-loop amplitudes} \\

%%%%%%%%%%%%%%%%%%%%%%%%%%%%%%%%%%%%%%%%%%%%%%%%%%%%%%%%%%
\vspace{1.0in} {\sc Yasuhiro Abe} \\
\vskip .12in {\it Cereja Technology Co., Ltd.\\
1-13-14 Mukai-Bldg. 3F, Sekiguchi \\
Bunkyo-ku, Tokyo 112-0014, Japan } \\
\vskip .07in {\tt abe@cereja.co.jp}\\
\vspace{1.3in}
%%%%%%%%%%%%%%%%%%%%%%%%%%%%%%%%%%%%%%%%%%%%%%%%%%%%%%%%%%%%
\centerline{\large\bf Abstract}
\end{center}
We consider generalization of the Cachazo-Svrcek-Witten (CSW) rules
to one-loop amplitudes of $\N = 4$ super Yang-Mills theory 
in a recently developed holonomy formalism in twistor space.
We first reconsider off-shell continuation of the Lorentz-invariant
Nair measure for the incorporation of loop integrals.
We then formulate an S-matrix functional for general amplitudes
such that it implements the CSW rules at quantum level.
For one-loop MHV amplitudes, the S-matrix functional correctly
reproduces the analytic expressions obtained in the
Brandhuber-Spence-Travaglini (BST) method.
Motivated by this result, we propose a novel regularization scheme
by use of an iterated-integral representation of polylogarithms
and obtain a set of new analytic expressions for one-loop NMHV
and N$^2$MHV amplitudes in a conjectural form.
We also briefly sketch how the extension to one-loop non-MHV
amplitudes in general can be carried out.

\end{titlepage}
%%%%%%%%%%%%%%%%%%%%%%%%%%%%%%%%%%%%%%%%%%%%%%%%%%%%%%%%%%%%%
\pagestyle{plain} \setcounter{page}{2} %\baselineskip =14pt

%\begin{small}
%\tableofcontents
%\end{small}

%%%%%%%%%%%%%%%%%%%%%%%%%%%%%%%%%%%%%%%%%%%%%%%%%%%%%%%%%%%%%%%%%%
\section{Introduction}

Recent years witness a lot of progress in
the study of loop amplitudes in $\N = 4$ super Yang-Mills theory.
At a relatively early stage, the progress is initiated by
the discovery of the so-called Cachazo-Svrcek-Witten (CSW) rules
for tree-level gluon amplitudes \cite{Cachazo:2004kj} (for related works,
see, {\it e.g.}, \cite{Georgiou:2004wu}-\cite{Abe:2004ep})
and the applicability of the CSW rules to one-loop MHV (Maximally
Helicity Violating) amplitudes in $\N = 4$ and less supersymmetric theories
\cite{Cachazo:2004zb,Cachazo:2004by,Cachazo:2004dr}.
A particularly important set of results in regard to the applicability
are the so-called Brandhuber-Spence-Travaglini (BST) method
\cite{Brandhuber:2004yw}-\cite{Brandhuber:2005kd}; see also
\cite{Luo:2004ss,Quigley:2004pw} and a recent review \cite{Brandhuber:2011ke}.
In the original BST paper \cite{Brandhuber:2004yw}, it is analytically shown that
the CSW-based loop calculation, together with
the use of an off-shell continuation of the Nair measure ({\it i.e.},
the Lorentz-invariant measure in terms of spinor momenta,
first given in \cite{Nair:1988bq}),
leads to the previously known results on the one-loop MHV amplitudes
in $\N = 4$ super Yang-Mills theory which
Bern, Dixon, Dunbar and Kosower (BDDK) have obtained many years before, utilizing
a unitary-cut method \cite{Bern:1994zx,Bern:1994cg}.
Although the success of the BST method is somewhat restricted
to the MHV amplitudes, the importance of the CSW rules
in gauge theories has been well-recognized at this stage and it has motivated
further developments in the unitary-cut method for various types of one-loop
amplitudes \cite{Britto:2004nc}-\cite{Bern:2004bt}.

At a later stage, the progress in loop calculation
integrates higher-loop {\it planar} amplitudes of the $\N = 4$ theory.
There are a series of important studies which guide this advance, including
a simple proportional relation between four-point
one- and two-loop planar amplitudes \cite{Anastasiou:2003kj},
a conjectured iterative relation (in loop order) among
MHV loop amplitudes \cite{Anastasiou:2003kj,Bern:2005iz}\footnote{
These relations, known as the ABDK/BDS relations, do not hold
at higher-loop level in general.
An analytic expression for such a discrepancy is called a reminder function.
Attentive studies of the reminder function at two-loop level are carried out
recently. For relevant works, see, {\it e.g.},
\cite{Bern:2008ap}-\cite{Gaiotto:2011dt}.
},
and a duality between gluon amplitudes and light-like Wilson loops
in strong coupling region of planar $\N = 4$ super Yang-Mills theory \cite{Alday:2007hr}
(see \cite{Alday:2008yw,Henn:2009bd} for a review of this duality).
A particularly important result in this context is
the discovery of dual superconformal symmetry \cite{Drummond:2008vq}.
There are many investigations on this new symmetry in the loop
amplitudes; see, {\it e.g.}, \cite{Elvang:2009ya}-\cite{Brandhuber:2009kh}
and a recent review \cite{Henn:2011xk}.
The discovery of Yangian symmetry in the $\N = 4$ amplitudes is
also reported in this context \cite{Drummond:2009fd};
for the study of Yangian symmetry in loop amplitudes,
see, {\it e.g.}, \cite{Beisert:2010gn,Drummond:2010uq}
and a review \cite{Bargheer:2011mm}.
More recently, largely motivated by the above progress, a powerful computational
framework for the planar amplitudes is proposed
\cite{ArkaniHamed:2009si,ArkaniHamed:2009dn,ArkaniHamed:2010kv}.
This is carried out by use of a so-called momentum twistor space and is
followed by a number of intensive works;
some of them can be traced in \cite{Drummond:2010mb}-\cite{Eden:2011ku}.
Along with these developments, we also notice that
many new results are obtained in regard to the unitary-cut method
for higher-loop calculations; see, {\it e.g.},
\cite{Bern:2009xq}-\cite{Schabinger:2011wh} and
a series of recent reviews \cite{Britto:2010xq}-\cite{Schabinger:2011kb}.

In the present paper, we introduce a new CSW- or BST-based calculation
for one-loop amplitudes of $\N = 4$ super Yang-Mills theory
in the framework of a recently developed holonomy formalism in twistor space
\cite{Abe:2009kn,Abe:2009kq}.
As shown in \cite{Abe:2009kn}, an S-matrix functional
for tree gluon amplitudes can be obtained from a holonomy
operator in supertwistor space.
The S-matrix functional contains a Wick-like contraction
operator that implements the CSW rules.
Thus, in terms of functional formulation, it would
and should be straightforward to
generalize the holonomy formalism to loop amplitudes.
We shall pursue this possibility in the present paper.

In carrying out practical calculations at loop level,
there are subtleties in regard to an off-shell property of the loop momentum.
Remember that the original CSW rules give a prescription for the non-MHV tree
amplitudes in terms of a combination of MHV vertices which are connected
one another by a scalar propagator having an off-shell momentum transfer.
Thus the issue of off-shell continuation in the CSW rules exists even at tree level.
For tree amplitudes, this can be circumvented by considering
in a momentum-space representation. In fact, it turns out
that the momentum-space tree amplitudes can be determined
up to the choice of the so-called reference spinors.
Starting at one-loop level, however, we need to seriously consider the
off-shell continuation of the Nair measure. This is analogous to
the calculation of quantum anomalies since, in either case,
quantum effects arise from an analysis of the integral measures
which behave nontrivially at loop level.

The first step toward such an analysis for one-loop MHV amplitudes is taken in
the BST method \cite{Brandhuber:2004yw}.
In the present paper, we shall follow their step in spirit
and consider natural generalization to the other (non-MHV) one-loop amplitudes.
It should be noticed that there exist no analytic confirmations
of the BST method to one-loop non-MHV amplitudes although it is
widely recognized in the literature that
the CSW generalization is applicable to any types of loop amplitudes.
In fact, quite interestingly, an S-matrix functional which realizes
the CSW generalization to multi-loop, non-MHV and {\it non-planar} amplitudes
of $\N = 4$ super Yang-Mills theory is recently proposed by
Sever and Vieira in \cite{Sever:2009aa}.\footnote{
A main purpose of ref. \cite{Sever:2009aa} is to show
superconformal invariance of the proposed S-matrix
(which, by the way, is structurally similar to the S-matrix of the holonomy formalism)
at loop level and, hence, no analytic expressions for any loop amplitudes are provided.}
Motivated by these considerations, in the present paper,
we reconsider the off-shell continuation of the Nair measure
and then introduce a novel realization of the CSW rules in the holonomy formalism
such that the CSW generalization to loop amplitudes becomes straightforward.
To be more specific, we shall reproduce one-loop MHV amplitudes
by reformulating the S-matrix functional in a
coordinate-space representation such that off-shell Nair measures
are naturally incorporated into the holonomy operator.
We then obtain a set of new analytic expressions for one-loop NMHV
amplitudes in a conventional functional method.
We also consider the generalization to other non-MHV amplitudes.
The results suggest that general one-loop amplitudes can be expressed in terms of
a set of polylogarithms, ${\rm Li}_k$ $(k \le m+2)$ for $n$-gluon
one-loop N$^{m}$MHV amplitudes ($m = 0, 1,2, \cdots , n-4$).
The main purpose of this paper is to show that such expressions
can systematically be obtained in the holonomy formalism.

This paper is organized as follows.
In section 2, we review the holonomy formalism at tree level.
We discuss how the CSW rules are implemented by use of a contraction operator.
As a simple example, we present explicit calculations of
the six-point NMHV tree amplitudes in a momentum-space representation.
In section 3, we first consider the off-shell continuation of the Nair measure.
We then reformulate the S-matrix functional in a coordinate-space
representation such that the off-shell Nair measures are naturally incorporated into
the holonomy formalism.
We discuss that an application of such a formulation correctly reproduces
the BST representation of one-loop MHV amplitudes.
In section 4, we present an alternative derivation of the BST
representation, making explicit use of the off-shell Nair measure.
This enable us to propose a new efficient regularization
scheme for the one-loop amplitudes.
In section 5, we apply the alternative method to NMHV amplitudes
and obtain a new analytic expression for them.
As an example, we present explicit calculations of the six-point
one-loop NMHV amplitudes in the momentum-space representation in section 6.
In section 7, we further consider the application to one-loop N$^2$MHV
amplitudes and obtain new analytic expressions for them in a conjectural form.
We also briefly sketch how the application to one-loop non-MHV amplitudes
in general can be carried out.
Lastly, we present some concluding remarks.

%%%%%%%%%%%%%%%%%%%%%%%%%%%%%%%%%%%%%%%%%%%%%%%%%%%%%%%%%%%
\section{Tree amplitudes}

In this section, we review how the original Cachazo-Svrcek-Witten (CSW)
rules for tree amplitudes \cite{Cachazo:2004kj}
are implemented in the holonomy formalism \cite{Abe:2009kn}.
We first review the CSW rules briefly and then explicitly write down
an S-matrix functional for the tree amplitudes of gluons
in terms of a holonomy operator that is suitably defined in supertwistor space.
As an example, we show how to calculate the six-point NMHV tree amplitude
in our formulation.
We also discuss the color structure of the tree amplitudes in some detail.

\noindent
\underline{MHV amplitudes and spinor momenta}

In the spinor-helicity formalism, the simplest way of
describing gluon amplitudes is to factorize the amplitudes
into a set of MHV amplitudes.
The MHV amplitudes are the scattering amplitudes of $(n-2)$
positive-helicity gluons and $2$ negative-helicity gluons or the other way around.
In a momentum-space representation, the MHV tree amplitudes of gluons
are expressed as
\beqar
    \A_{MHV(0)}^{(1_+ 2_+ \cdots a_{-} \cdots b_{-} \cdots n_+ )} (u, \bu)
    & \equiv &
    \A_{MHV(0)}^{(a_{-} b_{-})} (u, \bu)
    \nonumber \\
    & = & i g^{n-2}
    \, (2 \pi)^4 \del^{(4)} \left( \sum_{i=1}^{n} p_i \right) \,
    \widehat{A}_{MHV(0)}^{(a_{-} b_{-})} (u)
    \label{2-1}\\
    \widehat{A}_{MHV(0)}^{(a_{-} b_{-})} (u)
    &=&
    \!\!\! \sum_{\si \in \S_{n-1}} \!\!
    \Tr (t^{c_1} t^{c_{\si_2}} t^{c_{\si_3}} \cdots t^{c_{\si_n}}) \,
    \widehat{C}_{MHV(0)}^{(a_{-} b_{-})} (u; \si )
    \label{2-2} \\
    \widehat{C}_{MHV(0)}^{(a_{-} b_{-})} (u; \si )
    & = &
    \frac{ (u_a u_b )^4}{ (u_1 u_{\si_2})(u_{\si_2} u_{\si_3})
    \cdots (u_{\si_n} u_1)}
    \label{2-3}
\eeqar
where $u_i$ denotes a two-component
spinor momentum of the $i$-th gluon ($i=1,2, \cdots, n$) and
$\bu_i$ denotes its complex conjugate.
In (\ref{2-1}), $a$ and $b$ label the numbering indices of
the two negative-helicity gluons,
$g$ denotes the Yang-Mills coupling constant, and $t^{c_i}$'s
denote the generators of a gauge group in the fundamental representation.
In the present paper, we consider an $SU(N)$ gauge theory, with
$t^{c_i}$'s given by the fundamental representation of the $SU(N)$ algebra.
In (\ref{2-2}), the sum is taken over the permutations labeled by
\beq
    \si=\left(%
    \begin{array}{c}
      2 ~ 3 ~ \cdots ~ n ~ \\
      \si_2 \si_3 \cdots \si_n \\
    \end{array}%
    \right)  .
    \label{2-4}
\eeq
Notice that a set of the permutations forms a rank-$(n-1)$ symmetric group $\S_{n-1}$.
This is why we denote the sum by that of $\si \in \S_{n-1}$ in (\ref{2-2}).

The product of the spinor momenta $( u_i u_j )$ and its conjugate are defined as
\beq
    u_i \cdot u_j \equiv (u_i u_j) =   \ep_{AB} u_{i}^{A}u_{j}^{B} \, , ~~~~~
    \bu_i \cdot \bu_j \equiv [\bu_i \bu_j]  = \ep^{\Ad \Bd} \bu_{i \, \Ad}
    \bu_{j \, \Bd}
    \label{2-5}
\eeq
where $\ep_{AB}$ $\ep^{\Ad \Bd}$ are the rank-2 Levi-Civita tensors, with
the upper-case indices taking the value of $(1, 2)$.
These tensors can also be used to raise or lower the indices, {\it e.g.},
$u_B = \ep_{AB}u^A$ and $\bu^{\Bd} = \ep^{\Ad \Bd} \bu_\Ad$.
In terms of the spinor momenta, the four-momentum
$p^{A \Ad}$ of a massless gluon is parametrized as
\beq
    p^{A \Ad} \, = \, p^\mu ( \tau_\mu )^{A \Ad} \, = \, u^A \bu^\Ad
    \label{2-6}
\eeq
where $\mu$ $(= 0,1,2,3 )$ is the ordinary Minkowski index and
$\tau_\mu$ is given by $\tau_\mu = ( {\bf 1} , \vec{\tau} )$,
with ${\bf 1}$ and $\vec{\tau}$ being the $( 2 \times 2)$ identity matrix
and the Pauli matrices, respectively.
The spinor momentum $u^A$ or $\bu^\Ad$ can be determined by
the null momentum $p^\mu$, with $p^2 = 0$, up to a phase factor.
An explicit choice of $u^A$ can be given by
\beq
    u^A = {1 \over \sqrt{p_0 - p_3}} \left(
            \begin{array}{c}
              {p_1 - i p_2} \\
              {p_0 - p_3} \\
            \end{array}
          \right) .
    \label{2-7}
\eeq
The Lorentz symmetry of
$u^A$'s is given by an $SL(2, {\bf C})$ group.
The product $(u_i u_j)$ in (\ref{2-5}) is invariant under
the Lorentz transformations.
Similarly, the Lorentz group of $\bu^\Ad$ is given by another $SL(2, {\bf C})$.
The four-dimensional Lorentz symmetry is then given by
a combination of these, {\it i.e.}, by $SL(2, {\bf C}) \times SL(2, {\bf C})$.

\noindent
\underline{Non-MHV amplitudes, reference spinors and the CSW rules}

The non-MHV gluon amplitudes, or the general gluon amplitudes, can be expressed in terms of
the MHV amplitudes $\widehat{A}_{MHV(0)}^{(a_{-} b_{-})} (u)$.
Prescription for such expressions is called the CSW rules \cite{Cachazo:2004kj}.
For the next-to-MHV (NMHV) amplitudes, which contain three negative-helicity
gluons and $(n-3)$ positive-helicity gluons, the CSW rules can be expressed as
\beq
    \widehat{A}^{( a_- b_- c_- )}_{NMHV(0)} (u) \, = \,\sum_{(i,j)}
    \widehat{A}^{(i_+ \cdots a_- \cdots b_- \cdots j_+ l_+)}_{MHV(0)} (u)
    \, \frac{1}{q_{ij}^2} \,
    \widehat{A}^{( (-l)_{-} \, (j+1)_+ \cdots c_- \cdots (i-1)_{+} )}_{MHV(0)} (u)
    \label{2-8}
\eeq
where {\it the sum is taken over all possible choices for $(i, j)$ that
satisfy the ordering $i \le a < b \le j < c$ (mod $n$) such that
the number of indices for
each of the MHV amplitudes is more than or equal to three.}
The numbering indices for the negative-helicity gluons are now given by
$a$, $b$ and $c$.
The momentum transfer $q_{ij}$ between the two MHV vertices can be
expressed by a partial sum of the gluon four-momenta:
\beq
    q_{ij}^{A \Ad} = p_{i}^{A \Ad} + p_{i+1}^{A \Ad} +
    \cdots + p_{a}^{A \Ad} + \cdots + p_{b}^{A \Ad} + \cdots + p_{j}^{A \Ad} \, .
    \label{2-9}
\eeq
The non-MHV amplitudes are then obtained by iterative use of
the relation (\ref{2-8}), given that the spinor momentum for the
index $l$ or $-l$ is defined by
\beq
    u_{ij}^{A} \, \equiv \, q_{ij}^{A \Ad} {\bar \eta}_\Ad ~~
    ( \, = \, u_{l}^{A} \, = \, - u_{-l}^{A} \, )
    \label{2-10}
\eeq
where $u_{ij}^{A}$ can be treated as an on-shell spinor momenta
corresponding to the off-shell momentum $q_{ij}$ or
the four-momentum of the virtual gluon.
${\bar \eta}_\Ad$ is an arbitrarily fixed two-component spinor
which behaves like $\bu$'s. Once fixed, this so-called reference spinor
should be identical for any $q_{ij}$'s in the expression (\ref{2-8}).
We shall call this requirement {\it the single reference-spinor principle}
in the tree-level CSW rules.
Notice that the parametrization (\ref{2-10}) is always possible,
which can be understood directly by decomposing $q_{ij}^{A \Ad}$ as
\beq
    q_{ij}^{A \Ad} \, = \, u_{ij}^{A} \bu_{ij}^{\Ad} \, + \, w
    \, \eta^A {\bar \eta}^\Ad \,
    \label{2-11}
\eeq
where $w$ is a real number.
From this decomposition, we can easily find the relation
$u_{ij}^{A} = q_{ij}^{A} {\bar \eta}_{\Ad} / [ \bu_{ij} {\bar \eta} ]$.
Thus, thanks to the scale invariance of the spinor momenta,
this relation naturally reduces to the definition (\ref{2-10}).

\noindent
\underline{Single-trace color structure in non-MHV amplitudes}

Now we assign a $U(1)$ direction of the $SU(N)$ gauge group
to the color factor of the propagator such that
the full algebra of the amplitudes becomes the $U(N)$ algebra.
Notice that the $U(1)$ part gives an auxiliary
degree of freedom here since the external legs are all
assigned by the $SU(N)$ generators in the fundamental representation.
The completeness relation for the generators of the $U(N)$ group is
conventionally given by
\beq
    ( t^{\hat c} )_{ij} ( t^{\hat c} )_{kl} \, = \, \hf \, \del_{il} \del_{jk}
    \label{2-12}
\eeq
where we take a sum over the $U(N)$ color index ${\hat c} = 1,2, \cdots , N^2$.
We here distinguish the $U(N)$ index ${\hat c}$ from the $SU(N)$
index $c = 1, 2, \cdots, N^{2} -1$.
The normalization factor $\hf$ can be absorbed into the definition of the
reference spinor.
Alternatively, we can redefine $q_{ij}^2$ in (\ref{2-8}) as
a square of $q_{ij}^{A \Ad}$ rather than that of $q_{ij}^{\mu}$.
Since the latter is twice as large as the former, the factor $\hf$ cancels
in the expression (\ref{2-8}). Notice that
the product of four-vectors, say, $v^{A \Ad}$ and $v^{\prime}_{\Ad A}$,
is related to that of $v^{\mu}$ and $v^{\prime}_{\mu}$ by
\beq
    v^{A \Ad} v^{\prime}_{\Ad A} =
    2( v_{0} v^{\prime}_{0} - v_{1} v^{\prime}_{1} -
    v_{2} v^{\prime}_{2} - v_{3} v^{\prime}_{3} )=2 v^{\mu} v^{\prime}_{\mu}
    \label{2-13}
\eeq
in general regardless whether the four-vectors are null or not.
{\it In any respect, the completeness condition (\ref{2-12}) guarantees
the single-trace property of the tree amplitudes.
A naive extension
of the CSW rules to loop amplitudes also preserves this property.
We shall take this feature for granted throughout the present paper.}

Assigning the $U(1)$ color index to the internal numbering indices
$l$ and $-l$ in the CSW rules (\ref{2-8}),
we can {\it in principle} express N$^k$MHV tree amplitudes
($k = 0, 1, 2, \cdots , n-4$) in general as
\beqar
    \A^{(1_{h_1} 2_{h_2} \cdots n_{h_n})}_{N^{k}MHV(0)} (u, \bu )
    & = &
    i g^{n-2}
    \, (2 \pi)^4 \del^{(4)} \left( \sum_{i=1}^{n} p_i \right) \,
    \widehat{A}^{(1_{h_1} 2_{h_2} \cdots n_{h_n})}_{N^{k}MHV(0)} (u)
    \label{2-14}
    \\
    \widehat{A}^{(1_{h_1} 2_{h_2} \cdots n_{h_n})}_{N^{k}MHV(0)} (u)
    &=&
    \sum_{\si \in \S_{n-1}}
    \Tr (t^{c_1} t^{c_{\si_2}} t^{c_{\si_3}} \cdots t^{c_{\si_n}}) ~
    \widehat{C}^{(1_{h_1} 2_{h_2} \cdots n_{h_n})}_{N^{k}MHV(0)} (u; \si )
    \label{2-15}
\eeqar
where $h_i = \pm$ denotes the helicity of the $i$-th gluon,
with the total number of negative helicities being $k+2$.
$\widehat{C}^{(1_{h_1} 2_{h_2} \cdots n_{h_n})}_{N^{k}MHV(0)} (u; \si )$
denotes a function of the Lorentz-invariant scalar products $(u_i u_j)$.
The simplest form of this function is given by the MHV case (\ref{2-3}).
By use of the CSW rules, we can obtain
$\widehat{C} (u; \si )$'s of non-MHV helicity configurations
in terms of (\ref{2-3}) and a set of off-shell momenta defined in (\ref{2-9}).
Accordingly, the permutations of the numbering indices
that involve the non-MHV amplitudes decreases, which
means that many of $\widehat{C} (u; \si )$'s vanish
for particular choices of $\si$. This is related
to the well-known redundancy of the expression (\ref{2-15})
due to the so-called $U(1)$ decoupling identities.
For example, the NMHV tree amplitudes (\ref{2-8}) can alternatively
be written as
\beqar
    \widehat{A}^{( a_- b_- c_- )}_{NMHV(0)} (u)
    & = &
    \sum_{i = 1}^{n} \sum_{r = 1}^{n-3}
    \widehat{A}^{( i_{+} \cdots a_{-} \cdots b_{-} \cdots (i+r)_{+} l_{+} )}_{MHV(0)} (u)
    \, \frac{1}{q_{i \, i+r}^2} \,
    \widehat{A}^{( (-l)_{-} \, (i+r+1)_{+} \cdots c_{-} \cdots (i-1)_{+} )}_{MHV(0)} (u)
    \nonumber \\
    &=&
    \sum_{i = 1}^{n}
    \sum_{r = 1}^{n - 3 }
    \sum_{ \si^{(1)} \in \S_{r+1}}
    \sum_{ \si^{(2)} \in \S_{n-r-1}} \!\!\!
    \Tr ( t^{\si^{(1)}_{i}} \cdots t^{\si^{(1)}_{i+r}} \,
    t^{\si^{(2)}_{i+r+1}} \cdots t^{\si^{(2)}_{i-1}} )
    \nonumber \\
    &&
    \left.
    \widehat{C}^{( i_{+} \cdots a_{-} \cdots b_{-} \cdots (i+r)_{+} l_{+} )}_{MHV(0)} (u)
    \, \frac{1}{q_{i \, i+r}^2} \,
    \widehat{C}^{( (-l)_{-} \, (i+r+1)_{+} \cdots c_{-} \cdots (i-1)_{+} )}_{MHV(0)} (u)
    \right|_{l = q_{i \, i+r} \bar{\eta}}
    \label{2-15a}
\eeqar
where we replace index $j$ in (\ref{2-8}) by $i+r$ with $r= 1,2, \cdots , n-3$.
Here the $SU(N)$ generators $t^{c_{\si_i}}$'s are abbreviated by
$t^{\si_i}$'s. To avoid unnecessary complexity, we shall use this notation from here on.
The expression (\ref{2-15a}) explicitly shows that the color structure
of the NMHV tree amplitudes is decomposed into
the two sums over the permutations $\si^{(1)}$ and $\si^{(2)}$, respectively.
This is essentially the same as the sum
over the so-called cyclicly ordered permutations (COP's)
used in the unitary-cut method \cite{Bern:1990ux,Bern:1991aq},
except that the single-trace color structure is preserved here
thanks to our abelian choice of internal degrees of freedom.

Lastly, we would like to emphasize that there are ambiguities in the expression (\ref{2-14})
due to the dependence on the reference spinors.
Once a single reference spinor is fixed, it should be identical for all propagators.
In other words, the non-MHV amplitudes are determined up to the choice of
the reference spinor or equivalently the reference null-vector $\eta^{A \Ad} = \eta^A
\bar{\eta}^{\Ad}$.
This fact is referred to as the single reference-spinor
principle in the present paper.
As shown in \cite{Cachazo:2004kj}, if we choose $\bar{\eta}^\Ad$ to be equal to
one of the $\bu_{i}^{\Ad}$'s, then we can obtain manifestly
Lorentz covariant expressions for the non-MHV amplitudes.

\noindent
\underline{A holonomy operator in supertwistor space}

From here on, we briefly review the construction of a holonomy operator
in supertwistor space as proposed in \cite{Abe:2009kn}.
An S-matrix functional for tree amplitudes
can be obtained in terms of this holonomy operator.
The holonomy operator is defined by
\beq
    \Theta_{R, \ga}^{(A)} (u; x, \th ) \, = \, \Tr_{R, \ga} \, \Path \exp \left[
    \sum_{m \ge 2} \oint_{\ga} \underbrace{A \wedge A \wedge \cdots \wedge A}_{m}
    \right]
    \label{2-16}
\eeq
where $A$ is what we call a comprehensive gauge one-form and is defined by the
following set of equations.
\beqar
    A &=&  g \sum_{1 \le i < j \le n} A_{ij} \, \om_{ij}
    \label{2-17} \\
    \om_{ij} & = & d \log(u_i u_j) = \frac{d(u_i u_j)}{(u_i u_j)}
    \label{2-18} \\
    A_{ij} &=& \sum_{\hat{h}_{i}} a_{i}^{( \hat{h}_{i} )} (x, \th) \otimes a_{j}^{(0)}
    \label{2-19} \\
    a_{i}^{( \hat{h}_{i} )} (x, \th) & = &
    \left. \int d \mu (p_i) ~ a_{i}^{(\hat{h}_{i})} (\xi_i) ~  e^{ i x_\mu p_{i}^{\mu} }
    \right|_{\xi_{i}^{\al} = \th_{A}^{\al} u_{i}^{A} }
    \label{2-20} \\
    d \mu (p_i) & \equiv &
    \frac{d^3 p_i}{(2 \pi)^3} \frac{1}{2 p_{i  0}} ~ = ~
    \frac{1}{4} \left[
    \frac{u_i \cdot d u_i}{2 \pi i} \frac{d^2 \bu_i}{(2 \pi)^2} -
    \frac{\bu_i \cdot d \bu_i }{2 \pi i} \frac{d^2 u_i }{(2 \pi)^2}
    \right]
    \label{2-21}
\eeqar
$a_{i}^{(\hat{h}_{i})} (\xi_i)$'s are physical operators
that are defined in a four-dimensional $\N = 4$ chiral
superspace $(x, \th)$ where $x_{\Ad A}$ denotes coordinates
of four-dimensional spacetime and $\th_{A}^{\al}$ $(A = 1,2; \al = 1,2,3,4)$
denotes their chiral superpartners with $\N = 4$ extended supersymmetry.
As explicitly discussed in \cite{Abe:2009kn},
these coordinates emerges from homogeneous coordinates
of the supertwistor space $\cp^{3|4}$, denoted by $( u^A , v_\Ad , \xi^\al )$,
that satisfy the so-called supertwistor conditions
\beq
    v_\Ad \, = \, x_{\Ad A} u^A \, , ~~~
    \xi^\al \, = \, \th_{A}^{\al} u^A \, .
    \label{2-22}
\eeq
Notice that for the ordinary twistor space its homogeneous coordinate
is given by $Z_I = ( u^A , v_\Ad )$ ($I = 1,2,3,4$)
where $u^A$ and $v_\Ad$ are two-component
complex spinors, with $Z_I$ satisfying the scale invariance.
The spacetime coordinates emerges from the first condition in (\ref{2-22}).
As in (\ref{2-6}), the Minkowski coordinates $x_\mu$ can easily be related to $x_{\Ad A}$
via $x_{\Ad A} = x_\mu ( \tau^\mu )_{A \Ad}$.
In the spinor-helicity formalism in supertwistor space, we identify $u^A$ as
the spinor momentum defined in (\ref{2-7}) so that we can essentially describe
four-dimensional physics in terms of $u^A$'s with an imposition of
the supertwistor conditions (\ref{2-22}).

The physical operators $a_{i}^{(\hat{h}_{i})} (\xi_i)$ are relevant to
creations of gluons and their superpartners, having the helicity
$\hat{h}_{i} = (0, \pm \hf , \pm 1 )$.
In the two-component indices, we can properly define the Pauli-Lubanski spin vector
for the $i$-th gluon as
\beq
    S_{i}^{A \Ad} \, = \, p_{i}^{A \Ad}
    \left( 1 - \frac{1}{2} u_i^B \frac{\d}{\d u_i^B} \right) \, .
    \label{2-23}
\eeq
This means that the helicity $\hat{h}_{i}$ of the $i$-th particle or supermultiplet
can be determined by
\beq
    \hat{h}_{i} = 1 - \hf  u_{i}^{A} \frac{\d}{\d u_{i}^{A}} \, .
    \label{2-24}
\eeq
In other words, the helicity is essentially given by the degree of homogeneity
in $\xi_i^\al$'s for each of the physical operators $a_{i}^{(\hat{h}_{i})} (\xi_i)$.
Thus we can write down explicit forms of $a_{i}^{(\hat{h}_{i})} (\xi_i)$'s as below.
\beqar
    \nonumber
    a_{i}^{(+)} (\xi_i) &=& a_{i}^{(+)} \\ \nonumber
    a_{i}^{\left( + \frac{1}{2} \right)} (\xi_i) &=& \xi_{i}^{\al}
    \, a_{i \, \al}^{ \left( + \frac{1}{2} \right)} \\
    a_{i}^{(0)} (\xi_i) &=& \hf \xi_{i}^{\al} \xi_{i}^{\bt} \, a_{i \, \al \bt}^{(0)}
    \label{2-25}
    \\ \nonumber
    a_{i}^{\left( - \frac{1}{2} \right)} (\xi_i) &=&
    \frac{1}{3!} \xi_{i}^{\al}\xi_{i}^{\bt}\xi_{i}^{\ga}
    \ep_{\al \bt \ga \del} \, {a_{i}^{ \left( - \frac{1}{2} \right)}}^{ \del}
    \\ \nonumber
    a_{i}^{(-)} (\xi_i) &=& \xi_{i}^{1} \xi_{i}^{2} \xi_{i}^{3} \xi_{i}^{4} \, a_{i}^{(-)}
\eeqar
The color factor can be attached to each of the physical operators $a_{i}^{( \hat{h} )}$:
\beq
    a_{i}^{( \hat{h}_i )} \, = \, t^{c_i} a_{i}^{( \hat{h}_i ) c_i}
    \label{2-26}
\eeq
where, as in the case of (\ref{2-2}), $t^{c_i}$'s are given by
the generators of the $SU(N)$ gauge group in the fundamental representation.
The color part of the symbol $R$ in (\ref{2-16})
then refers to the fundamental representation of the $SU(N)$ algebra
in this paper.\footnote{
Notice that the symbol $R$ also refers to an irreducible representation
of the Iwahori-Hecke algebra in the holonomy formalism \cite{Abe:2009kq}.
As shown in (\ref{2-31}),
this can be materialized by a sum over permutations over the numbering indices.
In this sense, we can practically regard the symbol $R$ as the fundamental representation
of the gauge group.
For the emergence of the Iwahori-Hecke algebra as an irreducible representation
of the braid group, one may refer to \cite{Kohno:2002bk,Ueno:2008bk}.
}

The physical Hilbert space of the holonomy formalism
is given by $V^{\otimes n} = V_1 \otimes V_2 \otimes \cdots \otimes V_n$
where $V_ i$ $(i=1,2,\cdots,n)$ denotes a Fock space that creation operators
of the $i$-th particle with helicity $\pm$ act on.
The creation operators of gluons and their scalar partner form
an $SL(2, {\bf C})$ algebra:
\beq
    [ a_{i}^{(+)}, a_{j}^{(-)}] = 2 a_{i}^{(0)} \, \del_{ij}  \, , ~~~
    [ a_{i}^{(0)}, a_{j}^{(+)}] = a_{i}^{(+)} \, \del_{ij} \, , ~~~
    [ a_{i}^{(0)}, a_{j}^{(-)}] = - a_{i}^{(-)} \, \del_{ij}
    \label{2-27}
\eeq
where Kronecker's deltas show that
the non-zero commutators are obtained only when $i = j$.
The remaining of commutators, those expressed otherwise, all vanish.

\noindent
\underline{Non-supersymmetric part of the holonomy operator}

In the following, we consider the gluonic part of the holonomy operator in (\ref{2-16});
as we shall discuss later, contributions from gluinos and scalar partners to
the MHV vertices vanish upon Grassmann integrals.
To distinguish the gluon operators from the super multiplets,
we denote the former by $a_{i}^{( h_i )}$ with $h_{i} = \pm 1 = \pm $
instead of $a_{i}^{( \hat{h} )}$
with $\hat{h}_{i} = (0, \pm \hf , \pm 1 )$.
The physical configuration space for gluons can be defined
in terms of a set of $\cp^1$'s on which the gluonic spinor momenta are defined.
Since gluons are bosons, these $\cp^1$'s are symmetric under permutations.
The physical configuration space is therefore given by $\C = {\bf C}^n / \S_n$.
The fundamental homotopy group of $\C$ is given by the braid group,
$\Pi_1 (\C) = \B_n$.
The symbol $\ga$  in (\ref{2-16}) denotes a closed path defined in $\C$.
Linear transformations of the holonomy operator depends on the homotopy class
of the closed path $\ga$ on $\C = {\bf C}^n / \S_n$. Thus the holonomy
operator gives a linear representation of the braid group $\B_n$.

The symbol $\Path$ in (\ref{2-16}) denotes a ``path'' ordering
of the numbering indices along $\ga$.
The meaning of $\Path$ acting on the exponent
of (\ref{2-16}) can explicitly be written as
\beqar
    \nonumber
    \Path \sum_{m \ge 2}  \oint_{\ga} \underbrace{A \wedge A \wedge \cdots \wedge A}_{m}
    &=& \sum_{m \ge 2} \oint_{\ga}  A_{1 2} A_{2 3} \cdots A_{m 1}
    \, \om_{12} \wedge \om_{23} \wedge \cdots \wedge \om_{m 1} \\
    \nonumber
    &=& \sum_{m \ge 2}  \frac{1}{2^{m+1}} \sum_{(h_1, h_2, \cdots , h_m)}
    (-1)^{h_1 + h_2 + \cdots + h_m} \\
    && ~~~ \times \,
    a_{1}^{(h_1)} \otimes a_{2}^{(h_2)} \otimes \cdots \otimes a_{m}^{(h_m)}
    \, \oint_{\ga} \om_{12} \wedge \cdots \wedge \om_{m1}
    \label{2-28}
\eeqar
where we consider only the gluonic part so that
$h_{i} = \pm = \pm 1$ ($i=1,2,\cdots, m$) denotes the helicity of the $i$-th gluon.
In obtaining the above expression, we use an ordinary definition of commutators for
bialgebraic operators. For example, using the commutation relations (\ref{2-27}),
we can calculate $[ A_{12} , A_{23} ]$ as
\beqar
    [ A_{12}, A_{23}]
    &=& a_{1}^{(+)} \otimes a_{2}^{(+)} \otimes a_{3}^{(0)}
    - a_{1}^{(+)} \otimes a_{2}^{(-)} \otimes a_{3}^{(0)}
    \nonumber\\
    && \!\!\! + \, a_{1}^{(-)} \otimes a_{2}^{(+)} \otimes a_{3}^{(0)}
    - a_{1}^{(-)} \otimes a_{2}^{(-)} \otimes a_{3}^{(0)} \, .
    \label{2-29}
\eeqar
In (\ref{2-28}), we also define $a_{1}^{(\pm)} \otimes a_{2}^{(h_2)}
\otimes \cdots \otimes a_{m}^{(h_m)} \otimes a_{1}^{(0 )}$ as
\beqar
    a_{1}^{( \pm )} \otimes a_{2}^{(h_2)} \otimes \cdots \otimes a_{m}^{(h_m)} \otimes a_{1}^{(0 )}
    & \equiv & \hf [ a_{1}^{(0)} , a_{1}^{(\pm )} ] \otimes a_{2}^{(h_2 )}
    \otimes \cdots \otimes a_{m}^{(h_m)}
    \nonumber \\
    &=&
    \pm \hf a_{1}^{( \pm )} \otimes a_{2}^{(h_2)} \otimes \cdots \otimes a_{m}^{(h_m)}
    \label{2-30}
\eeqar
where we implicitly use an antisymmetric property for the indices due to the wedge products
in (\ref{2-28}).

The trace $\Tr_{R, \ga}$ in the definition (\ref{2-16}) means
a trace over the color factors of the gluons.
As discussed in \cite{Abe:2009kq},
this trace includes not only a trace over the $SU(N)$ generators
but also that of braid generators.
The latter, a so-called braid trace, is realized by
a sum over permutations of the numbering indices.
Thus the trace $\Tr_{R, \ga}$ over the non-supersymmetric
exponent of (\ref{2-16}) can be expressed as
\beq
    \Tr_{R, \ga} \Path \sum_{m \ge 2} \oint_{\ga}
    \underbrace{A \wedge A \wedge \cdots \wedge A}_{m}
    = \sum_{m \ge 2}
    \sum_{\si \in \S_{m-1}} \oint_{\ga}  A_{1 \si_{2}}
    A_{\si_{2} \si_{3}} \cdots A_{\si_{m} 1}
    \, \om_{1 \si_2} \wedge \om_{\si_{2} \si_{3}} \wedge \cdots \wedge
    \om_{\si_{m} 1}
    \label{2-31}
\eeq
where the sum of $\si \in \S_{m-1}$ is taken over the
permutations
$\si = \left(%
\begin{array}{c}
  2 ~ 3 ~ \cdots ~ m \\
  \si_{2} \si_{3} \cdots \si_{m} \\
\end{array}%
\right)$.

From (\ref{2-28}) and (\ref{2-31}), we find that the holonomy operator
can be described in terms of the logarithmic one-forms $\om_{i \,i+1}$
and the permutations of the numbering indices.
This fact suggests the dual conformal invariance \cite{Drummond:2008vq} of
the holonomy operator.
As mentioned above, the holonomy operator gives
a linear representation of the braid group $\B_n$.
Thus, by use of the Kohno-Drinfel'd monodromy theorem \cite{Chari:1994pz},
one can argue that the holonomy operator and hence
the tree amplitudes in general preserve the Yangian symmetry \cite{Drummond:2009fd}.
Interested readers may refer to \cite{Abe:2009kn} for details on this topic.

\noindent
\underline{An S-matrix functional for tree amplitudes}

In terms of the supersymmetric holonomy operator (\ref{2-16}),
we can construct an S-matrix functional for the MHV tree amplitudes as
\beq
    \F_{MHV} \left[ a^{(h)c} \right]
    ~ = ~ \exp \left[ \frac{i}{g^2} \int d^4 x d^8 \th
    ~ \Theta_{R, \ga}^{(A)} (u; x ,\th) \right]
    \label{2-32}
\eeq
where $a^{(h)c}$ denotes a generic expression for $a_{i}^{(h_i)c_i}$.
In terms of $\F_{MHV} \left[ a^{(h)c} \right]$,
the MHV tree amplitudes (\ref{2-2}) are generated as
\beqar
    \nonumber
    &&
    \frac{\del}{\del a_{1}^{(+) c_1} (x_{1}) } \otimes
    \cdots \otimes \frac{\del}{\del a_{a}^{(-) c_a} (x_{a})} \otimes \cdots
    \\
    \nonumber
    && ~~~~~~~~~~~~~~~~
    \left. \cdots \otimes \frac{\del}{\del a_{b}^{(-) c_b} (x_{b})} \otimes
    \cdots \otimes \frac{\del}{\del a_{n}^{(+) c_n} (x_{n})}
    ~ \F_{MHV} \left[ a^{(h)c} \right]
     \right|_{a^{(h)c} ( x ) =0}
    \\
    & = &
    i g^{n-2} \, \widehat{A}_{MHV(0)}^{(a_{-} b_{-})} (u)
    \label{2-33}
\eeqar
where $a^{(h)c} ( x )$'s denote $x$-space representation of
the gluon creation operators:
\beq
    a_{i}^{(h_i)} (x) =
    \int d \mu (p_i)
    \,  a_{i}^{(h_i)} \,  e^{i x_\mu p_{i}^{\mu} }
    \label{2-34}
\eeq
where $d \mu (p_i)$ is the Nair measure (\ref{2-21}).

The expression (\ref{2-33}) can easily be checked with the following two relations.
First, we choose the normalization of the spinor momenta as
\beq
    \oint_\ga d(u_1 u_2) \wedge d(u_2 u_3) \wedge \cdots \wedge d (u_m u_1) = 2^{m+1}
    \label{2-35}
\eeq
Under a permutation of the numbering indices, a sign factor arises in
the above expression. We omit this sign factor as well as the factor
$(-1)^{h_1 + h_2 + \cdots + h_n}$ in (\ref{2-28})
since physical quantities are given by the squares of the amplitudes.
Secondly, the Grassmann integral over $\th$'s vanishes unless we have
the following integrand:
\beq
    \left. \int d^8 \th  \, \xi_{r}^{1}\xi_{r}^{2}\xi_{r}^{3}\xi_{r}^{4}
    \, \xi_{s}^{1}\xi_{s}^{2}\xi_{s}^{3}\xi_{s}^{4}
    \right|_{\xi_{i}^{\al} = \th_{A}^{\al} u_{i}^{A} }
    = \, (u_r u_s )^4  \, .
    \label{2-36}
\eeq
This relation guarantees that only the MHV amplitudes are picked up
upon the execution of the Grassmann integral.

We can therefore naturally express the MHV S-matrix functional
$\F_{MHV}\left[ a^{(h)c} \right]$ in terms of
the supersymmetric holonomy operator $\Theta_{R, \ga}(u; x, \th)$.
We can also express an S-matrix functional for non-MHV tree amplitudes in general
by use of  $\F_{MHV} \left[ a^{(h)c} \right]$
if we manage to realize the CSW rules in a functional language.
Such a general S-matrix functional, denoted by $\F \left[ a^{(h)c} \right]$,
can be expressed as
\beqar
    \F \left[ a^{(h)c} \right] &= &
    \widehat{W}^{(A)} \, \F_{MHV} \left[ a^{(h)c} \right] \, ,
    \label{2-37}\\
    \widehat{W}^{(A)} &=& \exp \left[ -i
    \int d^4 x d^4 y \, \frac{1}{q^2} \,
    \frac{\del}{\del a_{l}^{(+)}(x)} \otimes
    \frac{\del}{\del a_{-l}^{(-)}(y)}
    \right] \, .
    \label{2-38}
\eeqar
In terms of $\F \left[ a^{(h)c} \right]$, the general tree amplitudes
$\widehat{A}^{(1_{h_1} 2_{h_2} \cdots n_{h_n})}_{N^{k}MHV(0)} (u )$
in (\ref{2-15}) can be generated as
\beqar
    &&
    \nonumber
    \left. \frac{\del}{\del a_{1}^{(h_1) c_1} (x_1) } \otimes
    \frac{\del}{\del a_{2}^{(h_2) c_2} (x_2)} \otimes
    \cdots \otimes \frac{\del}{\del a_{n}^{(h_n) c_n} (x_n)}
    ~ \F \left[  a^{(h)c}  \right] \right|_{a^{(h)c} (x)=0} \\
    &=&
    i g^{n-2} \widehat{A}^{(1_{h_1} 2_{h_2} \cdots n_{h_n})}_{N^{k}MHV(0)} (u )
    \label{2-39}
\eeqar
where a set of $h_i = \pm$ $(i = 1, 2, \cdots , n)$
are arbitrarily chosen on condition that they contain $k+2$ negative helicities in total.
The condition $a^{(h)} (x) = 0$ means that the remaining operators
(or, to be precise, source functions) should be
evaluated as zero in the end of the calculation.

The Wick-like contraction operator $\widehat{W}^{(A)}$ in (\ref{2-38}) is introduced
so that the CSW rules are realized in a functional method.
The functional derivatives in (\ref{2-39}), with a help of the
Grassmann integral (\ref{2-36}) automatically lead to the summation over
$(i,j)$ in (\ref{2-8}); this is why we denote the momentum transfer
by $q$, without the $(i, j)$ indices, in (\ref{2-38}).

The Grassmann integral (\ref{2-36}) guarantees that gluon amplitudes
vanish unless the helicity configuration can be factorized into
the MHV helicity configurations.
Thus the CSW rules are automatically satisfied by the Grassmann integral.
In this respect, we can specify the helicity index by
$h_i = (+ , -)$, rather than the supersymmetric version
$\hat{h}_i = (0, \pm \frac{1}{2} , \pm )$.
{\it In the functional formulation of the tree amplitudes,
the essence of the CSW rules therefore lies in the use of
Grassmann integral (\ref{2-36}).}
This feature would and should be held in loop calculations as well, since
the formula (\ref{2-39}) can naturally apply to loop amplitudes,
with an appropriate understanding of loop-integral measures.
We shall consider this issue in the next section.

\noindent
\underline{Six-point NMHV tree amplitudes}

\begin{figure} [htbp]
\begin{center}
\includegraphics[width=125mm]{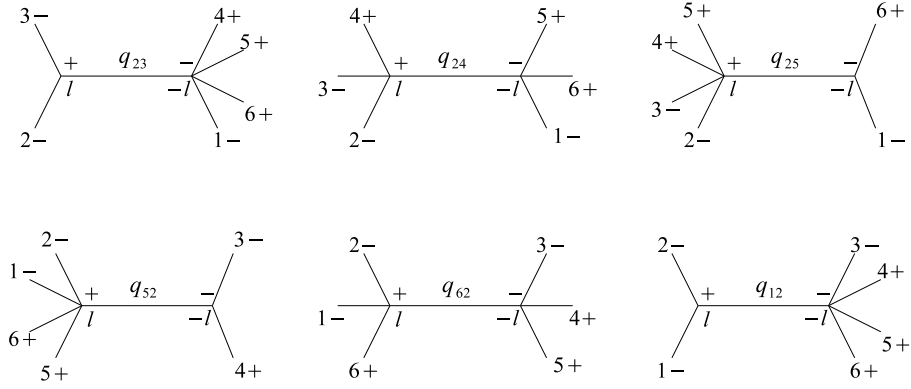}
\caption{Diagrams contributing to the six-gluon NMHV tree amplitude
$\widehat{A}^{(1_{-} 2_{-} 3_{-} 4_{+} 5_{+} 6_{+} )}_{NMHV} (u)$ }
\label{fighol0401}
\end{center}
\end{figure}

As an example of our formulation, we now calculate
the six-point NMHV tree amplitudes, the simplest non-MHV tree amplitudes.
All six-point NMHV amplitudes can be obtained by applying cyclic
permutations and reflections to the three distinct helicity
configurations $( 1_- 2_- 3_- 4_+ 5_+ 6_+ )$,
$( 1_- 2_+ 3_- 4_+ 5_- 6_+ )$ and $( 1_- 2_+ 3_+ 4_- 5_- 6_+ )$.
Using the expression (\ref{2-39}) for the first
configuration, we can explicitly carry out the
functional derivatives as
\beqar
    &&
    \frac{\del}{\del a_{1}^{(-) c_1} (x_1) } \otimes
    \frac{\del}{\del a_{2}^{(-) c_2} (x_2)} \otimes
    \frac{\del}{\del a_{3}^{(-) c_3} (x_3)}
    \nonumber \\
    && ~~~~~~~~~~~~~~~~~ \otimes
    \left.
    \frac{\del}{\del a_{4}^{(+) c_4} (x_4)} \otimes
    \frac{\del}{\del a_{5}^{(+) c_5} (x_5)} \otimes
    \frac{\del}{\del a_{6}^{(+) c_6} (x_6)}
    ~ \F \left[  a^{(h)c}  \right] \right|_{a^{(h)c} (x)=0}
    \nonumber \\
    &=&
    \frac{\del}{\del a_{1}^{(-) c_1} (x_1) } \otimes
    \cdots \otimes
    \frac{\del}{\del a_{6}^{(+) c_6} (x_6)}
    \nonumber \\
    &&
    \left.
    \, (-i)  \int_{x, y} \, \frac{1}{q^2} \,
    \frac{\del}{\del a_{k}^{(+)}(x)} \otimes
    \frac{\del}{\del a_{l}^{(-)}(y)} \,
    \exp \left[ \frac{i}{g^2} \int_{x , \th} \,
    \Theta_{R, \ga}^{(A)} (u; x ,\th) \right]
    \right|_{a^{(h)c} (x)=0}
    \nonumber \\
    &=&
    ig^{4}  \, \left[ ~
    \widehat{A}^{( 2_{-} 3_{-} l_{+} )}_{MHV(0)} (u)
    \, \frac{1}{q_{23}^2} \,
    \widehat{A}^{( (-l)_{-} 4_{+} 5_{+} 6_{+} 1_{-} )}_{MHV(0)} (u)
    \, + \,
    \widehat{A}^{( 2_{-} 3_{-} 4_{+} l_{+} )}_{MHV(0)} (u)
    \, \frac{1}{q_{24}^2} \,
    \widehat{A}^{( (-l)_{-} 5_{+} 6_{+} 1_{-} )}_{MHV(0)} (u)
    \right.
    \nonumber \\
    && ~~~~~~~~
    \, + \,
    \widehat{A}^{( 2_{-} 3_{-} 4_{+} 5_{+} l_{+} )}_{MHV(0)} (u)
    \, \frac{1}{q_{25}^2} \,
    \widehat{A}^{( (-l)_{-} 6_{+} 1_{-} )}_{MHV(0)} (u)
    \, + \,
    \widehat{A}^{( 5_{+} 6_{+} 1_{-} 2_{-} l_{+} )}_{MHV(0)} (u)
    \, \frac{1}{q_{52}^2} \,
    \widehat{A}^{( (-l)_{-} 3_{-} 4_{+} )}_{MHV(0)} (u)
    \nonumber \\
    && ~~~~~~~~  \left.
    \, + \,
    \widehat{A}^{( 6_{+} 1_{-} 2_{-} l_{+} )}_{MHV(0)} (u)
    \, \frac{1}{q_{62}^2} \,
    \widehat{A}^{( (-l)_{-} 3_{-} 4_{+} 5_{+} )}_{MHV(0)} (u)
    \, + \,
    \widehat{A}^{( 1_{-} 2_{-} l_{+} )}_{MHV(0)} (u)
    \, \frac{1}{q_{12}^2} \,
    \widehat{A}^{( (-l)_{-} 3_{-} 4_{+} 5_{+} 6_{+} )}_{MHV(0)} (u)
    ~ \right]
    \nonumber \\
    &=&
    i g^{4} \widehat{A}^{(1_{-} 2_{-} 3_{-} 4_{+} 5_{+} 6_{+} )}_{NMHV(0)} (u)
    \label{2-40}
\eeqar
where in the second lines we use an abbreviated notation for the integral measures over
the spacetime coordinates and the Grassmann variables.
In the third lines, we omit the supplemental conditions $l = q_{ij} {\bar \eta}$, which
can easily read off form each term.
Those terms that contribute to
$\widehat{A}^{(1_{-} 2_{-} 3_{-} 4_{+} 5_{+} 6_{+} )}_{NMHV} (u)$ are
diagrammatically shown in Figure \ref{fighol0401}.

%%%%%%%%%%%%%%%%%%%%%%%%% figure %%%%%%%%%%%%%%%%%%%%%%%%%
%%%%%%%%%%%%%%%%%%%%%%%%% figure %%%%%%%%%%%%%%%%%%%%%%%%%

Similarly, for the remaining helicity configurations
we can calculate the six-point tree amplitudes as
\beqar
    &&
    \!\!\!
    \widehat{A}^{(1_{-} 2_{+} 3_{-} 4_{+} 5_{-} 6_{+} )}_{NMHV(0)} (u)
    \nonumber \\
    &=&
    \!\!\!
    \widehat{A}^{( 1_{-} 2_{+} 3_{-} l_{+} )}_{MHV(0)} (u)
    \, \frac{1}{q_{13}^2} \,
    \widehat{A}^{( (-l)_{-} 4_{+} 5_{-} 6_{+}  )}_{MHV(0)} (u)
    \, + \,
    \widehat{A}^{( 1_{-} 2_{+} 3_{-} 4_{+} l_{+} )}_{MHV(0)} (u)
    \, \frac{1}{q_{14}^2} \,
    \widehat{A}^{( (-l)_{-} 5_{-} 6_{+}  )}_{MHV(0)} (u)
    \nonumber \\
    &&
    \!\!\!
    + \,
    \widehat{A}^{( 6_{+} 1_{-} 2_{+} 3_{-} l_{+} )}_{MHV(0)} (u)
    \, \frac{1}{q_{63}^2} \,
    \widehat{A}^{( (-l)_{-} 4_{+} 5_{-} )}_{MHV(0)} (u)
    \, + \,
    \widehat{A}^{( 5_{-} 6_{+} 1_{-} l_{+} )}_{MHV(0)} (u)
    \, \frac{1}{q_{51}^2} \,
    \widehat{A}^{( (-l)_{-} 2_{+} 3_{-} 4_{+} )}_{MHV(0)} (u)
    \nonumber \\
    &&
    \!\!\!
    + \,
    \widehat{A}^{( 5_{-} 6_{+} 1_{-} 2_{+} l_{+} )}_{MHV(0)} (u)
    \, \frac{1}{q_{52}^2} \,
    \widehat{A}^{( (-l)_{-} 3_{-} 4_{+} )}_{MHV(0)} (u)
    \, + \,
    \widehat{A}^{( 4_{+} 5_{-} 6_{+} 1_{-} l_{+} )}_{MHV(0)} (u)
    \, \frac{1}{q_{41}^2} \,
    \widehat{A}^{( (-l)_{-} 2_{+} 3_{-}  )}_{MHV(0)} (u)
    \nonumber \\
    &&
    \!\!\!
    + \,
    \widehat{A}^{(  3_{-} 4_{+} 5_{-} l_{+} )}_{MHV(0)} (u)
    \, \frac{1}{q_{35}^2} \,
    \widehat{A}^{( (-l)_{-} 6_{+} 1_{-} 2_{+} )}_{MHV(0)} (u)
    \, + \,
    \widehat{A}^{(  3_{-} 4_{+} 5_{-} 6_{+} l_{+} )}_{MHV(0)} (u)
    \, \frac{1}{q_{36}^2} \,
    \widehat{A}^{( (-l)_{-} 1_{-} 2_{+} )}_{MHV(0)} (u)
    \nonumber \\
    &&
    \!\!\!
    + \,
    \widehat{A}^{( 2_{+} 3_{-} 4_{+} 5_{-} l_{+} )}_{MHV(0)} (u)
    \, \frac{1}{q_{25}^2} \,
    \widehat{A}^{( (-l)_{-} 6_{+} 1_{-} )}_{MHV(0)} (u) \,  ,
    \label{2-41}\\
    &&
    \!\!\!
    \widehat{A}^{(1_{-} 2_{+} 3_{+} 4_{-} 5_{-} 6_{+} )}_{NMHV(0)} (u)
    \nonumber \\
    &=&
    \!\!\!
    \widehat{A}^{( 1_{-} 2_{+} 3_{+} 4_{-} l_{+} )}_{MHV(0)} (u)
    \, \frac{1}{q_{14}^2} \,
    \widehat{A}^{( (-l)_{-} 5_{-} 6_{+}  )}_{MHV(0)} (u)
    \, + \,
    \widehat{A}^{( 5_{-} 6_{+} 1_{-} l_{+} )}_{MHV(0)} (u)
    \, \frac{1}{q_{51}^2} \,
    \widehat{A}^{( (-l)_{-} 2_{+} 3_{+} 4_{-} )}_{MHV(0)} (u)
    \nonumber \\
    &&
    \!\!\!
    + \,
    \widehat{A}^{( 5_{-} 6_{+} 1_{-} 2_{+} l_{+} )}_{MHV(0)} (u)
    \, \frac{1}{q_{52}^2} \,
    \widehat{A}^{( (-l)_{-} 3_{+} 4_{-} )}_{MHV(0)} (u)
    \, + \,
    \widehat{A}^{( 4_{-} 5_{-} l_{+} )}_{MHV(0)} (u)
    \, \frac{1}{q_{45}^2} \,
    \widehat{A}^{( (-l)_{-} 6_{+} 1_{-} 2_{+} 3_{+} )}_{MHV(0)} (u)
    \nonumber \\
    &&
    \!\!\!
    + \,
    \widehat{A}^{( 4_{-} 5_{-} 6_{+} l_{+} )}_{MHV(0)} (u)
    \, \frac{1}{q_{46}^2} \,
    \widehat{A}^{( (-l)_{-} 1_{-} 2_{+} 3_{+} )}_{MHV(0)} (u)
    \, + \,
    \widehat{A}^{( 3_{+} 4_{-} 5_{-} l_{+} )}_{MHV(0)} (u)
    \, \frac{1}{q_{35}^2} \,
    \widehat{A}^{( (-l)_{-} 6_{+} 1_{-} 2_{+} )}_{MHV(0)} (u)
    \nonumber \\
    &&
    \!\!\!
    + \,
    \widehat{A}^{( 3_{+} 4_{-} 5_{-} 6_{+} l_{+} )}_{MHV(0)} (u)
    \, \frac{1}{q_{36}^2} \,
    \widehat{A}^{( (-l)_{-} 1_{-} 2_{+} )}_{MHV(0)} (u)
    \, + \,
    \widehat{A}^{( 2_{+} 3_{+} 4_{-} 5_{-} l_{+} )}_{MHV(0)} (u)
    \, \frac{1}{q_{25}^2} \,
    \widehat{A}^{( (-l)_{-} 6_{+} 1_{-} )}_{MHV(0)} (u)
    \label{2-42}
\eeqar
where, as in the case of (\ref{2-40}),
the spinor momenta $u_l$, $u_{-l}$ are
defined through the corresponding $q_{ij}$ as (\ref{2-9}).

From Figure \ref{fighol0401}, we can easily find that
the number of the contractions due to the operator
$\widehat{W}^{(A)}$ corresponds to the number of negative-helicity
states minus two.
Notice that the power of $g$ is independent of the helicity configuration;
it depends only on the total number of scattering gluons.
This feature is inherent in all tree amplitudes since
iterative use of contraction operator $\widehat{W}^{(A)}$
does not alter the power of $g$.
The above example (\ref{2-40}) illustrates that the functional formula
(\ref{2-39}) elegantly materializes
the CSW rules by use of the Grassmann integral (\ref{2-36}).

Lastly, we would like to emphasize the characteristic
feature of tree amplitudes in our formulation of the CSW rules,
namely, the single-trace color structure of the amplitudes
furnished with sums over distinct permutations of the numbering indices.
The sums arise from the braid trace
that is inherent in the definition of the holonomy
operator (\ref{2-16}) and, hence, in the S-matrix functional (\ref{2-37})
for general gluon amplitudes.
The number of such sums corresponds to the number of negative-helicity
gluons minus one.
The single-trace color structure is guaranteed by
assigning the $U(1)$ color factor to the internal propagators
that connect two MHV vertices.
As we discuss in the following sections, this feature
is pertinent to one-loop amplitudes as well.
Notice that there are similarity and difference
between our functional formulation and the known unitary-cut
method \cite{Bern:1990ux,Bern:1991aq}.
The similarity is that the sums over permutations
in the former is basically the same as the sum over what is called
the cyclicly ordered permutations (COP's) in the latter,
while the difference is that the former has a single-trace
property but the latter has multi-trace decomposition
in terms of the color structure of one-loop amplitudes.
These are essential similarity and difference between
a CSW-based method and a non-CSW method in the calculation
of gluon amplitudes.
{\it In this context, our formulation is
qualitatively different from the unitary-cut method.
Since the unitary-cut method borrows a computational technique from
string theory (or, to be more precise, four-dimensional
heterotic string theory), we can alternatively state that
our resultant amplitudes are qualitatively
different from bosonic open-string amplitudes in structure.}
% this suggests non-planarity of our amplitudes
% say later in sec 6 & 8

%%%%%%%%%%%%%%%%%%%%%%%%%%%%%%%%%%%%%%%%%%%%%%%%%%%%%%%%%%%
\section{One-loop MHV amplitudes: functional derivation}

In this section, we generalize the above formulation to one-loop MHV amplitudes.
We first consider an off-shell continuation of the Nair measure
$d \mu ( p )$ defined in (\ref{2-21}).
We then reformulate the S-matrix functional in an $x$-space representation
such that it incorporates the off-shell Nair measures.
We find that such reformulation naturally leads to
the Brandhuber-Spence-Travaglini (BST) representation of
one-loop MHV amplitudes \cite{Brandhuber:2004yw}
in a functional method.

\noindent
\underline{Off-shell continuation of the Nair measure}

The spinor momentum $u^A$ in (\ref{2-7}) can be
regarded as a homogeneous coordinate of the complex projective space $\cp^1$.
Thus it can be parametrized as
\beq
    u^A = \frac{1}{\sqrt{p_0 - p_3}} \left(
            \begin{array}{c}
              {p_1 - i p_2} \\
              {p_0 - p_3} \\
            \end{array}
          \right)
    = \al \left(
            \begin{array}{c}
              1 \\
              z \\
            \end{array}
          \right),
    ~~~~  \al \in {\bf C} - \{ 0 \}
    \label{3-1}
\eeq
where $z$ represents a local complex coordinate of $\cp^1$,
with $\al$ being a non-zero complex number.
In terms of $z$ and $\al$, the Nair measure (\ref{2-21}) can be expressed as
\beq
    d \mu (p)  \, = \,
    \frac{d^3 p}{(2 \pi)^3} \frac{1}{2 p_{0}} \, = \,
    \frac{1}{(2 \pi)^3} \frac{({\bar \al} \al) d ({\bar \al} \al)}{2}
    \frac{ dz d \bz}{(-2i)}
    \label{3-2}
\eeq
where we omit the numbering index for simplicity.
In terms of the null momentum components,
$z$ and ${\bar \al} \al$ are expressed as
\beq
    z = \frac{p_1 + i p_2}{p_0 + p_3} \, , ~~~
    {\bar \al} \al = p_0 + p_3 \, .
    \label{3-3}
\eeq
This means that the on-shell Nair measure can alternatively be written as
\beq
    d \mu (p)  \, = \,
    \frac{d p_{1}}{2 \pi} \frac{d p_{2}}{2 \pi}
    \frac{1}{4 \pi} \frac{d ( p_0 + p_3 ) } {( p_0 + p_3 ) }
    \, .
    \label{3-4}
\eeq
Upon suitable normalization along $p_1$ and $p_2$ directions, this
implies that the Nair measure is essentially
encoded by $d \log ( {\bar \al} \al )$, {\it i.e.},
\beq
    d \mu (p) \, \approx \,
    \frac{1}{4 \pi } d \log ( {\bar \al} \al )
    \label{3-5}
\eeq
where $\approx$ denotes that we use a conventional normalization
for spatial area.
Notice that this can also be interpreted as a projection of
single-particle trajectories onto a certain spacial direction.

We now consider an off-shell continuation of the form (\ref{3-4}).
Let $L_\mu$ and $l_\mu$ be off-shell and on-shell four-momenta, respectively.
Following the notation (\ref{2-11}), we relate these to each other by
\beq
    L_\mu \, = \, l_\mu + w \eta_\mu
    \label{3-6}
\eeq
where $\eta_\mu$ is a reference null-vector, satisfying $\eta^2 = 0$,
and $w$ is a real number. Since both $\eta_\mu$ and $w$
can arbitrarily be chosen, we can in fact
fix the scaling freedom for either $\eta_\mu$ or $w$.

According to the CSW rules, the physics and hence the
off-shell measure should be independent of the reference null-vector.
This implies that such off-shell measures are parametrized
in terms of $l_\mu$ and $w$.
Thus, in order to obtain one realization of the off-shell measures,
we can fix $\eta_\mu$ to a suitable null vector.
We here fix the reference vector by $\eta_\mu = (1, 0, 0, -1)$.
We can then calculate the off-shell Nair measure as
\beqar
    d \mu (L) & = &
    \frac{d L_{1}}{2 \pi} \frac{d L_{2}}{2 \pi}
    \frac{1}{4 \pi} \frac{d ( L_0 + L_3 ) } {( L_0 + L_3 ) }
    \nonumber \\
    &=& \frac{d l_{1}}{2 \pi} \frac{d l_{2}}{2 \pi}
    \frac{1}{4 \pi} \left[
    \frac{d (l_0 + l_3 )}{(l_0 + l_3 )} + \frac{d w^2}{w^2}
    \right]
    \label{3-7}
\eeqar
where we have used the off-shell condition
\beq
    L^2 = L_{0}^{2} - L_{1}^{2} - L_{2}^{2} - L_{3}^{2}
    = 2 w ( L_0 + L_3 )
    \label{3-8}
\eeq
to calculate the particular factor $d ( L_0 + L_3 )$.
Using the on-shell notation $l_0 + l_3 = {\bar \al} \al $ as before,
we can also express $d \mu (L)$ as
\beqar
    d \mu (L) & \approx &
    \frac{1}{4 \pi }  \left[ d \log ( {\bar \al} \al ) + d \log w^2 \right]
    \nonumber \\
    & \approx &  \frac{1}{4 \pi } d \log ( {\bar \al} \al )w^2
    \nonumber \\
    & \approx & d \mu (l) + \frac{1}{4 \pi }  \frac{d w^2}{w^2} \, .
    \label{3-9}
\eeqar
This is an off-shell analog of the expression (\ref{3-5})
and is a very interesting result in terms of the application
of the spinor-helicity formalism to massive theories.
The second-line expression, in particular, shows that
the off-shell continuation can simply be implemented by
the scaling of $\al \rightarrow w \al$ and
${\bar \al} \rightarrow w {\bar \al}$ ($w \in {\bf R} - \{ 0 \}$).
{\it This suggests that in the holonomy formalism
massive particles can effectively be treated
as massless ones by means of the scaling
$\al \rightarrow w \al$ and the use of the off-shell Nair measures.}
The off-shell momentum $L_\mu$ is parametrized as (\ref{3-6})
which is, of course, not proportional to the on-shell
momentum $l_\mu$.
In this sense, the above prescription is merely
effective and its usage should be clarified.
We shall relate it to the CSW prescription in a moment
but clarification on this matter is still missing.
We shall investigate this point in a future paper.

Notice that the $w$-dependence of
the resultant off-shell measure comes in only
by the factor of $d \log w^2$.
Thus there is a scaling ambiguity in defining $w^2$.
This ambiguity can in fact be absorbed into the definition of
the reference null-vector.
Therefore, without losing generality, we can limit the
range of $w^2$ by
\beq
    0 < w^2 < 1 \, .
    \label{3-10}
\eeq
If $w^2$ is more than one, we can always redefine $\eta_\mu$
such that the integral part of $w^2$ is absorbed into
the reference null-vector.

\noindent
\underline{An $x$-space representation, off-shell Nair measure and the CSW rules}

The CSW rules, which are realized by the S-matrix functional in (\ref{2-37}),
do lead to the correct tree amplitudes, however,
the appearance of $1/q^{2}$ in the contraction operator (\ref{2-38})
is rather abrupt.
Conventionally, this factor is interpreted as a
contribution from a massless scalar propagator that connects MHV vertices.
This interpretation is convenient and correct in writing
down the momentum-space tree amplitudes which do not require
the knowledge of off-shell measures.
As discussed above, however, by use of the off-shell Nair measure
we can deal with massive fields somewhat analogous to massless fields.
This suggests that we may find a more natural understanding of the
factor $1/q^{2}$ by introducing the off-shell Nair measure to
the S-matrix functional (or, more concretely, to
the contraction operator $\widehat{W}^{(A)}$) in an $x$-space representation.
We shall consider this possibility in what follows.

Notice that the $x$-space representation of the amplitudes is given by
\beq
    \A^{(1_{h_1} 2_{h_2} \cdots n_{h_n})}_{N^{k}MHV(0)} (x) \, = \,
    \prod_{i=1}^{n} \int d \mu (p_i) ~
    \A^{(1_{h_1} 2_{h_2} \cdots n_{h_n})}_{N^{k}MHV(0)} (u, \bu )
    \label{3-11}
\eeq
where $\A^{(1_{h_1} 2_{h_2} \cdots n_{h_n})}_{N^{k}MHV(0)} (u, \bu )$ is defined in (\ref{2-14}).
In terms of the MHV S-matrix functional $\F_{MHV} \left[  a^{(h)c} \right]$
in (\ref{2-32}), the $x$-space MHV amplitudes at tree level can be expressed as
\beqar
    \nonumber
    &&
    \left. \frac{\del}{\del a_{1}^{(+) c_1}} \otimes
    \cdots \otimes \frac{\del}{\del a_{a}^{(-) c_a}} \otimes
    \cdots \otimes \frac{\del}{\del a_{b}^{(-) c_b}} \otimes
    \cdots \otimes \frac{\del}{\del a_{n}^{(+) c_n}}
    ~ \F_{MHV} \left[  a^{(h)c} \right] \right|_{a^{(h)c}=0} \\
    &=&
    \A_{MHV(0)}^{(a_{-} b_{-})} (x)
    \label{3-12}
\eeqar
where $a^{(h)c}$'s denote the gluon creation operators
in the momentum-space representation, which are treated as source
functions here.
As in the previous section, the notation $(0)$
specifies that the amplitudes are at tree level.

Generalization of this expression or the $x$-space analog of
(\ref{2-39}), not necessarily limited to the case of tree
amplitudes, can be written as
\beq
    \left. \frac{\del}{\del a_{1}^{(h_1) c_1}  } \otimes
    \frac{\del}{\del a_{2}^{(h_2) c_2} } \otimes
    \cdots \otimes \frac{\del}{\del a_{n}^{(h_n) c_n} }
    \, \F \left[  a^{(h)c}  \right] \right|_{a^{(h)c} =0}
    \, = \,
    \A^{(1_{h_1} 2_{h_2} \cdots n_{h_n})}_{N^{k}MHV} ( x )
    \label{3-13}
\eeq
where the S-matrix functional is now represented
with an $x$-space contraction operator $W^{(A)} (x)$:
\beqar
    \F \left[  a^{(h)c}  \right]
    &=& W^{(A)} (x) \F_{MHV} \left[  a^{(h)c} \right]
    \label{3-14} \\
    \widehat{W}^{(A)} (x) & = &  \exp \left[ -
    \int d \mu (p) \left(
    \frac{\del}{\del a_{p}^{(+)}} \otimes \frac{\del}{\del a_{-p}^{(-)}}
    \right)
    e^{- ip (x -y) }
    \right]_{y \rightarrow x}
    \label{3-15}
\eeqar
where $p$ denotes the on-shell partner of the momentum transfer
$q$, corresponding to the one in (\ref{2-9}).
For simplicity, we here express the relation (\ref{2-11}) as
\beq
    q_\mu \, = \,  p_\mu + w \eta_\mu
    \label{3-16}
\eeq
where $\eta_\mu$ denotes the reference null-vector.
As in the expression (\ref{3-9}), the off-shell Nair measure for $q$ is given by
\beq
    d \mu ( q ) \, \approx \,  d \mu (p) + \frac{1}{4 \pi } \frac{d w^2}{w^2} \, .
    \label{3-17}
\eeq

The factor $e^{- ip (x - y) }$ in (\ref{3-15})
is necessary to guarantee the energy-momentum
conservation for $x - y \ne 0$.
At the end of calculation, we eventually take the limit $y \rightarrow x$.
We assume that the limit is taken such that
the time ordering $x^0 > y^0$ is preserved.
In other words, we shall take the limit
$y \rightarrow x$ with $x^0 - y^0 \rightarrow 0_{+}$.
Since we take this limit, the null property of $p$ is not
necessary in performing the integral and in assuring the
energy-momentum conservation.
Therefore we may also express
the contraction operator $\widehat{W}^{(A)} (x)$ in (\ref{3-15}) as
\beqar
    \widehat{W}^{(A)} (x) & = & \exp \left[-
    \int d \mu (q) \left(
    \frac{\del}{\del a_{p}^{(+)}} \otimes \frac{\del}{\del a_{-p}^{(-)}}
    \right)
    e^{- iq (x -y) }
    \right]_{y \rightarrow x}
    \nonumber  \\
    &=&
    \exp \left[-
    \int \frac{ d^4 q}{(2 \pi)^4} \frac{i}{q^2} \left(
    \frac{\del}{\del a_{p}^{(+)}} \otimes \frac{\del}{\del a_{-p}^{(-)}}
    \right)
    e^{- iq (x -y) }
    \right]_{y \rightarrow x}
    \label{3-18}
\eeqar
where we use a well-known identity in the calculation
of the Feynman propagator
\beq
    \int d \mu ( p ) \left[
    \th ( x^0 - y^0 ) e^{-i p (x - y) } + \th ( y^0 - x^0 ) e^{i p (x - y)}
    \right]
    \, = \,
    \int \frac{ d^4 p} {(2 \pi )^4} \,
    \frac{i}{ p^2 + i \ep }
    \, e^{- i p( x - y ) }
    \label{3-19}
\eeq
($p$ is a null momentum and $\ep$ is a positive infinitesimal).
Notice that there appears no propagator mass for $q$ in (\ref{3-18}).
This is a consequence of the aforementioned interpretation of
the off-shell momentum $q$ (as an effective null-momentum with
an incorporation of the scaling factor $w$)
and essentially embodies the CSW prescription in our formulation.

In the holonomy formalism, the CSW rules are encoded in
the definition of $\widehat{W}^{(A)} (x)$.
The CSW rules in our formulation are then stated as follows.
\begin{enumerate}
  \item There is no propagator mass for the virtual gluon represented by
  the off-shell momentum transfer $q$.
  \item The factor $e^{-i q (x - y)}$ can be replaced by $e^{-i p (x -y)}$ as
  we eventually take the limit $y \rightarrow x$, keeping the time ordering
  $x^0 > y^0$.
  \item We do not define creation operators for the virtual gluon,
  $a_{q}^{(+)}$ or $a_{-q}^{(-)}$, since these
  are irrelevant, if related, to the creation operators of gluons
  in terms of which the MHV S-matrix functional $\F_{MHV}$ is constructed.
\end{enumerate}
These rules are materialized by the following contraction operator.
\beqar
    \widehat{W}^{(A)} (x)
    & = & \exp \left[-
    \int d \mu (q) \left(
    \frac{\del}{\del a_{p}^{(+)}} \otimes \frac{\del}{\del a_{-p}^{(-)}}
    \right)
    e^{- ip (x -y) }
    \right]_{y \rightarrow x}
    \nonumber \\
    & \approx &  \exp \left[-
    \int d \mu (p) \left(
    \frac{\del}{\del a_{p}^{(+)}} \otimes \frac{\del}{\del a_{-p}^{(-)}}
    \right)
    e^{- ip (x -y) }
    \right]
    \nonumber \\
    && \times \left. \exp \left[-
    \frac{1}{4 \pi } \int \frac{d w^2}{w^2}  \left(
    \frac{\del}{\del a_{p}^{(+)}} \otimes \frac{\del}{\del a_{-p}^{(-)}}
    \right)
    e^{- ip (x -y) }
    \right] \right|_{y \rightarrow x}
    \label{3-20}
\eeqar
where we use (\ref{3-17}) and (\ref{3-18}) with a replacement of
the factor $e^{- iq (x -y) }$ by $e^{- ip (x -y) }$.

Apparently, the expressions (\ref{3-15}) and (\ref{3-20})
contradict each other. Quantum theoretically, however, the
two (and the other expressions in (\ref{3-18}))
can be regarded as equivalent, with the
former being identified as a regularized form of the latter.
The exponent of the second factor in (\ref{3-20})
is proportional to
\beq
    \int_{0}^{1} \frac{ d w^2 } { w^2 } = \infty
    \label{3-21}
\eeq
where we use the condition (\ref{3-10}).
Thus we can indeed obtain the expression
(\ref{3-15}) from (\ref{3-20}) by regularizing the factor
involving the above log divergence.

As we have seen in the previous section,
the above understanding of regularization on the contraction operators
is valid through tree amplitudes where the log divergence arises at each propagator.
As long as we resort to the functional derivation of gluon amplitudes,
this interpretation should be applicable to loop amplitudes in general.
In the following, we consider a derivation of one-loop MHV amplitudes in particular,
utilizing the $x$-space S-matrix functional (\ref{3-14}).

\noindent
\underline{Reproduction of the BST results in the holonomy formalism}
%\underline{Reproduction of the BST representation}

\begin{figure} [htbp]
\begin{center}
\includegraphics[width=125mm]{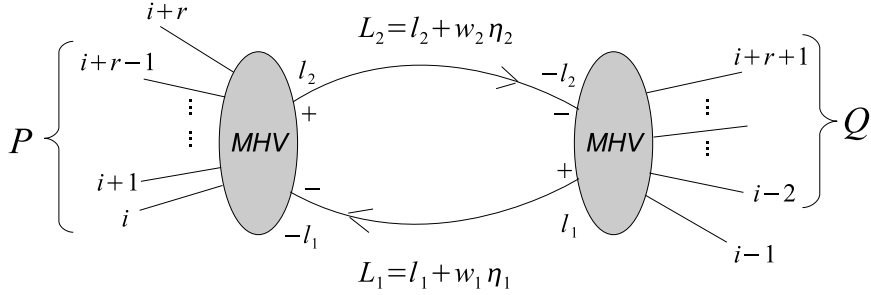}
\caption{One-loop MHV diagram --- The internal off-shell momenta are
labeled by $L_i = l_i + w_i \eta_i$ ($i = 1,2$), with $l_i$ and $\eta_i$
being null four-vectors. $w_i$ are real variables. The diagram corresponds to
the case where the negative-helicity indices
$( a_-  b_- )$ are split into the left and the right
MHV vertices. If the both indices are on the same side, which always occurs
as we rotate the index $i$, we need to flip the direction
of one of the propagators accordingly. When the left and the right
vertices have the same number of external legs, we have reflection symmetry.
This explains the factor of  $\left( 1 - \frac{1}{2} \del_{ \frac{n}{2}  ,r+1} \right)$
in the equation (\ref{3-26}).}
\label{fighol0402}
\end{center}
\end{figure}

In terms of the S-matrix functional (\ref{3-14}), the MHV amplitude
in the $x$-space representation is generated as a loop expansion:
\beqar
    &&
    \left. \frac{\del}{\del a_{1}^{(+) c_1}} \otimes
    \cdots \otimes \frac{\del}{\del a_{a}^{(-) c_a}} \otimes
    \cdots \otimes \frac{\del}{\del a_{b}^{(-) c_b}} \otimes
    \cdots \otimes \frac{\del}{\del a_{n}^{(+) c_n}}
    ~ \F \left[  a^{(h)c} \right] \right|_{a^{(h)c}=0} \nonumber \\
    &=& \A_{MHV}^{( a_{-} b_{-})} (x)
    \nonumber \\
    &=&
    \A_{MHV(0)}^{(a_{-} b_{-})} (x) \, + \, \A_{MHV(1)}^{(a_{-} b_{-})} (x)
    \, + \, \A_{MHV(2)}^{(a_{-} b_{-})} (x) \, + \, \cdots \, .
    \label{3-22}
\eeqar
The number of contractions by $\widehat{W}^{(A)} (x)$ corresponds
to the loop order of the $L$-loop amplitude $\A_{MHV(L)}^{(a_{-} b_{-})} (x)$.
In terms of the coupling constant $g$, the $L$-loop amplitude
can be expressed as
\beq
    \A_{MHV(L)}^{(a_{-} b_{-})} (x) \, \sim \, g^{n - 2 + 2L} \, .
    \label{3-23}
\eeq
Picking up the $L=1$ term in (\ref{3-22}), we can calculate
the one-loop MHV amplitudes as
\beqar
    \A_{MHV(1)}^{( a_-  b_- )} (x) & = &
    \left[-
    \int d\mu(L_1 ) \left(
    \frac{\del}{\del a_{l_1 }^{(+)}} \otimes \frac{\del}{\del a_{-l_1 }^{(-)}}
    \right)
    \right]
    \,
    \left[-
    \int d\mu(L_2 ) \left(
    \frac{\del}{\del a_{l_2 }^{(+)}} \otimes \frac{\del}{\del a_{-l_2 }^{(-)}}
    \right)
    \right]
    \nonumber \\
    &&
    \times \left.
    \frac{\del}{\del a_{1}^{(+) c_1}} \otimes
    \cdots \otimes \frac{\del}{\del a_{n}^{(+) c_n}}
    \F \left[  a^{(h)c} \right]
    \right|_{a^{(h)c}=0} \nonumber \\
    &=&
    \prod_{i = 1}^{n} \int d \mu (p_{i})
    \, \A_{MHV(1)}^{(a_{-} b_{-} )} (u, \bu )
    \label{3-24} \\
    \A_{MHV(1)}^{(a_{-} b_{-})} (u, \bu )
    &=& i g^{n-2}
    \, (2 \pi)^4 \del^{(4)} \left( \sum_{i=1}^{n} p_i \right) \,
    \widehat{A}_{MHV(1)}^{(a_{-} b_{-})} (u)
    \label{3-25}
\eeqar
where $L_1$ and $L_2$ denote loop momenta. $l_1$ and $l_2$ are
corresponding on-shell vectors, respectively.
A typical one-loop MHV diagram is shown in Figure \ref{fighol0402}.

%%%%%%%%%%%%%%%%%%%%%%%%% figure %%%%%%%%%%%%%%%%%%%%%%%%%
%%%%%%%%%%%%%%%%%%%%%%%%% figure %%%%%%%%%%%%%%%%%%%%%%%%%

The holomorphic part of the one-loop amplitudes
$\widehat{A}_{MHV(1)}^{(a_{-} b_{-})} (u)$ are then calculated as
\beqar
    \widehat{A}_{MHV(1)}^{(a_{-} b_{-})} (u)
    & = &
    i g^2 \,
    \sum_{i = 1}^{n} \sum_{r = 1}^{\left\lfloor \frac{n}{2} \right\rfloor -1 }
    \left( 1 - \frac{1}{2} \del_{ \frac{n}{2}  ,r+1} \right) \,
    \int d \mu ( L_1 ) d \mu ( L_2 )
    \nonumber \\
    &&
    \times ~
    \widehat{A}_{MHV(0)}^{( ( -l_1 )_{-} i_{+} \cdots a_{-} \cdots (i+r)_{+} l_{2+})} (u)
    \,
    \widehat{A}_{MHV(0)}^{( ( -l_2 )_{-} (i+r+1)_{+} \cdots b_{-} \cdots (i-1)_{+} l_{1+})} (u)
    \label{3-26}
\eeqar
where $\left\lfloor \frac{n}{2} \right\rfloor$
denotes the largest integer less than or equal to $\frac{n}{2}$.
Since $r$ is a positive integer, $\del_{ \frac{n}{2} ,r+1}$
vanishes unless $n = 2k$ ($k= 2, 3, \cdots $).
The factor $\left( 1 - \frac{1}{2} \del_{ \frac{n}{2} ,r+1} \right)$
arises in order to compensate the double counting due to the left-right
reflection symmetry.
There is a freedom of choice for the
positions of the internal indices $\{ l_1 , - l_1 , l_2 , - l_2 \}$.
Diagrammatically, this means that we can arbitrarily choose
the indices $i$ ($= 1,2, \cdots , n$) and
$r$ ($= 1,2, \cdots , \left\lfloor {n}/{2} \right\rfloor - 1)$
in Figure \ref{fighol0402}.
We have reflection symmetry when the left and the right MHV
vertices have the same number of external legs.
In the functional language, this means that $x$ and $y$
are exchangeable in carrying out the contraction with (\ref{3-20}).
Since we preserve the time ordering $x^0 > y^0$
in taking the $y \rightarrow x$ limit, there are only
half contributions from the diagrams with reflection symmetry.
Alternatively, we can also understand this fact from
the Taylor expansion of the MHV S-matrix
$\F_{MHV} \left[  a^{(h)c} \right]$ in (\ref{3-14}),
whose explicit form is defined by (\ref{2-32}),
since in the symmetric case we can treat $x$ and $y$ identical to each other.
Thus a factor $\frac{1}{2!}$ automatically arises in the functional method.
This naturally explains the factor
$\left( 1 - \frac{1}{2} \del_{ \frac{n}{2} ,r+1} \right)$ in (\ref{3-26}).

Notice also that the use of the functional derivatives in (\ref{3-22}) and
the Grassmann integral in (\ref{2-36}) automatically pick a correct choice of
the internal helicity configuration for given $i$ and $r$.
Thus, as far as formalism is concerned, our functional
formulation of the one-loop amplitudes is more efficient
and natural than a diagrammatic method.

The single-trace color structure of the one-loop MHV amplitudes
arises from the product of the holomorphic tree MHV  amplitudes
in (\ref{3-26}) and it can be calculated as
\beqar
    &&
    \widehat{A}_{MHV(0)}^{( (-l_1 )_{-} i_{+} \cdots a_{-}
    \cdots (i+r)_{+} l_{2+} )} (u)
    ~
    \widehat{A}_{MHV(0)}^{( (-l_2 )_{-}  (i+r+1)_{+}  \cdots b_{-}
    \cdots (i-1)_{+} l_{1+} )} (u)
    \nonumber \\
    &=&
    \sum_{ \si^{(1)} \in \S_{r+1}}
    \sum_{ \si^{(2)} \in \S_{n-r-1}} \!\!
    \Tr ( t^{\si^{(1)}_{i}} \cdots t^{\si^{(1)}_{i+r}} \, t^{\si^{(2)}_{i+r+1}}
    \cdots t^{\si^{(2)}_{i-1}} )
    \nonumber \\
    && ~ \times ~
    \widehat{C}_{MHV(0)}^{( (-l_1 )_{-} i_{+} \cdots a_{-}
    \cdots (i+r)_{+} l_{2+} )} (u; \si^{(1)})
    ~
    \widehat{C}_{MHV(0)}^{( (-l_2 )_{-}  (i+r+1)_{+}  \cdots b_{-}
    \cdots (i-1)_{+} l_{1+} )} (u; \si^{(2)})
    \nonumber \\
    &=&
    \Tr ( t^i \cdots t^{i+r} \, t^{i+r+1} \cdots t^{i-1}) ~
    \widehat{C}_{MHV(0)}^{( (-l_1 )_{-} a_{-} )} (u;{\bf 1}^{(1)})
    ~
    \widehat{C}_{MHV(0)}^{( (-l_2 )_{-}  b_{-} )} (u; {\bf 1}^{(2)})
    \nonumber \\
    &&
    + ~
    \P ( i \cdots i+r | i+r+1 \cdots i-1 )
    \label{3-27}
\eeqar
where $\P ( i \cdots i+r | i+r+1 \cdots i-1 )$ denotes the terms obtained
by the double permutations of $\si^{(1)}$ and $\si^{(2)}$.
${\bf 1}^{(1)}$ and ${\bf 1}^{(2)}$ denote the identity transpositions
for $\si^{(1)}$ and $\si^{(2)}$, respectively:
\beq
    {\bf 1}^{(1)} \, = \,
    \left(
      \begin{array}{c}
        i \cdots i+r \\
        i \cdots i+r \\
      \end{array}
    \right) , ~~~
    {\bf 1}^{(2)} \, = \,
    \left(
      \begin{array}{c}
        i+r+1 \cdots i-1 \\
        i+r+1 \cdots i-1 \\
      \end{array}
    \right) .
    \label{3-28}
\eeq
That the functional method has generality in the position
of negative-helicity indices $(a_- b_- )$ means that
the product of the colorless factors
in (\ref{3-27}) can be replaced by the following product:
\beqar
    &&
    \widehat{C}_{MHV(0)}^{( (-l_1 )_{-} i_{+} \cdots a_{-}
    \cdots (i+r)_{+} l_{2+} )} (u; {\bf 1}^{(1)})
    ~
    \widehat{C}_{MHV(0)}^{( (-l_2 )_{-}  (i+r+1)_{+}  \cdots b_{-}
    \cdots (i-1)_{+} l_{1+} )} (u; {\bf 1}^{(2)})
    \nonumber \\
    &\rightarrow&
    \widehat{C}_{MHV(0)}^{(l_{1+} i_{+} \cdots a_{-}
    \cdots b_{-} \cdots (i+r)_{+} l_{2+} )} (u; {\bf 1}^{(1)})
    ~
    \widehat{C}_{MHV(0)}^{( (-l_2 )_{-}  (i+r+1)_{+}  \cdots
    (i-1)_{+} (-l_{1})_{-} )} (u; {\bf 1}^{(2)})
    \nonumber \\
    & = &
    \R_{n:r;i}^{( l_{1}, l_{2} )}
    \widehat{C}_{MHV(0)}^{( i_{+} \cdots  a_{-} \cdots b_{-} \cdots
    (i-1)_{+} )} (u; {\bf 1}^{(1)} \otimes {\bf 1}^{(2)})
    \label{3-29}
\eeqar
where $\R_{n:r;i}^{( l_{1}, l_{2} )}$ is defined by
\beq
    \R_{n:r;i}^{( l_{1}, l_{2} )}
    ~ = ~
    \frac{( i+r ~ i+r+1)( l_1 ~ l_2 )}{( i+r ~ l_2 )( -l_1 ~ i )}
    \frac{( i-1 ~ i) ( l_2 ~ l_1 )}{( i-1 ~ l_1 )( -l_2 ~ i+r+1 )} \, .
    \label{3-30}
\eeq
Here, for simplicity, the holomorphic scalar products of spinor momenta are represented by
the numbering indices, {\it e.g.}, $( i+r ~ i+r+1)
= (u_{i+r} u_{i+r+1})$, $( l_1 ~ l_2 )= ( u_{l_1} u_{l_2} )$ and so on, with
the scalar products being defined in (\ref{2-5}).
Notice that the factor $\R_{n:r;i}^{( l_{1}, l_{2} )}$ is quadratic in either $l_1$ or $l_2$.
Thus it is independent of the signs of $l_1$, $l_2$.
The arrow in (\ref{3-29}) means that we take a limit of
$\th^\prime \rightarrow \th$ where
$\th^\prime$ and $\th$ are the $\N = 4$ chiral superpartners
of the four-dimensional space-time coordinates $y$ and $x$,
respectively. In this limit, we have the relation
\beqar
    &&
    (-l_{1} \, a)^4 (b \, -l_{2})^4
    \nonumber \\
    &=&
    \left.
    \int d^8 \th  \, \xi_{-l_{1}}^{1}\xi_{-l_{1}}^{2}
    \xi_{-l_{1}}^{3}\xi_{-l_{1}}^{4}
    \, \xi_{a}^{1}\xi_{a}^{2}\xi_{a}^{3}\xi_{a}^{4}
    \right|_{\xi_{i}^{\al} = \th_{A}^{\al} u_{i}^{A} }
    \left.
    \int d^8 \th^\prime
    \, \eta_{b}^{1}\eta_{b}^{2}\eta_{b}^{3}\eta_{b}^{4}
    \, \eta_{-l_{2}}^{1}\eta_{-l_{2}}^{2}
    \eta_{-l_{2}}^{3}\eta_{-l_{2}}^{4}
    \right|_{\eta_{i}^{\al} = \th_{A}^{\prime  \al} u_{i}^{A} }
    \nonumber \\
    & \rightarrow &
    (a \, b)^4 ( l_{1} \, l_{2} )^4
    ~~~~~~~ \mbox{($\th^\prime \rightarrow \th$)} \, .
    \label{3-31}
\eeqar
As discussed below (\ref{2-21}),
the holonomy operator is defined in $\N = 4$ chiral
superspace $(x , \th) = ( x_{\Ad  A}, \th_{A}^{\al} )$.
This means that taking the limit of $y \rightarrow x$ in the definition
of $\widehat{W}^{(A)} (x)$
should be understood as taking the limit of $( y, \th^\prime ) \rightarrow (x, \th )$.
Thus, in the supertwistor framework the relation (\ref{3-31}) is naturally
required, and so is the relation (\ref{3-29}).

The holomorphic one-loon MHV amplitudes (\ref{3-26}) are then calculated as
\beqar
    \widehat{A}_{MHV(1)}^{(a_{-} b_{-})} (u)
    & = &
    i g^2 \,
    \sum_{i = 1}^{n} \sum_{r = 1}^{\left\lfloor \frac{n}{2} \right\rfloor -1 }
    \left( 1 - \frac{1}{2} \del_{ \frac{n}{2}  ,r+1} \right)
    \nonumber \\
    &&
    \times \sum_{ \si^{(1)} \in \S_{r+1}}
    \sum_{ \si^{(2)} \in \S_{n-r-1}} \!\!
    \Tr ( t^{\si^{(1)}_{i}} \cdots t^{\si^{(1)}_{i+r}} \, t^{\si^{(2)}_{i+r+1}}
    \cdots t^{\si^{(2)}_{i-1}} )
    \nonumber \\
    &&
    \times
    ~ \L_{n:r; \si_i }^{( l_1 , l_2 )} ~
    \widehat{C}_{MHV(0)}^{( a_{-} b_{-} )} (u; \si^{(1)} \otimes \si^{(2)})
    \label{3-32}\\
    \L_{n:r; \si_i }^{( l_1 , l_2 )}
    & = &
    \int d \mu ( L_1 ) d \mu ( L_2 ) \,
    \R_{n:r; \si_i }^{( l_1 , l_2 )}
    \label{3-33} \\
    \R_{n:r; \si_i }^{( l_1 , l_2 )}
    & = &
    \frac{( \si_{i+r}^{(1)} ~ \si_{i+r+1}^{(2)} )( l_1 ~ l_2 )}
    {( \si_{i+r}^{(1)} ~ l_2 )( -l_1 ~ \si_{i}^{(1)} )}
    \frac{( \si_{i-1}^{(2)} ~ \si_{i}^{(1)}) ( l_2 ~ l_1 )}
    {( \si_{i-1}^{(2)} ~ l_1 )( -l_2 ~ \si_{i+r+1}^{(2)} )}
    \label{3-34}
\eeqar
where an explicit form
of $\widehat{C}_{MHV(1)}^{( a_{-} b_{-} )} (u; \si^{(1)} \otimes \si^{(2)})$
is given by
\beq
    \widehat{C}_{MHV(0)}^{( a_{-} b_{-} )} (u; \si^{(1 \otimes 2)})
    \, = \,
    \frac{(a ~ b)^4}{
    ( \si_{i}^{(1)} \, \si_{i+1}^{(1)} ) ( \si_{i+1}^{(1)} \, \si_{i+2}^{(1)} )
    \cdots ( \si_{i+r}^{(1)} \, \si_{i+r+1}^{(2)} ) ( \si_{i+r+1}^{(2)} \, \si_{i+r+2}^{(2)} )
    \cdots ( \si_{i-1}^{(2)} \, \si_{i}^{(1)} )
    } \, .
    \label{3-35}
\eeq
Here we abbreviate $\si^{(1)} \otimes \si^{(2)}$ by $\si^{(1 \otimes 2)}$.
Notice that the negative-helicity indices can take any values here, {\it i.e.},
$a , b \in \{ i , i+1 \cdots , i-1 \} = \{ 1, 2, \cdots , n \}$.
As in the expression (\ref{3-27}),
by use of the notation $\P ( i \cdots i+r | i+r+1 \cdots i-1 )$
for the double-permutation terms, we can attribute
the calculation of (\ref{3-32}) to that of
\beq
    \L_{n:r; i }^{( l_1 , l_2 )}
    ~ = ~
    \int d \mu ( L_1 ) d \mu ( L_2 ) \,
    \R_{n:r; i }^{( l_1 , l_2 )}
    \label{3-36}
\eeq
where $\R_{n:r; i }^{( l_1 , l_2 )}$ is defined by (\ref{3-30}).
Now, using the Schouten identities
\beqar
    ( i+r ~ i+r +1 ) ( l_1 ~ l_2 )
    &=& ( i+r ~ l_2 ) ( l_1 ~ i+r+1 ) + ( i+r ~ l_1 )( i+r+1 ~ l_2 ) \, ,
    \nonumber \\
    ( i - 1 ~ i ) ( l_2 ~ l_1 )
    &=& ( i-1 ~ l_1 )( l_2 ~ i ) + ( i-1 ~ l_2 ) ( i ~ l_1 ) \, ,
    \nonumber
\eeqar
we can express $\R_{n:r;i}^{( l_{1} l_{2} )}$ as
\beq
    \R_{n:r;i}^{( l_{1}, l_{2} )}
    \, = \,
    \widehat{R}^{( l_{1}, l_{2} )}_{(i, i+r+1)}
    \, - \,
    \widehat{R}^{( l_{1}, l_{2} )}_{(i, i+r)}
    \, - \,
    \widehat{R}^{( l_{1}, l_{2} )}_{(i-1, i+r+1)}
    \, + \,
    \widehat{R}^{( l_{1}, l_{2} )}_{(i-1, i+r)}
    \label{3-37}
\eeq
where $\widehat{R}^{( l_{1}, l_{2} )}_{(i, j)}$ is defined by
\beq
    \widehat{R}^{( l_{1}, l_{2} )}_{(i, j)}
    \, \equiv \,
    \frac{( l_2 ~ i )( l_1 ~ j )}{( l_1 ~ i )( l_2 ~ j )} \, .
    \label{3-38}
\eeq
We here suppress the total number suffix $n$ for simplicity.

It is explicitly shown in \cite{Brandhuber:2004yw,Brandhuber:2005kd} that
the expression (\ref{3-36}) leads to the BST results for one-loop MHV amplitudes
by rewriting the off-shell measures as
\beq
    d \mu ( L_i ) \, \longrightarrow \, \frac{d^4 L_i}{{L_i}^2} ~~~~ \mbox{(for $i = 1,2$)}
    \label{3-39}
\eeq
up to a certain normalization factor.
In our formulation, this prescription can naturally be understood from
the expression (\ref{3-18}).
This means that the BST representation of
one-loop MHV amplitudes, which we show in a moment, can be reproduced
from the S-matrix functional (\ref{3-14}) with a suitable
choice of normalization for the off-shell measures.

\noindent
\underline{The BST representation of one-loop MHV amplitudes}

For the completion of our discussion,
in the following we briefly review the results of
the BST method for one-loop MHV amplitudes \cite{Brandhuber:2004yw}.
The essence of the BST results is that the calculation of
the integral (\ref{3-36}) with the prescription (\ref{3-39})
correctly leads to the previously known unitary-cut
results obtained by
Bern, Dixon, Dunbar and Kosower (BDDK) \cite{Bern:1994zx,Bern:1994cg}.

The BDDK representation of one-loop MHV amplitudes
is given in terms of color-stripped holomorphic amplitudes.
In our notation, this can be expressed in terms of
the colorless holomorphic tree MHV amplitudes
and the corresponding one-loop amplitudes:
\beq
    \begin{array}{rcl}
    \widehat{C}_{MHV(0)}^{( a_-  b_- )} (u)
    & \equiv &
    \widehat{C}_{MHV(0)}^{( a_-  b_- )} (u; {\bf 1}) \, ,
    \\
    \widehat{C}_{MHV(1)}^{( a_-  b_- )} (u)
    & \equiv &
    \widehat{C}_{MHV(1)}^{( a_-  b_- )} (u; {\bf 1}^{(1)} \otimes {\bf 1}^{(2)})
    \end{array}
    \label{3-40}
\eeq
where $\widehat{C}_{MHV(0)}^{( a_-  b_- )} (u; \si)$ is defined by (\ref{2-3}) and
$\widehat{C}_{MHV(1)}^{( a_-  b_- )} (u; \si^{(1)} \otimes \si^{(2)})$ is defined
by the expression
\beq
    \widehat{A}_{MHV(1)}^{(a_{-} b_{-})} (u)
    \, = \,
    \sum_{ \si^{(1)} \in \S_{r+1}}
    \sum_{ \si^{(2)} \in \S_{n-r-1}} \!\!\!\!\!
    \Tr ( t^{\si^{(1)}_{i}} \cdots t^{\si^{(1)}_{i+r}} \, t^{\si^{(2)}_{i+r+1}}
    \cdots t^{\si^{(2)}_{i-1}} )
    \,
    \widehat{C}_{MHV(1)}^{( a_{-} b_{-} )} (u; \si^{(1)} \otimes \si^{(2)}) \, .
    \label{3-41}
\eeq
In terms of (\ref{3-40}), the BDDK representation is then expressed as \cite{Bern:1994zx}
\beqar
    \widehat{C}_{MHV(1)}^{( a_-  b_- )} (u)
    &=&
    i g^2 \, c_{\Ga} \,  V_n \, \widehat{C}_{MHV (0)}^{( a_-  b_- )} (u)
    \label{3-42} \\
    c_{\Ga} &=& \frac{1}{(4 \pi )^{2-\ep}}
    \frac{\Ga (1 + \ep ) \Ga^2 (1 - \ep)}{\Ga (1 - 2 \ep )}
    \, \rightarrow \, \frac{1}{(4 \pi)^2} ~~~ ( \ep \rightarrow 0 )
    \label{3-43}
\eeqar
where $\ep = (4 - D) / 2 $ is the dimensional regularization
parameter. $V_n$ is independent of the negative-helicity indices
$(a_-  b_- )$ and is given by
\beq
    V_n \, = \,
    \sum_{i = 1}^{n} \sum_{r = 1}^{\left\lfloor \frac{n}{2} \right\rfloor -1 }
    \left( 1 - \frac{1}{2} \del_{ \frac{n}{2} ,r+1} \right)
    F^{\rm 2m \, e}_{n:r;i} \, .
    \label{3-44}
\eeq
$F^{\rm 2m \, e}_{n:r;i}$ is known as the (easy two-mass scalar)
box function and is given by
\beqar
    F^{\rm 2m \, e}_{n:r;i} & \equiv & F ( s, t, P^2 , Q^2 )
    \nonumber \\
    &=& - \frac{1}{\ep^2}
    \left[
    (-s)^{-\ep} + (-t)^{-\ep} - ((-P)^2)^{- \ep}- ((-Q)^2)^{- \ep}
    \right]
    \nonumber \\
    &&
    + \,
    {\rm Li}_2 \left( 1 - \frac{P^2}{s} \right)
    +
    {\rm Li}_2 \left( 1 - \frac{P^2}{t} \right)
    +
    {\rm Li}_2 \left( 1 - \frac{Q^2}{s} \right)
    +
    {\rm Li}_2 \left( 1 - \frac{Q^2}{t} \right)
    \nonumber \\
    &&
    - \,
    {\rm Li}_2 \left( 1 - \frac{P^2 Q^2}{st} \right)
    +
    \frac{1}{2} \log^2 \left( \frac{s}{t} \right)
    \label{3-45}
\eeqar
in terms of the following parametrization
\beqar
    P &=&  q_{i \, i+r-1} \, = \, p_i + p_{i+1} + \cdots p_{i+r-1}
    \label{3-46} \\
    Q &=& q_{i+r+1 \, i-2} \, = \, p_{i+r+1} + p_{i+r+2} + \cdots p_{i-2}
    \label{3-47} \\
    s &=& (p_{i-1} + P )^2 \, = \, ( p_{i+r} + Q )^2
    \label{3-48} \\
    t &=& ( p_{i+r} + P )^2 \, = \, ( p_{i-1} + Q )^2
    \label{3-49}
\eeqar
with $p_{i-1} + P + p_{i + r} + Q = 0$ where
$p_i$'s denote the $i$-th out-going gluon momenta $(i = 1,2, \cdots, n)$.
In (\ref{3-46}) and (\ref{3-47}) we use the notation (\ref{2-9}),
expressing $P$ and $Q$ as $P = q_{i \, i+r-1}$
and $Q = q_{i+r+1 \, i-2}$, respectively.
Diagrammatic relation of these variables can be seen in Figure \ref{fighol0402}.

Comparing the expressions
(\ref{3-32}) and (\ref{3-41})-(\ref{3-44}), we can easily find
that the box function (\ref{3-45}) corresponds to
the integral
$\L_{n:r; i }^{( l_1 , l_2 )}$
in (\ref{3-36}).
In the BST paper \cite{Brandhuber:2004yw}, this correspondence is
analytically confirmed, including the $\ep$-dependent part,
by careful evaluation of the integral.
The upshot of the BST results is given by the BST
representation of the box function:
\beqar
    F ( s, t, P^2 , Q^2 ) &=& - \frac{1}{\ep^2}
    \left[
    (-s)^{-\ep} + (-t)^{-\ep} - ((-P)^2)^{- \ep}- ((-Q)^2)^{- \ep}
    \right] \, + \, B (s, t, P^2 , Q^2 )
    \nonumber \\
    \label{3-50} \\
    B (s,t,P^2 , Q^2 ) &=&
    {\rm Li}_2 ( 1 - c_{r;i} P^2 )
    + {\rm Li}_2 ( 1 - c_{r;i} Q^2 )
    - {\rm Li}_2 ( 1 - c_{r;i} s )
    - {\rm Li}_2 ( 1 - c_{r;i} t )
    \nonumber \\
    \label{3-51}
\eeqar
where
\beq
    c_{r;i} \, = \, \frac{(P^2 + Q^2 ) - ( s + t )}{P^2 Q^2 - st }
    \, = \, \frac{( p_{i-1} + p_{i+r} )^2}{P^2 Q^2 - st } \, .
    \label{3-52}
\eeq
At the limit of $\ep \rightarrow 0$,
the finite part of the  box function $F (s, t , P^2 , Q^2 )$
is given by $B (s,t,P^2 , Q^2 )$.

%%%%%%%%%%%%%%%%%%%%%%%%%%%%%%%%%%%%%%%%%%%%%%%%%%%%%%%%%%%
\section{One-loop MHV amplitudes: polylog regularization}

In this section, we propose an alternative derivation of
the BST representation from the expression (\ref{3-36})
in a rather intuitive way.

{\it Our strategy to obtain the one-loop amplitudes
is to use the off-shell continuation of the
contraction operator (\ref{3-20}) such
that the finite part of the amplitude arises from the $w$-part of the off-shell
Nair measure, namely, contributions form the second factor in (\ref{3-20}).}

Instead of relying on the BST prescription (\ref{3-39}), we
now explicitly use the off-shell Nair measure of the form (\ref{3-17}).
The integral measure of $\L_{n:r;i}^{( l_{1} , l_{2} )}$ in (\ref{3-36}) is then expanded as
\beq
    d \mu ( L_1 ) d \mu ( L_2 ) \, = \,
     d \mu ( l_1 ) d \mu ( l_2 )
    \, + \,
    d \mu ( l_1 ) \frac{1}{4 \pi} \frac{d W_2}{W_2}
    \, + \,
    \frac{1}{4 \pi} \frac{d W_1}{W_1}  d \mu ( l_2 )
    \, + \,  \frac{1}{( 4 \pi )^2}
    \frac{d W_1}{W_1}  \frac{d W_2}{W_2}
    \label{4-1}
\eeq
where $W_i \equiv w_{i}^{2}$ ($i = 1,2$).

Since $\R_{n:r;i}^{( l_{1} , l_{2} )}$ in (\ref{3-36}) is a dimensionless quantity,
the integral over $d \mu ( l_1 )$ or $d \mu ( l_2 )$ leads to
quadratic divergence.
In order to obtain finite quantities from the measures
involving these, we need some regularization scheme.
In the unitary-cut method, dimensional regularization is
utilized to express the one-loop MHV amplitudes in terms of
the box functions.
Similarly, we may calculate
the integral over $d \mu ( l_1 ) d \mu ( l_2 )$, the
first term in (\ref{4-1}), by use of dimensional regularization.
The resultant amplitudes would correspond to
a certain collinear sector of the one-loop MHV amplitudes.
The collinear sector can be specified by assuming
that external legs for one of the MHV vertices, say,
the left-hand side MHV vertex, are all collinear.

The second and the third terms in (\ref{4-1}) also
include the quadratically divergent $d \mu ( l_i )$ measure.
Thus, for the same reasons above, we need to resort to some regularization
scheme unless the $W_i$-measure vanishes to cancel the divergence.
Such a cancelation, however, never happens because of the following.
The vanishing of the $W_i$ measure means that $W_i$ is fixed.
But what does fixed $W_i$ or $w_i$ mean
in the off-shell definition of $L_i = l_i + w_i \eta_i$
where $\eta_i$ is an arbitrary null vector?
Since $w_i$ couples to $\eta_i$, we can always shift
$w_i$ unless it is zero.
In this context, the fixed $w_i$ literally means
that we fix $w_i$ to zero, which of course contradicts the
off-shell definition.
Thus the $W_i$ measure never vanishes and
the second and third terms in (\ref{4-1}) lead to infinity.

Another interesting argument for the divergence can be made as follows.
In terms of the integral measure {\it per se},
the single-$W_i$ measure is equivalent to the one
that appears in the tree amplitudes in (\ref{3-21}).
As discussed there, the integral has log divergence
unless we change the range of $W_i$ from $0 < W_i < 1$.
As we shall discuss later, such an alternation
may occur for multiple $W_i$'s,
particularly for the double-$W_i$ measure in (\ref{4-1}),
but for the single-$W_i$ case that will not happen since
the single-$W_i$ measure is independent of the other $W_i$'s
and can effectively be treated as a tree-level integral measure.
Thus, as in the tree-level case, we can interpret the integrals
over the second and third terms in (\ref{4-1}) as unphysical.

\noindent
\underline{Iterated integral representation of polylogarithms}

In the rest of this section, we argue that finite physical quantities
of the one-loop MHV amplitudes, {\it i.e.},
the function $B(s,t, P^2 , Q^2 )$ in (\ref{3-51}), can be obtained from
an integral over the last term in (\ref{4-1}). We shall carry out the
analysis by paying attention to the range of $W_i$'s.
For this purpose, it is convenient to consider the double-$W_i$ measure
in terms of differential forms as we
have defined the logarithmic one-form $\om_{ij} = d \log (u_i u_j )$ in (\ref{2-18}).
Using the local coordinate parametrization (\ref{3-1}),
the one-form $\om_{ij}$ is expressed as
\beq
    \om_{ij} = d \log ( z_i - z_j ) = \frac{ d z_i - d z_j }{ z_i - z_j }
    \label{4-2}
\eeq
where $i, j$ denote the numbering indices, satisfying
$1 \le i < j \le n$.
Motivated by this form, we define the following one-forms
\beqar
    \om_{1}^{(1)} &\equiv& - d \log (W^{(1)} - W_1 ) ~ = ~ \frac{d W_1}{1 - W_1} \, ,
    \label{4-3}\\
    \om_{2}^{(0)} &\equiv& d \log (W^{(0)} - W_2 ) ~ = ~ \frac{ d W_2 }{W_2}
    \label{4-4}
\eeqar
where $W_1$ and $W_2$ are positive real variables and
we set $W^{(0)}$ and $W^{(1)}$ by
\beq
    W^{(0)} \equiv  0 \, , ~~~ W^{(1)} \equiv  1 \, .
    \label{4-5}
\eeq

It is well-known (see, {\it e.g.}, \cite{Kohno:2002bk})
that the dilogarithm function can be represented as an iterated integral:
\beqar
    {\rm Li}_2 (x) &=& \int_{0}^{x} \om^{(1)} \om^{(0)}
    \label{4-6}\\
    \om^{(0)} & \equiv & \frac{dt}{t} \, , ~~~ \om^{(1)} ~\equiv ~ \frac{dt}{1-t}
    \label{4-7}
\eeqar
where $t \in [ 0, x ]$ with $x \in {\bf R}- \{ 0 , 1 \}$.
To be more rigorous, it is known that $x$ can be analytically continued
to $x \in {\bf C} - \{ 0, 1 \}$ but, for our discussion, it is enough to
restrict $t$ (and $x$) to be real.
Generalization of this expression provides a neat set of representations
for the polylogarithm functions:
\beq
    {\rm Li}_k (x) \, = \, \int_{0}^{x} \om^{(1)}
    \underbrace{\om^{(0)} \om^{(0)} \cdots \om^{(0)} }_{k-1}
    \label{4-8}
\eeq
for $k = 2, 3, \cdots $.
This can be understood from the iterative relation
\beq
    {\rm Li}_{k} (x) = \int_{0}^{x} \frac{ {\rm Li}_{k-1} (t)}{t} dt \, .
    \label{4-9}
\eeq
With a suitable ordering of $W_i$'s, say,
\beq
    0 < W_1 \le W_2 < x \, , ~~~ x \in {\bf R}- \{ 0 , 1 \}\, ,
    \label{4-10}
\eeq
one can then express the dilogarithm function as an iterated integral
over (\ref{4-3}) and (\ref{4-4}):
\beq
    \int_{0}^{x} \om_{1}^{(1)} \om_{2}^{(0)} \, = \, {\rm Li}_2 (x) \, .
    \label{4-11}
\eeq
An analog of (\ref{4-8}) is then written as
\beq
    {\rm Li}_k (x) \, = \, \int_{0}^{x} \om_{1}^{(1)}
    \om_{2}^{(0)} \om_{3}^{(0)} \cdots \om_{k}^{(0)}
    \label{4-12}
\eeq
where, as in the case of (\ref{4-4}), $\om_{k}^{(0)}$ is
defined by
\beq
    \om_{k}^{(0)} \equiv d \log (W^{(0)} - W_k ) ~ = ~ \frac{ d W_k }{W_k}
    \label{4-13}
\eeq
for $k = 2,3, \cdots$. In this general case, the range of $W_i$'s is
given by
\beq
    0 < W_1 \le W_2 \le \cdots \le W_k < x
    \, , ~~~ x \in {\bf R}- \{ 0 , 1 \}\, .
    \label{4-14}
\eeq
For details of these iterated integral expressions,
one may also refer to \cite{Abe:2010sa}.

In what follows, we identify the
integral over $\frac{ d W_1 }{W_1} \frac{ d W_2 }{W_2} $
as the iterated integral (\ref{4-11}).
With such an identification, the problem of integral can be
reduced to that of determining the range of $W_i$'s which
corresponds to the value of $x \in {\bf R} - \{ 0 ,1 \}$.
We shall consider this problem in the following.

\noindent
\underline{Prescription for the BST representation}

In terms of the iterated integral, the integral
$\L_{n:r; i }^{( l_1 , l_2 )}$
in (\ref{3-36}) can be expressed as
\beqar
    \int \om_{1}^{(1)} \om_{2}^{(0)} \,
    \R^{( l_1 , l_2 )}_{n:r;i}
    & = &
    \widehat{R}^{( l_{1}, l_{2} )}_{(i, i+r+1)} \,
    \int \om_{1}^{(1)} \om_{2}^{(0)}
    \, - \,
    \widehat{R}^{( l_{1}, l_{2} )}_{(i, i+r)} \,
    \int \om_{1}^{(1)} \om_{2}^{(0)}
    \nonumber \\
    &&
    - \,
    \widehat{R}^{( l_{1}, l_{2} )}_{(i-1, i+r+1)} \,
    \int \om_{1}^{(1)} \om_{2}^{(0)}
    \, + \,
    \widehat{R}^{( l_{1}, l_{2} )}_{(i-1, i+r)} \,
    \int \om_{1}^{(1)} \om_{2}^{(0)}
    \label{4-15}
\eeqar
where we use the expansion
of $\R^{( l_1 , l_2 )}_{n:r;i}$ in (\ref{3-37}).
Notice that $\R^{( l_1 , l_2 )}_{n:r;i}$ is not a function of
$( W_1 , W_2 )$ but that of $( l_1 , l_2 )$.
The paths of iterated integrals are to be determined.
The expression (\ref{4-15}) lead to
the BST box function (\ref{3-51}), {\it if} we impose the following prescription:
\beq
    \widehat{R}^{( l_{1}, l_{2} )}_{(i, j)} \,
    \int \om_{1}^{(1)} \om_{2}^{(0)}
    ~ \rightarrow ~ \int_{0}^{1 - \Delta_{(i, j)}} \om_{1}^{(1)} \om_{2}^{(0)}
    ~ =~ {\rm Li}_2 ( 1 - \Delta_{(i,j)} )
    \label{4-16}
\eeq
where $\Delta_{(i,i+r+1)}$, $\Delta_{(i,i+r)}$, $\Delta_{(i-1,i+r+1)}$ and
$\Delta_{(i-,i+r)}$ are defined by
\beqar
    \Delta_{(i,i+r+1)} & \equiv & c_{r;i} \, Q^2 ~=~ c_{r;i} \, ( P + p_{i-1} + p_{i+r})^2
    \label{4-17}\\
    \Delta_{(i,i+r)} & \equiv & c_{r;i} \, s ~=~ c_{r;i} \, ( P + p_{i-1} )^2
    \label{4-18}\\
    \Delta_{(i-1,i+r+1)} & \equiv & c_{r;i} \, t ~=~ c_{r;i} \, ( P + p_{i+r} )^2
    \label{4-19}\\
    \Delta_{(i-1,i+r)} & \equiv & c_{r;i} \, P^2
    \label{4-20}
\eeqar
with $c_{r;i}$ given in (\ref{3-52}).
In terms of $P = p_{i} + p_{i+1} + \cdots + p_{i+r-1}$,
$p_{i-1}$ and $p_{i+r}$,  the factor $c_{r;i}$ can be written as
\beq
    c_{r;i} \, = \,
    \frac{( p_{i-1} + p_{i+r} )^2}{P^2 (P + p_{i-1} + p_{i+r} )^2
    - (P+ p_{i-1} )^2 (P + p_{i+r} )^2} \, .
    \label{4-21}
\eeq
As pointed out in \cite{Brandhuber:2004yw}, this is invariant under
the transformations
\beq
    P \, \rightarrow \,
    P + \al p_{i-1} + \bt p_{i+r} \, \equiv \, P_{( \al , \bt )}
    \label{4-22}
\eeq
where $\al$ and $\bt$ are real numbers.
Using $P_{( \al , \bt )}$, we can uniformly express the
definitions (\ref{4-17})-(\ref{4-20}) as
\beq
    \Delta_{(i - 1 + \al , i + r + \bt )} \, = \, c_{r; i} \, P_{( \al , \bt )}^{\, 2}
    \label{4-23}
\eeq
for $( \al , \bt ) = (1,1), \, (1,0), \, (0,1), \, (0,0)$.
Notice that the BST prescription (\ref{4-16}) is symmetric under the exchange
of $l_1$ and $l_2$ indices since the original integration measure
is given by (\ref{4-1}). This means that the action of
$\widehat{R}^{( l_{2}, l_{1} )}_{(i, j)} = \widehat{R}^{( l_{1}, l_{2} )-1}_{(i, j)}$
is effectively the same as that of $\widehat{R}^{( l_{1}, l_{2} )}_{(i, j)}$
in terms of the resultant dilogarithms.

At the preset, it is not clear how to derive the prescription (\ref{4-16}).\footnote{
We have tried to figure out the derivation considerably in preparing the present paper.
However, many approaches (including an idea of zero-modes in the holonomy
formalism \cite{Abe:2010sa}) turn out to be unsuccessful.
We expect that some clues will be given by the invariance of $c_{r;i}$
under the transformations (\ref{4-22}) but can not make a convincing argument yet.
}
{\it What is interesting here is, however, that the same analytic
results as the BST method can be obtained from the $W$-part integrals more directly.
In other words, by use of the iterated-integral representation of
the polylogarithm functions, we can extract finite physical
quantities of one-loop MHV amplitudes more efficiently than the BST method.}
In this sense, it is reasonable to
interpret our prescription (\ref{4-16})-(\ref{4-20}) as a new regularization
scheme for one-loop calculation.
In what follows, we call this scheme ``polylog regularization'' for simplicity.

Use of the iterated-integral representation of polylogarithms in (\ref{4-12})
suggests that the prescription, if generalized to non-MHV amplitudes,
would lead to a systematic production
of polylogarithms in the computation of one-loop amplitudes.
Polylogarithm contributions are generally observed in higher-loop calculations
and corresponding reminder functions.
Thus we do not think that the polylog regularization is just a computational trick
which is applicable to one-loop MHV amplitudes.
Bearing in mind that there are no CSW- or BST-based
analytic expressions for one-loop non-MHV amplitudes, we
find it worth trying how far we can generalize the
polylog regularization scheme to the one-loop calculations;
this is exactly what we shall carry out in the rest of the present paper.

%%%%%%%%%%%%%%%%%%%%%%%%%%%%%%%%%%%%%%%%%%%%%%%%%%%%%%%%%%%
\section{One-loop NMHV amplitudes: formalism}

In the following sections, we apply the above formulation
to one-loop N$^{m}$MHV amplitudes ($m = 1, 2, \cdots n-4$) where
the number of negative-helicity gluons is $(m+2)$, with $n$ being the
total number of scattering gluons.
We do this in a deductive fashion, focusing on
on the case of $m=1$ in the present section.

\noindent
\underline{The $x$-space representation}

As in the MHV amplitudes (\ref{3-22}), the NMHV amplitudes in the $x$-space
representation are generated as
\beqar
    &&
    \left. \frac{\del}{\del a_{1}^{(+) c_1}} \otimes
    \cdots \otimes \frac{\del}{\del a_{a}^{(-) c_a}} \otimes
    \cdots \otimes \frac{\del}{\del a_{b}^{(-) c_b}} \otimes
    \cdots \otimes \frac{\del}{\del a_{c}^{(-) c_c}} \otimes
    \cdots \otimes \frac{\del}{\del a_{n}^{(+) c_n}}
    ~ \F \left[  a^{(h)c} \right] \right|_{a^{(h)c}=0} \nonumber \\
    &=& \A_{NMHV}^{( a_{-} b_{-} c_{-})} (x)
    \nonumber \\
    &=&
    \A_{NMHV(0)}^{(a_{-} b_{-} c_{-})} (x) \, + \, \A_{NMHV(1)}^{(a_{-} b_{-} c_{-})} (x)
    \, + \, \A_{NMHV(2)}^{(a_{-} b_{-} c_{-})} (x) \, + \, \cdots \, .
    \label{5-1}
\eeqar
For tree NMHV amplitudes $\A_{NMHV(0)}^{(a_{-} b_{-} c_{-})} (x)$, there is a single
contraction by the operator $\widehat{W}^{(A)} (x)$ in (\ref{3-20}).
Thus, to this order, the amplitudes are expresses as
\beqar
    \A_{NMHV(0)}^{(a_{-} b_{-} c_{-})} (x)
    &=& \!\!
    \frac{\del}{\del a_{1}^{(+) c_1}} \otimes
    \cdots \otimes \frac{\del}{\del a_{a}^{(-) c_a}} \otimes
    \cdots \otimes \frac{\del}{\del a_{b}^{(-) c_b}} \otimes
    \cdots \otimes \frac{\del}{\del a_{c}^{(-) c_c}} \otimes
    \cdots \otimes \frac{\del}{\del a_{n}^{(+) c_n}}
    \nonumber \\
    &&  \left.  \left[-
    \int d\mu(q) \left(
    \frac{\del}{\del a_{l}^{(+)}} \otimes \frac{\del}{\del a_{-l}^{(-)}}
    \right) e^{-il (x - y)}
    \right]_{y \rightarrow x}
    \, \F \left[  a^{(h)c} \right] \right|_{a^{(h)c}=0}
    \nonumber \\
    &=&
    \left. \frac{\del}{\del a_{1}^{(+) c_1}} \otimes
    \cdots \otimes \frac{\del}{\del a_{n}^{(+) c_n}}
    \, \int d\mu(q) \left(
    \frac{\del}{\del a_{l}^{(+)}} \otimes \frac{\del}{\del a_{-l}^{(-)}}
    \right)
    \, \F \left[  a^{(h)c} \right] \right|_{a^{(h)c}=0} \nonumber \\
    &=&
    \prod_{i = 1}^{n} \int d \mu (p_{i})
    \, \A_{NMHV(0)}^{(a_{-} b_{-} c_{-})} (u, \bu )
    \label{5-2}\\
    \A_{NMHV(0)}^{(a_{-} b_{-} c_{-})} (u, \bu )
    &=& i g^{n-2}
    \, (2 \pi)^4 \del^{(4)} \left( \sum_{i=1}^{n} p_i \right) \,
    \widehat{A}_{NMHV(0)}^{(a_{-} b_{-} c_{-})} (u)
    \label{5-3}
\eeqar
\beqar
    &&
    \widehat{A}_{NMHV(0)}^{(a_{-} b_{-} c_{-})} (u)
    \nonumber \\
    & = &
    -i \sum_{i = 1}^{n} \sum_{r = 1}^{n-3}
    \left.
    \int d \mu ( q_{i \, i+r} )
    \widehat{A}^{( i_{+} \cdots a_{-} \cdots b_{-} \cdots (i+r)_{+} l_{+} )}_{MHV(0)} (u)
    \,
    \widehat{A}^{( (-l)_{-} \, (i+r+1)_{+} \cdots c_{-} \cdots (i-1)_{+} )}_{MHV(0)} (u)
    \right|_{l = q_{i \, i+r} \bar{\eta}}
    \nonumber \\
    & = &
    \sum_{i = 1}^{n} \sum_{r = 1}^{n -3 }
    \left.
    \int \frac{d^4 q_{i\, i+r}}{(2 \pi )^4} \,
    \widehat{A}^{( i_{+} \cdots a_{-} \cdots b_{-} \cdots (i+r)_{+} l_{+} )}_{MHV(0)} (u)
    \, \frac{1}{q_{i \, i+r}^2} \,
    \widehat{A}^{( (-l)_{-} \, (i+r+1)_{+} \cdots c_{-} \cdots (i-1)_{+} )}_{MHV(0)} (u)
    \right|_{l = q_{i \, i+r} \bar{\eta}}
    \nonumber \\
    \label{5-4}
\eeqar
where the expression (\ref{5-4}) is identical to the tree-level CSW-results
(\ref{2-15a}) in section 2 except that the off-shell integral measure
$\frac{d^4 q_{i\, i+r}}{(2 \pi )^4}$ is inserted here.
The difference arises simply because we here consider the amplitudes
in the coordinate-space representation while in section 2 we have formulated
the CSW rules in the momentum-space representation.
As discussed in (\ref{3-21}), the off-shell Nair measure $d \mu (q_{i \, i+r})
= d \mu (l) + \frac{1}{4 \pi} \frac{ d w^2}{w^2}$ (with $q_{i \, i+r} = l + w \eta$)
may be replaced by the on-shell Nair measure $d \mu (l)$ upon regularizing
the log divergence.

Similarly, one-loop NMHV amplitudes $\A_{NMHV(1)}^{(a_{-} b_{-} c_{-})} (x)$
can be generated as
\beqar
    \A_{NMHV(1)}^{(a_{-} b_{-} c_{-})} (x)
    &=&
    \frac{\del}{\del a_{1}^{(+) c_1}} \otimes
    \cdots \otimes \frac{\del}{\del a_{a}^{(-) c_a}} \otimes
    \cdots \otimes \frac{\del}{\del a_{b}^{(-) c_b}} \otimes
    \cdots \otimes \frac{\del}{\del a_{c}^{(-) c_c}} \otimes
    \cdots \otimes \frac{\del}{\del a_{n}^{(+) c_n}}
    \nonumber \\
    &&  \!\!\! \times
    \left[-
    \int d\mu(L_1) \left(
    \frac{\del}{\del a_{l_1}^{(+)}} \otimes \frac{\del}{\del a_{-l_1}^{(-)}}
    \right) e^{-il_1 (x - y_1 )}
    \right]_{y_1 \rightarrow x}
    \nonumber \\
    && \!\!\! \times
    \left[-
    \int d\mu(L_2) \left(
    \frac{\del}{\del a_{l_2}^{(+)}} \otimes \frac{\del}{\del a_{-l_2}^{(-)}}
    \right) e^{-il_2 (x - y_2 )}
    \right]_{y_2 \rightarrow x}
    \nonumber \\
    && \!\!\!  \left. \times
    \left[-
    \int d\mu(L_3) \left(
    \frac{\del}{\del a_{l_3}^{(+)}} \otimes \frac{\del}{\del a_{-l_3}^{(-)}}
    \right) e^{-il_3 (x - y_3 )}
    \right]_{y_3 \rightarrow x}
    \F \left[  a^{(h)c} \right] \right|_{a^{(h)c}=0}
    \label{5-5}
\eeqar
where we make the limit $y \rightarrow x$ explicit for each of the contractions.
Up to the cyclicity of the indices, there are basically two ways of
taking these limits:
\beqar
    && \mbox{Type-I:} ~~~~ y_1 \rightarrow y_2 \, , ~  y_2 \rightarrow y_1 \, , ~
     y_3 \rightarrow y_2 \, ,
    \label{5-6} \\
    && \mbox{Type-II:} ~~~ y_1 \rightarrow y_2 \, , ~  y_2 \rightarrow y_3 \, , ~
     y_3 \rightarrow y_1
     \label{5-7}
\eeqar
which lead to the one-loop amplitudes.

%%%%%%%%%%%%%%%%%%%%%%%%% figure %%%%%%%%%%%%%%%%%%%%%%%%%
\begin{figure} [htbp]
\begin{center}
\includegraphics[width=135mm]{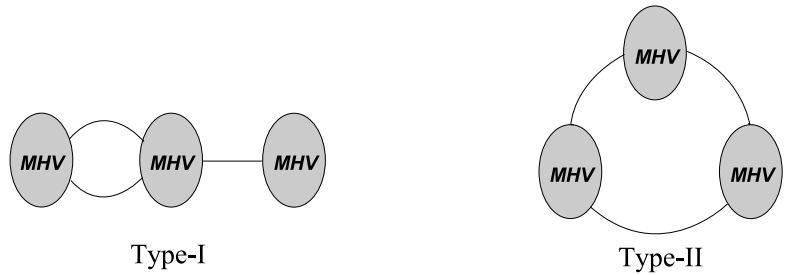}
\caption{MHV diagrams, Type-I (left) and Type-II (right),
contributing to one-loop NMHV amplitudes ---
Each of the MHV clusters has more than or equal to two external legs.}
\label{fighol0404}
\end{center}
\end{figure}
%%%%%%%%%%%%%%%%%%%%%%%%% figure %%%%%%%%%%%%%%%%%%%%%%%%%

\noindent
\underline{Type-I case: dilogarithm contributions}

In the type-I case, (\ref{5-5})
can be calculated as
\beqar
    \A_{NMHV(1; {\rm Li}_2 )}^{(a_{-} b_{-} c_{-})} (x)
    &=&
    \left[-
    \int d\mu(L_1 ) \left(
    \frac{\del}{\del a_{l_1 }^{(+)}} \otimes \frac{\del}{\del a_{-l_1 }^{(-)}}
    \right)
    \right]
    \,
    \left[-
    \int d\mu(L_2 ) \left(
    \frac{\del}{\del a_{l_2 }^{(+)}} \otimes \frac{\del}{\del a_{-l_2 }^{(-)}}
    \right)
    \right]
    \nonumber \\
    &&
    \!\!\!\!\!\!\!\!\!\!\!\!
    \times \left.
    \left[-
    \frac{\del}{\del a_{1}^{(+) c_1}} \otimes
    \cdots \otimes \frac{\del}{\del a_{n}^{(+) c_n}}
    \int d\mu(L_3 ) \left(
    \frac{\del}{\del a_{l_3 }^{(+)}} \otimes \frac{\del}{\del a_{-l_3 }^{(-)}}
    \right)
    \F \left[  a^{(h)c} \right]
    \right]
    \right|_{a^{(h)c}=0} \nonumber \\
    &=&
    \prod_{i = 1}^{n} \int d \mu (p_{i})
    \, \A_{NMHV(1; {\rm Li}_2 )}^{(a_{-} b_{-} c_{-})} (u, \bu ) \, .
    \label{5-8}
\eeqar
The expression in the second line, inside the squared parentheses,
is the same as the tree NMHV amplitudes (\ref{5-2}).
Thus, using (\ref{5-3}) and (\ref{5-4}), we can explicitly
write down the momentum-space amplitudes
$\A_{NMHV(1; {\rm Li}_2 )}^{(a_{-} b_{-} c_{-})} (u, \bu )$ as
\beqar
    \A_{NMHV(1; {\rm Li}_2 )}^{(a_{-} b_{-} c_{-})} (u, \bu )
    \! &=& \!
    i g^{n-2}
    \, (2 \pi)^4 \del^{(4)} \left( \sum_{i=1}^{n} p_i \right) \,
    \widehat{A}_{NMHV(1; {\rm Li}_2 )}^{(a_{-} b_{-} c_{-})} (u)
    \label{5-9}\\
    \widehat{A}_{NMHV(1; {\rm Li}_2 )}^{(a_{-} b_{-} c_{-})} (u)
    \! &=& \!
    \sum_{i = 1}^{n} \sum_{r = 1}^{n-3}
    \int \frac{d^4 q_{i\, i+r}}{(2 \pi )^4} \,
    \widehat{A}^{( i_{+} \cdots a_{-} \cdots b_{-} \cdots (i+r)_{+} l_{3+} )}_{MHV(1)} (u)
    \nonumber \\
    &&
    \times
    \left.
    \, \frac{1}{q_{i \, i+r}^2} \,
    \widehat{A}^{( (-l_3)_{-} \, (i+r+1)_{+} \cdots c_{-} \cdots (i-1)_{+} )}_{MHV(0)} (u)
    \right|_{l_3 = q_{i \, i+r} \bar{\eta}}
    \nonumber \\
    \label{5-10}
\eeqar
where the one-loop MHV amplitudes
$\widehat{A}^{( i_{+} \cdots a_{-} \cdots b_{-} \cdots (i+r)_{+} l_{3+} )}_{MHV(1)} (u)$
can be obtained from (\ref{3-32}) by replacing $n$ with $r+1$.
We have an extra internal index $l_3$ here.
Since we assign a $U(1)$ color direction to the internal index,
this is an auxiliary index in practical calculation.
Notice that the internal index has positive helicity, otherwise
one of the MHV clusters become no more MHV but NMHV due to the freedom
of the choice for the signs of internal null-momenta $l_1$ and $l_2$.
The amplitudes
$\widehat{A}^{( i_{+} \cdots a_{-} \cdots b_{-} \cdots (i+r)_{+} l_{3+} )}_{MHV(1)} (u)$
can then be written explicitly as
\beqar
    &&
    \widehat{A}^{( i_{+} \cdots a_{-} \cdots b_{-} \cdots (i+r)_{+} l_{3+} )}_{MHV(1)} (u)
    \nonumber \\
    & = &
    i g^2 \,
    \sum_{j = i}^{i+r}
    \sum_{t = 1}^{\left\lfloor \frac{r+1}{2} \right\rfloor -1 }
    \left( 1 - \frac{1}{2} \del_{ \frac{r+1}{2} , t+1} \right)
    \sum_{ \si^{(1)} \in \S_{t+1}}
    \sum_{ \si^{(2)} \in \S_{r-t}} \!\!\!
    \Tr ( t^{\si^{(1)}_{j}} \cdots t^{\si^{(1)}_{j+t}} \, t^{\si^{(2)}_{j+t+1}}
    \cdots t^{\si^{(2)}_{j+r}} t^{l_3} )
    \nonumber \\
    &&
    \, \L_{r+1: t ; \si_j }^{( l_1 , l_2 )} \,
    \widehat{C}_{MHV(0)}^{(j_{+} \cdots a_{-} \cdots b_{-} \cdots (j+r)_{+} l_{3+} )}
    (u; \si^{(1)} \otimes \si^{(2)})
    \nonumber \\
    &=&
    i g^2 \,
    \sum_{j = i}^{i+r}
    \sum_{t = 1}^{\left\lfloor \frac{r+1}{2} \right\rfloor -1 }
    \left( 1 - \frac{1}{2} \del_{ \frac{r+1}{2} , t+1} \right)
    \, \Biggl[ \,
    \Tr ( t^{j} \cdots t^{j+t} \, t^{j+t+1} \cdots t^{j+r} t^{l_3} )
    \, \L_{r+1: t ; j }^{( l_1 , l_2 )}
    \nonumber \\
    &&
    \times \,
    \widehat{C}_{MHV(0)}^{(j_{+} \cdots a_{-} \cdots b_{-} \cdots (j+r)_{+} l_{3+} )}
    (u; {\bf 1}^{(1)} \otimes {\bf 1}^{(2)})
    \, + \,
    \P ( j \cdots j+t | j+t+1 \cdots j+r )
    \, \Biggr]
    \label{5-11}
\eeqar
where
\beq
    \si^{(1)} \, = \,
    \left(
      \begin{array}{c}
        j ~~ \cdots ~~ j+t \\
        \si^{(1)}_{j} \cdots ~ \si^{(1)}_{j+t} \\
      \end{array}
    \right) , ~~
    \si^{(2)} \, = \,
    \left(
      \begin{array}{c}
        j+t+1 ~~ \cdots ~~ j+r \\
        \si^{(2)}_{j+t+1} ~ \cdots ~ \si^{(2)}_{j+r} \\
      \end{array}
    \right) .
    \label{5-12}
\eeq
The expression (\ref{5-11}) suggests that the one-loop
MHV diagram can be treated as a unit cluster in a broad sense.
Namely, the position of the internal index $l_3$ in the one-loop MHV diagram
can be chosen arbitrarily; only the fact that the one-loop MHV diagram is
connected to the other MHV cluster is essential in drawing the MHV diagram of
interest; see the type-I diagram in Figure \ref{fighol0404}.
In application of the polylog regularization, $\L_{r+1: t ; j }^{( l_1 , l_2 )}$ is
given by
\beqar
    \L_{r+1: t ; j }^{( l_1 , l_2 )}
    &=&
    \frac{1}{(4 \pi )^2}
    \int \om_{1}^{(1)} \om_{2}^{(0)} \,
    \R_{r+1: t ; j }^{( l_1 , l_2 )}
    \nonumber \\
    &=& \frac{1}{ (4 \pi)^2 }
    \biggl[ \,
    {\rm Li}_2 ( 1 - \Delta_{(j, j+t+1)})
    \, - \, {\rm Li}_2 ( 1 - \Delta_{(j, j+t)})
    \nonumber \\
    && ~
    \, - \, {\rm Li}_2 ( 1 - \Delta_{( j+r , j+t+1 )})
    \, + \, {\rm Li}_2 ( 1 - \Delta_{( j+r , j+t )}) \,
    \biggr]
    \label{5-13}
\eeqar
where $\Delta$'s are defined by
\beq
    \begin{array}{rcl}
    \Delta_{(j,j+t+1)} & = & c_{t;j} \, ( P + p_{j+r} + p_{j+t})^2
    \\
    \Delta_{(j, j+t)} & = & c_{t;j} \, ( P + p_{j+r} )^2
    \\
    \Delta_{( j+r , j+t+1)} & = & c_{t;j} \, ( P + p_{j+t} )^2
    \\
    \Delta_{( j+r , j+t )} & = & c_{t;j} \, P^2
    \\
    \end{array}
    \label{5-14}
\eeq
with $P = q_{j \, j+t-1} = p_j + p_{j+1} + \cdots + p_{j+t-1}$, and
$c_{t;j}$ being defined as
\beq
    c_{t;j} \, = \,
    \frac{( p_{j+r} + p_{j+t} )^2}{P^2 (P + p_{j+r} + p_{j+t} )^2
    - (P + p_{j+r} )^2 ( P + p_{j+t} )^2} \, .
    \label{5-15}
\eeq

The color factor of
$\widehat{A}_{NMHV(1; {\rm Li}_2 )}^{(a_{-} b_{-} c_{-})} (u)$
becomes single-trace, with the internal indices
$ \pm l_3$  contracted by each other.
Explicitly, we can write down the full amplitudes as
\beqar
    &&
    \widehat{A}_{NMHV(1; {\rm Li}_2 )}^{(a_{-} b_{-} c_{-})} (u)
    \nonumber \\
    &=&
    i g^2 \,
    \sum_{i = 1}^{n}
    \sum_{r = 1}^{n-3}
    \sum_{j = i}^{i+r}
    \sum_{t = 1}^{\left\lfloor \frac{r+1}{2} \right\rfloor -1 }
    \left( 1 - \frac{1}{2} \del_{ \frac{r+1}{2} , t+1} \right)
    \nonumber \\
    &&
    \sum_{ \si^{(1)} \in \S_{t+1}} \,
    \sum_{ \si^{(2)} \in \S_{r-t}} \,
    \sum_{ \si^{(3)} \in \S_{n-r-1}} \!\!
    \Tr ( t^{\si^{(1)}_{j}} \cdots t^{\si^{(1)}_{j+t}} \, t^{\si^{(2)}_{j+t+1}}
    \cdots t^{\si^{(2)}_{j+r}} \, t^{\si^{(3)}_{i+r+1}} \cdots t^{\si^{(3)}_{i-1}}  )
    \nonumber \\
    &&
    \int \frac{d^4 q_{i\, i+r}}{(2 \pi )^4} ~
    \L_{r+1: t ; \si_j }^{( l_1 , l_2 )} ~
    \widehat{C}_{MHV(0)}^{( j_{+} \cdots a_{-} \cdots b_{-} \cdots (j+r)_{+} l_{3+} )}
    (u; \si^{(1)} \otimes \si^{(2)})
    \nonumber \\
    &&
    \times
    \left.
    \, \frac{1}{q_{i \, i+r}^2} \,
    \widehat{C}^{( (-l_3)_{-} \, (i+r+1)_{+} \cdots c_{-} \cdots (i-1)_{+} )}_{MHV(0)}
    (u; \si^{(3)})
    \right|_{l_3 = q_{i \, i+r} \bar{\eta}}
    \label{5-16}
\eeqar
where $\si^{(3)}$ is the transposition corresponding to
the subamplitude
$\widehat{A}^{( (-l_3)_{-} \, (i+r+1)_{+} \cdots c_{-} \cdots (i-1)_{+} )}_{MHV(0)} (u)$
in (\ref{5-10}):
\beq
    \si^{(3)} \, = \,
    \left(
      \begin{array}{c}
        i+r+1 ~~ \cdots ~~ i-1 \\
        \si^{(3)}_{i+r+1} ~ \cdots ~ \si^{(3)}_{i-1} \\
      \end{array}
    \right) \, .
    \label{5-17}
\eeq
Since the amplitudes
$\widehat{A}_{NMHV(1; {\rm Li}_2 )}^{(a_{-} b_{-} c_{-})} (u)$
include a single-line propagator, as in the case of tree NMHV amplitudes,
these are determined up to the choice of reference spinors.
This fact is reflected in the $\pm l_3$-dependence in the expression (\ref{5-16}).
Notice that $\pm l_3$ enter only in $\widehat{C}_{MHV(0)}$'s. Thus
the sign of $l_3$ is irrelevant in the final results as expected
from our functional derivation.

\noindent
\underline{Type-II case: trilogarithm contributions}

In the type-II case (\ref{5-7}), on the other hand, we can no more utilize
the results of the previous sections.
This is because the three contraction operators in (\ref{5-5}) are
now treated equally so that we need to consider the three
as one set of operators.
The situation can readily be perceived
from the MHV diagrams in Figure \ref{fighol0404} contributing
to one-loop NMHV amplitudes.

\begin{figure} [htbp]
\begin{center}
\includegraphics[width=90mm]{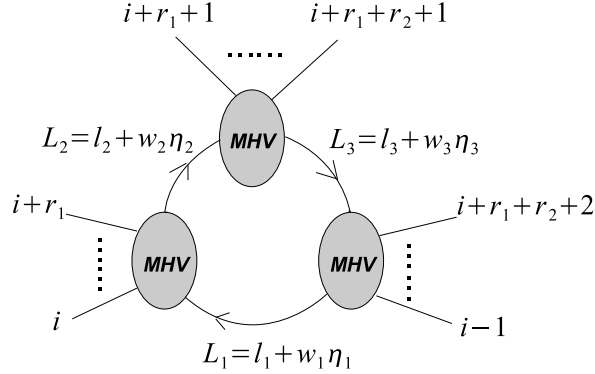}
\caption{The type-II MHV diagram that contributes to one-loop NMHV amplitudes}
\label{fighol0405}
\end{center}
\end{figure}

In the type-II case, the functional derivatives in (\ref{5-5})
can be calculated as
\beqar
    \A_{NMHV(1; {\rm Li}_3 )}^{(a_{-} b_{-} c_{-})} (x)
    &=&
    \left[ -
    \int d\mu(L_1 ) \left(
    \frac{\del}{\del a_{l_1 }^{(+)}} \otimes \frac{\del}{\del a_{-l_1 }^{(-)}}
    \right)
    \right] \,
    \left[ -
    \int d\mu(L_2 ) \left(
    \frac{\del}{\del a_{l_2 }^{(+)}} \otimes \frac{\del}{\del a_{-l_2 }^{(-)}}
    \right)
    \right]
    \nonumber \\
    && \!\!\!\!\!\!\!\!\!\!\!\!  \times
    \left.
    \left[-
    \int d\mu(L_3 ) \left(
    \frac{\del}{\del a_{l_3 }^{(+)}} \otimes \frac{\del}{\del a_{-l_3 }^{(-)}}
    \right)
    \right]
    \frac{\del}{\del a_{1}^{(+) c_1}} \otimes
    \cdots \otimes \frac{\del}{\del a_{n}^{(+) c_n}}
    \F \left[  a^{(h)c} \right]
    \right|_{a^{(h)c}=0 }
    \nonumber \\
    &=&
    \prod_{i = 1}^{n} \int d \mu (p_{i})
    \, \A_{NMHV(1; {\rm Li}_3 )}^{(a_{-} b_{-} c_{-})} (u, \bu )
    \label{5-18}
\eeqar
where we label the numbering indices of the type-II MHV diagram
as shown in Figure \ref{fighol0405}.

%%%%%%%%%%%%%%%%%%%%%%%%% figure %%%%%%%%%%%%%%%%%%%%%%%%%
%%%%%%%%%%%%%%%%%%%%%%%%% figure %%%%%%%%%%%%%%%%%%%%%%%%%

The amplitudes of interest,
$\widehat{A}_{NMHV(1; {\rm Li}_3 )}^{(a_{-} b_{-} c_{-})} (u, \bu )$
in (\ref{5-18}), are then straightforwardly calculated as
\beqar
    \A_{NMHV(1; {\rm Li}_3 )}^{(a_{-} b_{-} c_{-})} (u, \bu )
    &=& i g^{n-2}
    \, (2 \pi)^4 \del^{(4)} \left( \sum_{i=1}^{n} p_i \right) \,
    \widehat{A}_{NMHV(1; {\rm Li}_3 )}^{(a_{-} b_{-} c_{-})} (u)
    \label{5-19}\\
    \widehat{A}_{NMHV(1; {\rm Li}_3 )}^{(a_{-} b_{-} c_{-})} (u)
    &=&
     g^2 \,
    \sum_{i = 1}^{n}
    \sum_{r_1 = 1}^{\left\lfloor \frac{n}{3}  \right\rfloor -1 }
    \sum_{r_2 = 1}^{\left\lfloor \frac{n}{3}  \right\rfloor -1 }
    \left( 1 - \frac{2}{3} \del_{ \frac{n}{3}, r_{1} +1} \del_{ \frac{n}{3}, r_{2} +1} \right)
    \int d \mu (L_1 ) d \mu (L_2 ) d \mu (L_3 )
    \nonumber \\
    &&
    \widehat{A}_{MHV(0)}^{( (-l_{1} )_{-} i_{+} (i+1)_{+}
    \cdots a_{-} \cdots (i+r_{1})_{+} l_{2+} )}(u)
    \,
    \widehat{A}_{MHV(0)}^{( (-l_{2} )_{-} (i+r_1 +1)_{+} \cdots b_{-}
    \cdots (i+r_{1}+r_2 + 1)_{+} l_{3+} )} (u)
    \nonumber \\
    &&
    \widehat{A}_{MHV(0)}^{( (-l_{3} )_{-} (i+r_1 + r_2 + 2)_{+}
    \cdots c_{-} \cdots (i-1 )_{+} l_{1+} )} (u)
    \label{5-20}
\eeqar
where the factor
$\left( 1 - \frac{2}{3} \del_{ \frac{n}{3}, r_{1} +1} \del_{ \frac{n}{3}, r_{2} +1} \right)$
arises to compensate the redundant counting when
the type-II MHV diagram preserves a cyclic symmetry.
The cyclic symmetry is an analog of the reflection symmetry
considered in the one-loop MHV diagram.
As in the MHV case, this factor is related to a factor of $\frac{1}{3!}$
coming from the Taylor expansion of the MHV S-matrix functional (\ref{2-33})
under the cyclic symmetry. In the above expression, however, the calculation
is made in a rotational fashion in terms of the numbering indices. Thus the
``reflection'' part of the factor $\frac{1}{3!}$, {\it i.e.}, the factor of $\frac{1}{2}$,
is irrelevant and only the factor of $\frac{1}{3}$ survives to take care of the redundancy
in (\ref{5-20}). The negative-helicity indices $(a_{-} b_{-} c_{-})$ satisfy the ordering
\beq
    i \le a \le i+r_1 < b \le i+r_1 + r_2 + 1 < c ~~~ \mbox{(mod $n$)} \, .
    \label{5-21}
\eeq
This choice is the same as the tree NMHV amplitudes in (\ref{2-8}),
with an identification of $j = i + r_1 + r_2 + 1$.
In the expression (\ref{5-20}), we specify the positions and the signs
of $( \pm l_1 , \pm l_2 , \pm l_3 )$ in accord with the arrows in
Figure \ref{fighol0405}.
This specification, however, is superficial because, as in the
one-loop MHV amplitudes,
these positions can arbitrarily be chosen
(thanks to the functional derivatives) and
the holomorphic part of the amplitudes (in terms of the spinor momenta)
is independent of the signs of these indices.
This feature is already discussed in (\ref{3-29})-(\ref{3-31})
and is used to show that the one-loop MHV amplitudes are proportional
to the color-stripped tree MHV amplitudes; see (\ref{3-32}).
Analogously, in the present case, we expect that the type-II one-loop
NMHV amplitudes are proportional to some holomorphic tree amplitudes
which are independent of the internal indices $( l_1 , l_2 , l_3 )$.
This can be carried out as follows.

For simplicity, we first denote the key indices as
\beqar
    &&
    a_1 = i \, , ~~ a_2 = i + r_1 \, ,
    \nonumber \\
    &&
    b_1 = i + r_1 + 1 \, , ~~ b_2 = i + r_1 + r_2 + 1 \, ,
    \label{5-22}\\
    &&
    c_1 = i + r_1 + r_2 + 2 \, , ~~ c_2 = i - 1 \, .
    \nonumber
\eeqar
The tree-part of the amplitude in (\ref{5-20}) is then expressed as
\beqar
    &&
    \widehat{A}_{MHV(0)}^{( (-l_{1} )_{-} a_{1+}
    \cdots a_{-} \cdots a_{2+} l_{2+} )}(u)
    \,
    \widehat{A}_{MHV(0)}^{( (-l_{2} )_{-} b_{1+} \cdots b_{-}
    \cdots b_{2+} l_{3+} )} (u)
    \,
    \widehat{A}_{MHV(0)}^{( (-l_{3} )_{-} c_{1+}
    \cdots c_{-} \cdots c_{2+} l_{1+} )} (u)
    \nonumber \\
    &=& \!\!\!
    \sum_{ \si^{(1)} \in \S_{r_{1}+1}} \,
    \sum_{ \si^{(2)} \in \S_{r_{2}+1}} \,
    \sum_{ \si^{(3)} \in \S_{r_{3}+1}} \!\!
    \Tr ( t^{\si^{(1)}_{a_1}} \cdots t^{\si^{(1)}_{a_2}} \, t^{\si^{(2)}_{b_1}}
    \cdots t^{\si^{(2)}_{b_2}} \, t^{\si^{(3)}_{c_1}} \cdots t^{\si^{(3)}_{c_2}}  )
    \nonumber \\
    && \!\!\!\!\!\!\!\!\!
    \times ~
    \widehat{C}_{MHV(0)}^{( (-l_{1} )_{-} a_{1+}
    \cdots a_{-} \cdots a_{2+} l_{2+} )}(u; \si^{(1)})
    \,
    \widehat{C}_{MHV(0)}^{( (-l_{2} )_{-} b_{1+} \cdots b_{-}
    \cdots b_{2+} l_{3+} )} (u; \si^{(2)})
    \,
    \widehat{C}_{MHV(0)}^{( (-l_{3} )_{-} c_{1+}
    \cdots c_{-} \cdots c_{2+} l_{1+} )} (u; \si^{(3)})
    \nonumber \\
    &=&
    \Tr ( t^{a_1} \cdots t^{a_2} \, t^{b_1}
    \cdots t^{b_2} \, t^{c_1} \cdots t^{c_2}  ) \,
    \widehat{C}_{MHV(0)}^{( (-l_{1} )_{-} a_{1+}
    \cdots a_{-} \cdots a_{2+} l_{2+} )}(u; {\bf 1}^{(1)})
    \nonumber \\
    &&
    \times ~
    \,
    \widehat{C}_{MHV(0)}^{( (-l_{2} )_{-} b_{1+} \cdots b_{-}
    \cdots b_{2+} l_{3+} )} (u; {\bf 1}^{(2)})
    \,
    \widehat{C}_{MHV(0)}^{( (-l_{3} )_{-} c_{1+}
    \cdots c_{-} \cdots c_{2+} l_{1+} )} (u; {\bf 1}^{(3)})
    \nonumber \\
    &&
    + \,
    \P ( a_1 \cdots a_2 | b_1 \cdots b_2 | c_1 \cdots c_2 )
    \label{5-23}
\eeqar
where we introduce an auxiliary parameter $r_3 = n - (r_1 + r_2 + 3)$.
$\P ( a_1 \cdots a_2 | b_1 \cdots b_2 | c_1 \cdots c_2 )$
denotes the terms obtained by the triple permutations of
$\si^{(1)}$, $\si^{(2)}$ and $\si^{(3)}$. The above expressions is
the type-II NMHV version of the MHV expression (\ref{3-27}).
Using the relation (\ref{3-29}), we then find the following relation:
\beqar
    &&
    \widehat{C}_{MHV(0)}^{( (-l_{1} )_{-} a_{1+}
    \cdots a_{-} \cdots a_{2+} l_{2+} )}(u; {\bf 1}^{(1)})
    \,
    \widehat{C}_{MHV(0)}^{( (-l_{2} )_{-} b_{1+} \cdots b_{-}
    \cdots b_{2+} l_{3+} )} (u; {\bf 1}^{(2)})
    \nonumber \\
    &=&
    \widehat{C}_{MHV(0)}^{( (-l_{1} )_{-} a_{1+}
    \cdots a_{-} \cdots a_{2+} l_{2+} )}(u; {\bf 1}^{(1)})
    \,
    \widehat{C}_{MHV(0)}^{( (-l_{2} )_{-} b_{1+} \cdots b_{-}
    \cdots b_{2+} l_{1+} )} (u; {\bf 1}^{(2)})
    \,
    \frac{( b_2 ~ l_3 ) ( l_3 ~ (-l_{2}) )}{( b_2 ~ l_1 ) ( l_1 ~ (-l_{2}) )}
    \nonumber \\
    & \rightarrow &
    \R_{r_1 + r_2 + 2: r_1; a_1 }^{( l_1 , l_2 )}
    \widehat{C}_{MHV(0)}^{( a_{1+} \cdots a_{-} \cdots b_{-} \cdots b_{2+} )}
    (u; {\bf 1}^{(1)} \otimes {\bf 1}^{(2)})
    \,
    \frac{( b_2 ~ l_3 ) ( l_3 ~ (-l_{2}) )}{( b_2 ~ l_1 ) ( l_1 ~ (-l_{2}) )}
    \label{5-24}
\eeqar
where the total number of indices involved
in $\R_{r_1 + r_2 + 2: r_1; a_1 }^{( l_1 , l_2 )}$ is now $(r_1 + r_2 + 2)$.
In terms of (\ref{5-22}), it is written as
\beqar
    \R_{r_1 + r_2 + 2: r_1; a_1 }^{( l_1 , l_2 )}
    &=&
    - \frac{( a_{2} ~ b_{1}) ( l_{1} ~ l_{2})}{( a_{2} ~ l_{2})( -l_{1} ~ a_{1})}
    \frac{( b_{2} ~ a_{1}) ( l_{2} ~ l_{1})}{( b_{2} ~ l_{1}) ( -l_{2} ~ b_{1})}
    \nonumber \\
    &=&
    \frac{( a_{1} ~ b_{2}) ( b_{1} ~ a_{2} ) ( l_{1} ~ l_{2})^2}
    {( l_{1} ~ a_{1})( l_{1} ~ b_{2})( l_{2} ~ a_{2})( l_{2} ~ b_{1})} \, .
    \label{5-25}
\eeqar

Similarly, we find
\beqar
    &&
    \widehat{C}_{MHV(0)}^{( (-l_{2} )_{-} b_{1+} \cdots b_{-}
    \cdots b_{2+} l_{3+} )} (u; {\bf 1}^{(2)})
    \,
    \widehat{C}_{MHV(0)}^{( (-l_{3} )_{-} c_{1+}
    \cdots c_{-} \cdots c_{2+} l_{1+} )} (u; {\bf 1}^{(3)})
    \nonumber\\
    & \rightarrow &
    \R_{r_2 + r_3 + 2: r_2; b_1}^{( l_2 , l_3 )}
    \widehat{C}_{MHV(0)}^{( b_{1+} \cdots b_{-} \cdots c_{-} \cdots c_{2+} ) }
    (u; {\bf 1}^{(2)} \otimes {\bf 1}^{(3)})
    \,
    \frac{( c_2 ~ l_1 ) ( l_1 ~ (-l_{3}) )}{( c_2 ~ l_2 ) ( l_2 ~ (-l_{3}) )}
    \, ,
    \label{5-26}\\
    &&
    \widehat{C}_{MHV(0)}^{( (-l_{3} )_{-} c_{1+} \cdots c_{-}
    \cdots c_{2+} l_{1+} )} (u; {\bf 1}^{(3)})
    \,
    \widehat{C}_{MHV(0)}^{( (-l_{1} )_{-} a_{1+}
    \cdots a_{-} \cdots a_{2+} l_{2+} )} (u; {\bf 1}^{(1)})
    \nonumber \\
    & \rightarrow &
    \R_{r_3 + r_1 + 2: r_3; c_1}^{( l_3 , l_1 )}
    \widehat{C}_{MHV(0)}^{( c_{1+} \cdots c_{-} \cdots a_{-} \cdots a_{2+} ) }
    (u; {\bf 1}^{(3)} \otimes {\bf 1}^{(1)})
    \,
    \frac{( a_2 ~ l_2 ) ( l_2 ~ (-l_{1}) )}{( a_2 ~ l_3 ) ( l_3 ~ (-l_{1}) )}
    \, .
    \label{5-27}
\eeqar
From the definition (\ref{3-30}), we can readily write down
$\R^{( l_2 , l_3 )}_{r_2 + r_3 + 2: r_2; b_1}$ and
$\R^{( l_3 , l_1 )}_{r_3 + r_1 + 2: r_3; c_1}$ as
\beqar
    \R^{( l_2 , l_3 )}_{r_2 + r_3 + 2: r_2; b_1}
    &=&
    \frac{( b_{1} ~ c_{2}) ( c_{1} ~ b_{2}) ( l_{2} ~ l_{3})^2}
    {( l_{2} ~ b_{1})( l_{2} ~ c_{2})( l_{3} ~ b_{2})( l_{3} ~ c_{1})} \, ,
    \label{5-28} \\
    \R^{( l_3 , l_1 )}_{r_3 + r_1 + 2: r_3; c_1}
    &=&
    \frac{( c_{1} ~ a_{2}) ( a_{1} ~ c_{2} ) ( l_{3} ~ l_{1})^2}
    {( l_{3} ~ c_{1})( l_{3} ~ a_{2})( l_{1} ~ c_{2})( l_{1} ~ a_{1})} \, .
    \label{5-29}
\eeqar

From these relations, we can express the square of
the holomorphic factor in (\ref{5-23}) as
\beqar
    &&
    \!\!\!\!\!\!\!\!
    \left( \,
    \widehat{C}_{MHV(0)}^{( (-l_{1} )_{-} a_{1+}
    \cdots a_{-} \cdots a_{2+} l_{2+} )} (u; {\bf 1}^{(1)}) \,
    \widehat{C}_{MHV(0)}^{( (-l_{2} )_{-} b_{1+} \cdots b_{-}
    \cdots b_{2+} l_{3+} )} (u; {\bf 1}^{(2)}) \,
    \widehat{C}_{MHV(0)}^{( (-l_{3} )_{-} c_{1+}
    \cdots c_{-} \cdots c_{2+} l_{1+} )} (u; {\bf 1}^{(3)}) \,
    \right)^2
    \nonumber \\
    &&
    \!\!\!\!\!\!\!\!
    = \,
    \R^{( l_1 , l_2 )}_{r_1 + r_2 +2:r_1; a_1} \,
    \R^{( l_2 , l_3 )}_{r_2 + r_3 +2:r_2; b_1} \,
    \R^{( l_3 , l_1 )}_{r_3 + r_1 +2:r_3; c_1} \,
    \R^{( l_1 , l_2 , l_3 )}_{(a_2 , b_2 , c_2)}
    \,
    \widehat{C}_{MHV(0)}^{( a_{1+}  \cdots a_{-} \cdots b_{-} \cdots b_{2+} )}
    (u; {\bf 1}^{(1)} \otimes {\bf 1}^{(2)}) \,
    \nonumber \\
    &&
    \times \,
    \widehat{C}_{MHV(0)}^{( b_{1+}  \cdots b_{-} \cdots c_{-} \cdots c_{2+} )}
    (u; {\bf 1}^{(2)} \otimes {\bf 1}^{(3)}) \,
    \widehat{C}_{MHV(0)}^{( c_{1+}  \cdots c_{-} \cdots a_{-} \cdots a_{2+} )}
    (u; {\bf 1}^{(3)} \otimes {\bf 1}^{(1)})
    \label{5-30}
\eeqar
where $\R^{( l_1 , l_2 , l_3 )}_{(a_2 , b_2 , c_2)}$ is defined by
\beq
    \R^{( l_1 , l_2 , l_3 )}_{(a_2 , b_2 , c_2)}
    \, = \, -
    \frac{( l_3 ~ b_2 )( l_1 ~ c_2 )( l_2 ~ a_2 )}{( l_1 ~ b_2 )( l_2 ~ c_2 )( l_3 ~ a_2 )}
    \, .
    \label{5-31}
\eeq
From (\ref{5-20}), (\ref{5-23}) and (\ref{5-30}), we find that the
amplitudes of interest can be expressed as
\beqar
    &&
    \widehat{A}_{NMHV(1; {\rm Li}_3 )}^{(a_{-} b_{-} c_{-})} (u)
    \nonumber \\
    &=&
    i g^2\,
    \sum_{i = 1}^{n}
    \sum_{r_1 = 1}^{\left\lfloor \frac{n}{3}  \right\rfloor -1 }
    \sum_{r_2 = 1}^{\left\lfloor \frac{n}{3}  \right\rfloor -1 }
    \left( 1 - \frac{2}{3} \del_{ \frac{n}{3}, r_{1} +1} \del_{ \frac{n}{3}, r_{2} +1} \right)
    \int d \mu (L_1 ) d \mu (L_2 ) d \mu (L_3 )
    \nonumber \\
    &&
    \!\!\!
    \Tr ( t^{a_1} \cdots t^{a_2} \, t^{b_1}
    \cdots t^{b_2} \, t^{c_1} \cdots t^{c_2}  ) \,
    \sqrt{-
    \R^{( l_1 , l_2 )}_{r_1 + r_2 +2:r_1; a_1} \,
    \R^{( l_2 , l_3 )}_{r_2 + r_3 +2:r_2; b_1} \,
    \R^{( l_3 , l_1 )}_{r_3 + r_1 +2:r_3; c_1} \,
    \R^{( l_1 , l_2 , l_3 )}_{(a_2 , b_2 , c_2)}
    }
    \nonumber \\
    &&
    \!\!\!
    \times \,
    \sqrt{
    \widehat{C}_{MHV(0)}^{( a_{1+}  \cdots a_{-} \cdots b_{-} \cdots b_{2+} )}
    (u; {\bf 1}^{(1 \otimes 2)}) \,
    \widehat{C}_{MHV(0)}^{( b_{1+}  \cdots b_{-} \cdots c_{-} \cdots c_{2+} )}
    (u; {\bf 1}^{(2 \otimes 3)}) \,
    \widehat{C}_{MHV(0)}^{( c_{1+}  \cdots c_{-} \cdots a_{-} \cdots a_{2+} )}
    (u; {\bf 1}^{(3 \otimes 1)})
    }
    \nonumber \\
    &&
    \!\!\!
    + \,
    \P ( a_1 \cdots a_2 | b_1 \cdots b_2 | c_1 \cdots c_2 )
    \label{5-32}
\eeqar
where we abbreviate ${\bf 1}^{(1)} \otimes {\bf 1}^{(2)}$
by ${\bf 1}^{(1 \otimes 2)}$, and so on.
Precisely speaking,
there exists a sign ambiguity in defining the above amplitudes
which arises from taking the square root of the quantity (\ref{5-30}).
This ambiguity always exists in the holonomy formalism as discussed
below (\ref{2-35}). The physical quantities are, however, independent of
this ambiguity so that, in this sense, we can fix the sign of (\ref{5-32}) uniquely.
The indices $\{ a_1 , a_2 , b_1 , b_2 , c_1 , c_2 \}$ depend on the summing
indices $i$, $r_1$ and $r_2$ as specified in (\ref{5-22}).
The holomorphic part of the amplitudes are given by
the square root in the forth line of (\ref{5-32}).
Notice that this factor is in accord with the conventional relation (\ref{2-24})
between the helicity configurations and the degrees of homogeneity
in spinor momenta.
The square root in the third line can explicitly be calculated as
\beqar
    &&
    \sqrt{-
    \R^{( l_1 , l_2 )}_{r_1; a_1} \,
    \R^{( l_2 , l_3 )}_{r_2; b_1} \,
    \R^{( l_3 , l_1 )}_{r_3; c_1} \,
    \R^{( l_1 , l_2 , l_3 )}_{(a_2 , b_2 , c_2)}
    }
    \nonumber \\
    &=&
    \frac{
    ( l_1 ~ l_2 ) ( l_2 ~ l_3 ) ( l_3 ~ l_1 )
    \sqrt{
    ( a_1 ~ b_2 ) ( b_1 ~ a_2 )
    ( b_1 ~ c_2 ) ( c_1 ~ b_2 )
    ( c_1 ~ a_2 ) ( a_1 ~ c_2 )
    }
    }{
    ( l_1 ~ a_1 ) ( l_1 ~ b_2 )
    ( l_2 ~ b_1 ) ( l_2 ~ c_2 )
    ( l_3 ~ c_1 ) ( l_3 ~ a_2 )
    }
    \, .
    \label{5-33}
\eeqar

We now simplify the expression (\ref{5-33}), using the single
reference-spinor principle in the CSW rules. Namely,
we shall impose the condition
\beq
    \eta_1 \, = \, \eta_2 \, = \, \eta_3 \, .
    \label{5-34}
\eeq
In Figure \ref{fighol0405},
this implies that at least
one of the external legs for each MHV cluster is
collinear to one another. For example, we can impose the
collinearity
\beq
    u_{a_1} \, \parallel \, u_{b_1} \parallel \, u_{c_1} \, .
    \label{5-35}
\eeq
Notice that this condition is a natural consequence of
the CSW generalization rather than a convenient assumption.
By use of the Schouten identities, we can then simplify (\ref{5-33}) as
\beqar
    &&
    \left.
    \sqrt{-
    \R^{( l_1 , l_2 )}_{r_1; a_1} \,
    \R^{( l_2 , l_3 )}_{r_2; b_1} \,
    \R^{( l_3 , l_1 )}_{r_3; c_1} \,
    \R^{( l_1 , l_2 , l_3 )}_{(a_2 , b_2 , c_2)}
    }
    \right|_{ a_1 \parallel b_1 \parallel c_1}
    \nonumber \\
    &=&
    \left.
    \frac{
    ( l_1 ~ l_2 ) ( l_2 ~ l_3 ) ( l_3 ~ l_1 )
    ( a_1 ~ a_2 )
    ( b_1 ~ b_2 )
    ( c_1  ~ c_2 )
    }{
    ( l_1 ~ a_1 ) ( l_1 ~ b_2 )
    ( l_2 ~ b_1 ) ( l_2 ~ c_2 )
    ( l_3 ~ c_1 ) ( l_3 ~ a_2 )
    }
    \right|_{ a_1 \parallel b_1 \parallel c_1}
    \nonumber \\
    &=&
    -
    \widehat{R}^{( l_1 , l_2 )}_{( b_2 , b_1 )}
    \widehat{R}^{( l_2 , l_3 )}_{( c_2 , c_1 )}
    \widehat{R}^{( l_3 , l_1 )}_{( a_2 , a_1 )}
    +
    \widehat{R}^{( l_1 , l_2 )}_{( b_2 , c_2 )}
    \widehat{R}^{( l_3 , l_1 )}_{( a_2 , a_1 )}
    +
    \widehat{R}^{( l_3 , l_1 )}_{( a_2 , b_2 )}
    \widehat{R}^{( l_2 , l_3 )}_{( c_2 , c_1 )}
    -
    \widehat{R}^{( l_3 , l_1 )}_{( a_2 , b_2 )}
    \widehat{R}^{( l_1 , l_2 )}_{( c_1 , c_2 )}
    \nonumber \\
    &&
    +
    \widehat{R}^{( l_2 , l_3 )}_{( c_2 , a_2 )}
    \widehat{R}^{( l_1 , l_2 )}_{( b_2 , b_1 )}
    -
    \widehat{R}^{( l_1 , l_2 )}_{( b_2 , c_2 )}
    \widehat{R}^{( l_2 , l_3 )}_{( a_1 , a_2 )}
    -
    \widehat{R}^{( l_2 , l_3 )}_{( c_2 , a_2 )}
    \widehat{R}^{( l_3 , l_1 )}_{( b_1 , b_2 )}
    +
    \widehat{R}^{( l_3 , l_1 )}_{( b_1 , b_2 )}
    \widehat{R}^{( l_1 , l_2 )}_{( c_1 , c_2 )}
    \widehat{R}^{( l_2 , l_3 )}_{( a_1 , a_2 )}
    \label{5-36}
\eeqar
where we make the condition (\ref{5-35}) explicit by
the abbreviated notation $a_1 \parallel b_1 \parallel c_1$.
Note that $\widehat{R}$'s are defined as before, {\it e.g.},
\beq
    \widehat{R}^{( l_{1}, l_{2} )}_{( a_1 , a_2 )}
    \, = \,
    \frac{( l_2 ~ a_1 )( l_1 ~ a_2 )}{( l_1 ~ a_1 )( l_2 ~ a_2 )} \, .
    \label{5-37}
\eeq
Under the collinearity condition (\ref{5-35}), the indices $a_1$, $b_1$ and $c_1$
can arbitrarily be chosen in the suffix of $\widehat{R}$'s.
In the above expression, we make such choices that are
convenient for later purposes.

In the present type-II limit (\ref{5-7}), we can further impose
the cyclicity of the $( l_1 , l_2 , l_3 )$-indices.
Since these indices enter only in the factor of (\ref{5-36}), this
condition can be realized by rewriting (\ref{5-36}) as
\beqar
   \R_{n: r_1 , r_2 ;i}^{( l_{1}, l_{2} , l_{3} )}
    & \equiv &
    \left.
    \sqrt{-
    \R^{( l_1 , l_2 )}_{r_1; a_1} \,
    \R^{( l_2 , l_3 )}_{r_2; b_1} \,
    \R^{( l_3 , l_1 )}_{r_3; c_1} \,
    \R^{( l_1 , l_2 , l_3 )}_{(a_2 , b_2 , c_2)}
    }
    \right|_{ a_1 \parallel b_1 \parallel c_1  , \, cycl.}
    \nonumber \\
    &=&
    - \,
    \widehat{R}^{( l_1 , l_2 )}_{( b_2 , b_1 )}
    \widehat{R}^{( l_2 , l_3 )}_{( c_2 , c_1 )}
    \widehat{R}^{( l_3 , l_1 )}_{( a_2 , a_1 )}
    + \,
    \widehat{R}^{( l_1 , l_2 )-1}_{( b_2 , b_1 )}
    \widehat{R}^{( l_2 , l_3 )-1}_{( c_2 , c_1 )}
    \widehat{R}^{( l_3 , l_1 )-1}_{( a_2 , a_1 )}
    \nonumber \\
    &&
    + \,
    \widehat{R}^{( l_1 , l_2 )-1}_{( b_2 , a_2 )}
    \left(
    \widehat{R}^{( l_3 , l_1 )-1}_{( c_1 , c_2 )}
    -
    \widehat{R}^{( l_2 , l_3 )-1}_{( c_2 , c_1 )}
    \right)
    \, + \,
    \widehat{R}^{( l_2 , l_3 )-1}_{( c_2 , b_2 )}
    \left(
    \widehat{R}^{( l_1 , l_2 )-1}_{( a_1 , a_2 )}
    -
    \widehat{R}^{( l_3 , l_1 )-1}_{( a_2 , a_1 )}
    \right)
    \nonumber \\
    &&
    + \,
    \widehat{R}^{( l_3 , l_1 )-1}_{( a_2 , c_2 )}
    \left(
    \widehat{R}^{( l_2 , l_3 )-1}_{( b_1 , b_2 )}
    -
    \widehat{R}^{( l_1 , l_2 )-1}_{( b_2 , b_1 )}
    \right)
    \label{5-38}
\eeqar
where we make $\widehat{R}$'s in appropriate forms
such that we can relate them to those that appear in (\ref{3-37})
with the cyclicity of indices.

\noindent
\underline{Polylog regularization: emergence of trilogarithms}

Applying the results in section 4, we can then express the
integral  in (\ref{5-32}) as an iterated integral:
\beqar
    \L_{n: r_1 , r_2 ;i}^{( l_{1}, l_{2} , l_{3} )}
    & \equiv &
    \int d \mu (L_1 ) d \mu (L_2 ) d \mu (L_3 ) \,
    \R_{n: r_1 , r_2 ;i}^{( l_{1}, l_{2} , l_{3} )}
    \nonumber \\
    &=&
    \frac{1}{( 4 \pi)^3 } \int \om_{1}^{(1)} \om_{2}^{(0)} \om_{3}^{(0)} \,
    \R_{r_1 , r_2 ;i}^{( l_{1}, l_{2} , l_{3} )}
    \label{5-39}
\eeqar
where $\om_{1}^{(1)}$ and $\om_{k}^{(0)}$ ($k = 2,3$) are defined by
(\ref{4-3}) and (\ref{4-13}), respectively.
As mentioned below (\ref{4-23}), the action of
$\widehat{R}^{( l_1 , l_2 )-1}_{( b_2 , b_1 )}$
leads to the same result as that of
$\widehat{R}^{( l_1 , l_2 )}_{( b_2 , b_1 )}$
in implementing the BST prescription (\ref{4-16}).
Thus, applying the BST prescription to the integral (\ref{5-39}),
with the knowledge of the iterated-integral representation of polylogarithms
(\ref{4-12}), we find that the following NMHV analog of the
BST prescription:
\beqar
    &&
    \int \om_{1}^{(1)} \om_{2}^{(0)} \om_{3}^{(0)} \,
    \R_{r_1 , r_2 ;i}^{( l_{1}, l_{2} , l_{3} )}
    \nonumber \\
    & \longrightarrow &
    ~~~
    {\rm Li}_3 \left(
    1 - \Delta^{(1)}_{( b_2, a_2 )}
    - \Delta^{(3)}_{( c_1 , c_2 )}
    \right)
    \, - \,
    {\rm Li}_3 \left(
    1 - \Delta^{(1)}_{( b_2, a_2 )}
    - \Delta^{(2)}_{( c_2 , c_1 )}
    \right)
    \nonumber \\
    &&
    \, + \,
    {\rm Li}_3 \left(
    1 - \Delta^{(2)}_{( c_2, b_2 )}
    - \Delta^{(1)}_{( a_1 , a_2 )}
    \right)
    \, - \,
    {\rm Li}_3 \left(
    1 - \Delta^{(2)}_{( c_2, b_2 )}
    - \Delta^{(3)}_{( a_2 , a_1 )}
    \right)
    \nonumber \\
    &&
    \, + \,
    {\rm Li}_3 \left(
    1 - \Delta^{(3)}_{( a_2, c_2 )}
    - \Delta^{(2)}_{( b_1 , b_2 )}
    \right)
    \, - \,
    {\rm Li}_3 \left(
    1 - \Delta^{(3)}_{( a_2, c_2 )}
    - \Delta^{(1)}_{( b_2 , b_1 )}
    \right)
    \label{5-40}
\eeqar
where $\Delta^{(1)}$'s, $\Delta^{(2)}$'s and $\Delta^{(3)}$'s are defined
as follows.
\beqar
    \begin{array}{rcl}
    \Delta^{(1)}_{( a_1 , b_1 )}
    & = & c_{r_{1} ; a_1}^{(1)} \, ( P^{(a)} + p_{b_2} + p_{a_2} )^2
    \\
    \Delta^{(1)}_{( a_1 , a_2 )}
    & = & c_{r_{1} ; a_1}^{(1)} \, ( P^{(a)} + p_{b_2} )^2
    \\
    \Delta^{(1)}_{( b_2 , b_1 )}
    & = & c_{r_{1} ; a_1}^{(1)} \, ( P^{(a)} + p_{a_2} )^2
    \\
    \Delta^{(1)}_{( b_2 , a_2 )}
    & = & c_{r_{1} ; a_1}^{(1)} \,  P^{(a)  2}
    \end{array}
    &&
    \begin{array}{rcl}
    \Delta^{(2)}_{( b_1 , c_1 )}
    & = & c_{r_{2} ; b_1}^{(2)} \, ( P^{(b)} + p_{c_2} + p_{b_2} )^2
    \\
    \Delta^{(2)}_{( b_1 , b_2 )}
    & = & c_{r_{2} ; b_1}^{(2)} \, ( P^{(b)} + p_{c_2} )^2
    \\
    \Delta^{(2)}_{( c_2 , c_1 )}
    & = & c_{r_{2} ; b_1}^{(2)} \, ( P^{(b)} + p_{b_2} )^2
    \\
    \Delta^{(2)}_{( c_2 , b_2 )}
    & = & c_{r_{2} ; b_1}^{(2)} \,  P^{(b)  2}
    \end{array}
    \label{5-41}\\
    \begin{array}{rcl}
    \Delta^{(3)}_{( c_1 , a_1 )}
    & = & c_{r_{3} ; c_1}^{(3)} \, ( P^{(c)} + p_{a_2} + p_{c_2} )^2
    \\
    \Delta^{(3)}_{( c_1 , c_2 )}
    & = & c_{r_{3} ; c_1}^{(1)} \, ( P^{(c)} + p_{a_2} )^2
    \\
    \Delta^{(3)}_{( a_2 , a_1 )}
    & = & c_{r_{3} ; c_1}^{(3)} \, ( P^{(c)} + p_{c_2} )^2
    \\
    \Delta^{(3)}_{( a_2 , c_2 )}
    & = & c_{r_{3} ; c_1}^{(3)} \,  P^{(c)  2}
    \end{array}
    &&
    \begin{array}{rcl}
    P^{(a)} &=&  p_{a_1} + p_{a_{1}+1} + \cdots + p_{a_{2}-1}
    \\
    P^{(b)} &=&  p_{b_1} + p_{b_{1}+1} + \cdots + p_{b_{2}-1}
    \\
    P^{(c)} &=&  p_{c_1} + p_{c_{1}+1} + \cdots + p_{c_{2}-1}
    \end{array}
    \label{5-42}
\eeqar
\beqar
    c_{r_{1} ; a_1}^{(1)} & = &
    \frac{( p_{b_2} + p_{a_2} )^2}{
    P^{(a)  2} (P^{(a)} + p_{b_2} + p_{a_2} )^2
    - (P^{(a)} + p_{b_2} )^2 (P^{(a)} + p_{a_2} )^2}
    \nonumber \\
    c_{r_{2} ; b_1}^{(2)} & = &
    \frac{( p_{c_2} + p_{b_2} )^2}{
    P^{(b)  2} (P^{(b)} + p_{c_2} + p_{b_2} )^2
    - (P^{(b)} + p_{c_2} )^2 (P^{(b)} + p_{b_2} )^2}
    \label{5-43}\\
    c_{r_{3} ; c_1}^{(3)} & = &
    \frac{( p_{a_2} + p_{c_2} )^2}{
    P^{(c)  2} (P^{(c)} + p_{a_2} + p_{c_2} )^2
    - (P^{(c)} + p_{a_2} )^2 (P^{(c)} + p_{c_2} )^2}
    \nonumber
\eeqar
Here we include $\Delta^{(1)}_{( a_1 , b_1 )}$,
$\Delta^{(2)}_{( b_1 , c_1 )}$ and $\Delta^{(3)}_{( c_1 , a_1 )}$,
which do not appear in (\ref{5-40}), simply for the completion of the arguments.

There is one caveat about an ambiguity of the expression (\ref{5-38}).
Namely, the second term in (\ref{5-38}),
$\widehat{R}^{( l_1 , l_2 )-1}_{( b_2 , b_1 )}
\widehat{R}^{( l_2 , l_3 )-1}_{( c_2 , c_1 )}
\widehat{R}^{( l_3 , l_1 )-1}_{( a_2 , a_1 )}$,
can alternatively be written as
$\widehat{R}^{( l_2 , l_3 )}_{( b_1 , b_2 )}
\widehat{R}^{( l_3 , l_1 )}_{( c_1 , c_2 )}
\widehat{R}^{( l_1 , l_2 )}_{( a_1 , a_2 )}$, using the
cyclicity of the indices.
A naive application of the NMHV prescription (\ref{5-40})
leads to different results,
$\Delta^{(1)}_{( b_2 , b_1 )} + \Delta^{(2)}_{( c_2 , c_1 )} +
\Delta^{(3)}_{( a_2 , a_1 )}$
and
$\Delta^{(1)}_{( a_1 , a_2 )} + \Delta^{(2)}_{( b_1 , b_2 )} +
\Delta^{(3)}_{( c_1 , c_2 )}$
in terms of the arguments of trilogarithm functions.
Apparently, this causes discrepancy in our prescription
but, as far as the amplitudes (\ref{5-32}) are concerned,
the two expressions lead to the same result
thanks to the split-permutations of the numbering indices.
To be more specific, in (\ref{5-32}) we can change the
ordering of the indices from
$( a_1 \cdots a_2 \, b_1 \cdots b_2 \, c_1 \cdots c_2 )$
to
$( a_2 a_1 \cdots (a_{2} -1) \, b_2 b_1 \cdots ( b_{2}-1 ) \,
c_2 c_1 \cdots ( c_{2} -1 )  )$ thanks to the sum over
split-permutations represented by $\P ( a_1 \cdots a_2 |
b_1 \cdots b_2 | c_1 \cdots c_2 )$. In the MHV diagram,
this corresponds to the replacements of $a_2$ with $b_2$,
$b_2$ with $c_2$, and $c_2$ with $a_2$.
This means that we have transformations
$P^{(a)} + p_{a_2} \rightarrow P^{(a)} + p_{b_2}$,
$P^{(b)} + p_{b_2} \rightarrow P^{(b)} + p_{c_2}$, and
$P^{(c)} + p_{c_2} \rightarrow P^{(c)} + p_{a_2}$, respectively.
As mentioned in (\ref{4-22}), $c^{(1)}_{r_1 ; a_1}$,
$c^{(2)}_{r_2 ; b_1}$ and $c^{(3)}_{r_3 ; c_1}$
are invariant under these transformations.
Thus, under the above change of index orderings
we can replace
$\Delta^{(1)}_{( b_2 , b_1 )} + \Delta^{(2)}_{( c_2 , c_1 )} +
\Delta^{(3)}_{( a_2 , a_1 )}$
by
$\Delta^{(1)}_{( a_1 , a_2 )} + \Delta^{(2)}_{( b_1 , b_2 )} +
\Delta^{(3)}_{( c_1 , c_2 )}$
so that the resultant amplitudes are identical
regardless the choice of the expression
$\widehat{R}^{( l_1 , l_2 )-1}_{( b_2 , b_1 )}
\widehat{R}^{( l_2 , l_3 )-1}_{( c_2 , c_1 )}
\widehat{R}^{( l_3 , l_1 )-1}_{( a_2 , a_1 )}$ or
$\widehat{R}^{( l_2 , l_3 )}_{( b_1 , b_2 )}
\widehat{R}^{( l_3 , l_1 )}_{( c_1 , c_2 )}
\widehat{R}^{( l_1 , l_2 )}_{( a_1 , a_2 )}$ in (\ref{5-38}).
Eventually, however, this term cancels out with the first term in (\ref{5-38}).
Thus the above argument on the solution of ambiguity is in fact digressive.

The NMHV prescription (\ref{5-40}) is not derived rigorously but is obtained
somewhat in a conjectural fashion by use of the BST prescription (\ref{4-16})
which is, on the other hand, confirmed analytically thanks to the BST method.\footnote{
Notice that we can also derive the BST result (\ref{3-37}) in a squared form
as in (\ref{5-38}), using the fact that the identity can be replaced by
$\widehat{R}^{( l_1 , l_2 )}_{( a_1 , b_1 )}$
under the condition of $u_{a_1} \parallel u_{b_1}$.}
In obtaining the above expressions, we therefore
keep the structure of the BST prescription as much as possible,
using the proper cyclicity and the single reference-spinor principle of the CSW rules.
One may wonder why the arguments of ${\rm Li}_3$'s are additive rather than
multiplicative.
This is a reasonable question but can be answered in
a few interrelated ways.
First, the MHV clusters that involve a one-loop diagram
have always two and only two internal lines.
This suggests that the deformation of the integral path can be analyzed
by an iterative use of the BST prescription.
Secondly, the original path for the iterated integral
can be defined by a real line segment $[0,1]$ regardless the number of $w_i$'s.
Thus the way that the integral paths deform is
independent of the number of MHV clusters in a one-loop diagram.
In other words, the BST prescription can be applied, in an iterative fashion,
to the type-II one-loop NMHV amplitudes
as long as the deformations of integral paths are concerned.
Lastly, consider for example the cases of $r_1 = 1$ $r_2 > 1$ and $r_3 > 1$ where
$\Delta^{(1)}_{( b_2 , a_2 )} = 0$ but
$\Delta^{(2)}$'s and $\Delta^{(3)}$'s are not zero.
If the arguments of ${\rm Li}_3$'s were to be multiplicative,
we would have trivial arguments (identities) for such cases regardless
the choice of $r_2$ and $r_3$. This is not consistent with the above
picture of iterative contributions by the MHV clusters.
Thus, although we have only circumstantial evidences,
it is natural to require the arguments of ${\rm Li}_3$'s be additive.

\noindent
\underline{Summary}

To recapitulate the results of this section, we now write down the complete form of
the one-loop NMHV amplitudes in the {\it momentum-space} representation:
\beqar
    \A_{NMHV(1)}^{(a_{-} b_{-} c_{-})} (u, \bu)
    & = & i g^{n-2}
    \, (2 \pi)^4 \del^{(4)} \left( \sum_{i=1}^{n} p_i \right) \,
    \widehat{A}_{NMHV(1)}^{(a_{-} b_{-} c_{-})} (u) \, ,
    \label{5-44}\\
    \widehat{A}_{NMHV(1)}^{(a_{-} b_{-} c_{-})} (u )
    & = &
    \widehat{A}_{NMHV(1;{\rm Li}_2 )}^{(a_- b_- c_- )} (u)
    \, + \,
    \widehat{A}_{NMHV(1;{\rm Li}_3 )}^{(a_- b_- c_- )} (u) \, .
    \label{5-45}
\eeqar
In the momentum-space, the type-I and type-II subamplitudes are
written as
\beqar
    &&
    \widehat{A}_{NMHV(1; {\rm Li}_2 )}^{(a_{-} b_{-} c_{-})} (u)
    \nonumber \\
    &=&
    i g^2 \,
    \sum_{i = 1}^{n}
    \sum_{r = 1}^{n-3}
    \sum_{j = i}^{i+r}
    \sum_{t = 1}^{\left\lfloor \frac{r+1}{2} \right\rfloor -1 }
    \left( 1 - \frac{1}{2} \del_{ \frac{r+1}{2} , t+1} \right)
    \nonumber \\
    &&
    \!\!\!\!\!\!
    \sum_{ \si^{(1)} \in \S_{t+1}} \,
    \sum_{ \si^{(2)} \in \S_{r-t}} \,
    \sum_{ \si^{(3)} \in \S_{n-r-1}} \!\!
    \Tr ( t^{\si^{(1)}_{j}} \cdots t^{\si^{(1)}_{j+t}} \, t^{\si^{(2)}_{j+t+1}}
    \cdots t^{\si^{(2)}_{j+r}} \, t^{\si^{(3)}_{i+r+1}} \cdots t^{\si^{(3)}_{i-1}}  ) \,
    \L_{r+1: t ; \si_j }^{( l_1 , l_2 )}
    \nonumber \\
    && \!\!\!\!\!\!
    \left.
    \widehat{C}_{MHV(0)}^{( j_{+} \cdots a_{-} \cdots b_{-} \cdots (j+r)_{+} l_{3+} )}
    (u; \si^{(1 \otimes 2)}) \,
    \frac{1}{q_{i \, i+r}^2} \,
    \widehat{C}^{( (-l_3)_{-} \, (i+r+1)_{+} \cdots c_{-} \cdots (i-1)_{+} )}_{MHV(0)}
    (u; \si^{(3)})
    \right|_{l_3 = q_{i \, i+r} \bar{\eta}}
    \label{5-46}
\eeqar
where $\si^{(1)}, \si^{(2)}$ and $\si^{(3)}$ are defined in (\ref{5-12}) and (\ref{5-17}) and
$\L_{r+1: t ; \si_i }^{( l_1 , l_2 )} $ can be defined through (\ref{5-13})-(\ref{5-15}).
\beqar
    &&
    \widehat{A}_{NMHV(1;{\rm Li}_3 )}^{(a_- b_- c_- )} (u)
    \nonumber \\
    &=&
    i g^2 \,
    \sum_{i = 1}^{n}
    \sum_{r_1 = 1}^{\left\lfloor \frac{n}{3}  \right\rfloor -1 }
    \sum_{r_2 = 1}^{\left\lfloor \frac{n}{3}  \right\rfloor -1 }
    \left( 1 - \frac{2}{3} \del_{ \frac{n}{3}, r_{1} +1} \del_{ \frac{n}{3}, r_{2} +1} \right)
    \nonumber \\
    &&
    \sum_{ \si^{(1)} \in \S_{r_{1}+1}} \,
    \sum_{ \si^{(2)} \in \S_{r_{2}+1}} \,
    \sum_{ \si^{(3)} \in \S_{r_{3}+1}} \!\!
    \Tr ( t^{\si^{(1)}_{a_1}} \cdots t^{\si^{(1)}_{a_2}} \, t^{\si^{(2)}_{b_1}}
    \cdots t^{\si^{(2)}_{b_2}} \, t^{\si^{(3)}_{c_1}} \cdots t^{\si^{(3)}_{c_2}}  )
    \,
    \L_{n:r_1 , r_2 ; \si_i}^{(l_1 , l_2 , l_3 )}
    \nonumber \\
    &&
    \sqrt{
    \widehat{C}_{MHV(0)}^{( a_{1+}  \cdots a_{-} \cdots b_{-} \cdots b_{2+} )}
    (u; \si^{(1 \otimes 2)}) \,
    \widehat{C}_{MHV(0)}^{( b_{1+}  \cdots b_{-} \cdots c_{-} \cdots c_{2+} )}
    (u; \si^{(2 \otimes 3)}) \,
    \widehat{C}_{MHV(0)}^{( c_{1+}  \cdots c_{-} \cdots a_{-} \cdots a_{2+} )}
    (u; \si^{(3 \otimes 1)})
    }
    \nonumber \\
    \label{5-47}
\eeqar
where the indices $\{ a_1 , a_2 , b_1 , b_2 , c_1 , c_2 \}$
are defined by (\ref{5-22}) and
$\L_{n:r_1 , r_2 ; \si_i}^{(l_1 , l_2 , l_3 )}$ can be defined
through (\ref{5-39})-(\ref{5-43}).

As discussed in the previous section, the type-I or dilogarithm contributions
are a direct consequence of the BST results for one-loop MHV amplitudes
so that the validity of the expression (\ref{5-46}) is well confirmed, while
the type-II or trilogarithm contributions in (\ref{5-47}) are obtained by use of the
somewhat conjectural polylog regularization.
The latter contributions, however, provide what we consider
the most natural generalization of the BST representation to one-loop NMHV amplitudes.

%%%%%%%%%%%%%%%%%%%%%%%%%%%%%%%%%%%%%%%%%%%%%%%%%%%%%%%
\section{One-loop NMHV amplitudes: examples}
%\noindent
%\underline{Six-point one-loop NMHV amplitudes}

\begin{figure} [htbp]
\begin{center}
\includegraphics[width=135mm]{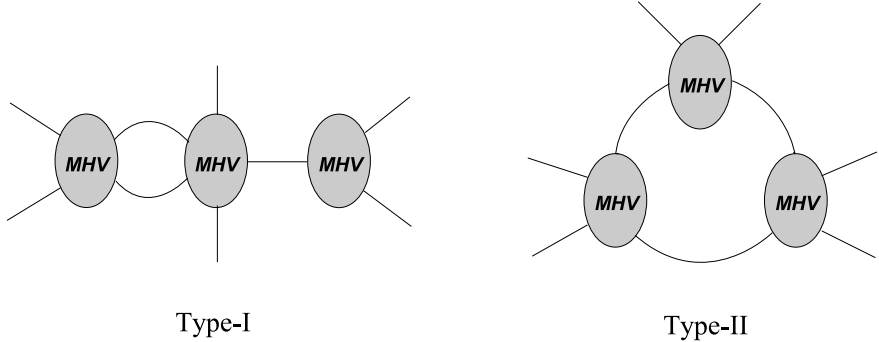}
\caption{The type-I and type-II
MHV diagrams that contribute to the six-point one-loop NMHV amplitude}
\label{fighol0406}
\end{center}
\end{figure}

As a simple example of our formulation, we calculate
the six-point one-loop NMHV amplitudes in this section.
The MHV diagrams contributing to the six-point one-loop NMHV amplitudes
are shown in Figure \ref{fighol0406}.

%%%%%%%%%%%%%%%%%%%%%%%%% figure %%%%%%%%%%%%%%%%%%%%%%%%%
%%%%%%%%%%%%%%%%%%%%%%%%% figure %%%%%%%%%%%%%%%%%%%%%%%%%

As discussed earlier, there are three distinct helicity
configurations $( 1_- 2_- 3_- 4_+ 5_+ 6_+ )$,
$( 1_- 2_+ 3_- 4_+ 5_- 6_+ )$ and $( 1_- 2_+ 3_+ 4_- 5_- 6_+ )$
for six-point amplitudes.
The corresponding tree amplitudes are obtained in (\ref{2-40})-(\ref{2-42}).
Applying the formulae of the previous section, we can then
easily write down the amplitudes of interest in the momentum-space representation.
The results are listed below.

\noindent
\underline{Six-point type-I amplitudes}

The type-I part of the $( 1_- 2_- 3_- 4_+ 5_+ 6_+ )$ amplitudes can be expanded as
\beqar
    &&
    \widehat{A}_{NMHV(1; {\rm Li}_2 )}^{( 1_{-} 2_{-} 3_{-} 4_{+} 5_{+} 6_{+} )} (u)
    \nonumber \\
    &=&
    \widehat{A}_{MHV(1)}^{( 2_- 3_- 4_+ 5_+ l_{3+} )} (u)
    \frac{1}{q_{24}^2}
    \widehat{A}_{MHV(0)}^{( ( -l_3 )_{-} 6_+ 1_- )} (u)
    \, + \,
    \widehat{A}_{MHV(1)}^{( 5_+ 6_+ 1_- 2_- l_{3+} )} (u)
    \frac{1}{q_{52}^2}
    \widehat{A}_{MHV(0)}^{( ( -l_3 )_{-} 3_- 4_+ )} (u)
    \label{6-1}
\eeqar
where
\beqar
    &&
    \widehat{A}_{MHV(1)}^{( 2_- 3_- 4_+ 5_+ l_{3+} )} (u)
    \nonumber \\
    &=&
    \frac{ig^2}{( 4 \pi )^2} \,
    \Biggl\{ \,
    \Tr ( t^{2} t^{3}t^{4} t^{5} t^{l_3}  )
    \left[
    {\rm Li}_2 \left( 1
    + \frac{( p_{5} \cdot p_{3} )}{( p_{2} \cdot p_{3} )}
    + \frac{( p_{5} \cdot p_{3} )}{( p_{2} \cdot p_{5} )}
    + \frac{( p_{5} \cdot p_{3} )^2}{( p_{2} \cdot p_{3} )( p_{2} \cdot p_{5} )}
    \right)
    \right.
    \nonumber  \\
    &&
    \left.
    - {\rm Li}_2 \left( 1 + \frac{( p_{5} \cdot p_{3} )}{( p_{2} \cdot p_{5} )} \right)
    - {\rm Li}_2 \left( 1 + \frac{( p_{5} \cdot p_{3} )}{( p_{2} \cdot p_{3} )} \right)
    + \frac{ \pi^2}{6}
    \right]
    \widehat{C}_{MHV(0)}^{( 2_- 3_- 4_+ 5_+ l_{3+} )} (u; {\bf 1})
    \, + \, \P(23|45)
    \,
    \Biggr\}
    \nonumber \\
    &&
    +
    \left(
    \begin{array}{c}
      2 \\
      3 \\
      4 \\
      5
    \end{array}
    \right)
    \leftrightarrow
    \left(
    \begin{array}{c}
      3 \\
      4 \\
      5 \\
      2
    \end{array}
    \right)
    \, ,
    \label{6-2}\\
    &&
    \widehat{A}_{MHV(1)}^{( 5_+ 6_+ 1_- 2_- l_{3+} )} (u)
    \nonumber \\
    &=&
    \frac{ig^2}{( 4 \pi )^2} \,
    \Biggl\{ \,
    \Tr ( t^{5} t^{6}t^{1} t^{2} t^{l_3}  )
    \left[
    {\rm Li}_2 \left( 1
    + \frac{( p_{2} \cdot p_{6} )}{( p_{5} \cdot p_{6} )}
    + \frac{( p_{2} \cdot p_{6} )}{( p_{5} \cdot p_{2} )}
    + \frac{( p_{2} \cdot p_{6} )^2}{( p_{5} \cdot p_{2} )( p_{5} \cdot p_{6} )}
    \right)
    \right.
    \nonumber  \\
    &&
    \left.
    - {\rm Li}_2 \left( 1 + \frac{( p_{2} \cdot p_{6} )}{( p_{5} \cdot p_{2} )} \right)
    - {\rm Li}_2 \left( 1 + \frac{( p_{2} \cdot p_{6} )}{( p_{5} \cdot p_{6} )} \right)
    + \frac{ \pi^2}{6}
    \right]
    \widehat{C}_{MHV(0)}^{( 5_+ 6_+ 1_- 2_- l_{3+} )} (u; {\bf 1})
    \, + \, \P(56|12)
    \,
    \Biggr\}
    \nonumber \\
    &&
    +
    \left(
    \begin{array}{c}
      5 \\
      6 \\
      1 \\
      2
    \end{array}
    \right)
    \leftrightarrow
    \left(
    \begin{array}{c}
      6 \\
      1 \\
      2 \\
      5
    \end{array}
    \right)
    \, .
    \label{6-3}
\eeqar
The tree subamplitudes in (\ref{6-1}) can be written as
\beq
    \begin{array}{rcl}
    \widehat{A}_{MHV(0)}^{( ( -l_3 )_{-} 6_+ 1_- )} (u)
    & = &
    \Tr ( t^{(-l_3)} t^6 t^1 ) \,
    \widehat{C}_{MHV(0)}^{( ( -l_3 )_{-} 6_+ 1_- )} (u; {\bf 1})
    \, + \, \P (61) \, , \\
    \widehat{A}_{MHV(0)}^{( ( -l_3 )_{-} 3_- 4_+ )} (u)
    & = &
    \Tr ( t^{(-l_3)} t^3 t^4 ) \,
    \widehat{C}_{MHV(0)}^{( ( -l_3 )_{-} 3_- 4_+ )} (u; {\bf 1})
    \, + \, \P (34)  \, . \\
    \end{array}
    \label{6-4}
\eeq
Thus, upon the contraction of color indices, the amplitudes
$\widehat{A}_{NMHV(1; {\rm Li}_2 )}^{( 1_{-} 2_{-} 3_{-} 4_{+} 5_{+} 6_{+} )} (u)$
become single-trace amplitudes.

Similarly, for the helicity configurations
$( 1_- 2_+ 3_- 4_+ 5_- 6_+ )$ and $( 1_- 2_+ 3_+ 4_- 5_- 6_+ )$
the type-I amplitudes are given by
\beqar
    &&
    \widehat{A}_{NMHV(1; {\rm Li}_2 )}^{( 1_{-} 2_{+} 3_{-} 4_{+} 5_{-} 6_{+} )} (u)
    \nonumber \\
    &=&
    \widehat{A}_{MHV(1)}^{( 1_- 2_+  3_- 4_+ l_{3+} )} (u)
    \frac{1}{q_{14}^2}
    \widehat{A}_{MHV(0)}^{( ( -l_3 )_{-} 5_- 6_+ )} (u)
    \, + \,
    \widehat{A}_{MHV(1)}^{( 6_+ 1_- 2_+ 3_- l_{3+} )} (u)
    \frac{1}{q_{63}^2}
    \widehat{A}_{MHV(0)}^{( ( -l_3 )_{-} 4_+ 5_- )} (u)
    \nonumber \\
    &&
    \!\!\!
    + \,
    \widehat{A}_{MHV(1)}^{( 5_- 6_+ 1_- 2_+ l_{3+} )} (u)
    \frac{1}{q_{52}^2}
    \widehat{A}_{MHV(0)}^{( ( -l_3 )_{-} 3_- 4_+ )} (u)
    \, + \,
    \widehat{A}_{MHV(1)}^{( 4_+ 5_- 6_+ 1_- l_{3+} )} (u)
    \frac{1}{q_{41}^2}
    \widehat{A}_{MHV(0)}^{( ( -l_3 )_{-} 2_+ 3_- )} (u)
    \nonumber \\
    &&
    \!\!\!
    + \,
    \widehat{A}_{MHV(1)}^{( 3_- 4_+ 5_- 6_+ l_{3+} )} (u)
    \frac{1}{q_{36}^2}
    \widehat{A}_{MHV(0)}^{( ( -l_3 )_{-} 1_- 2_+ )} (u)
    \, + \,
    \widehat{A}_{MHV(1)}^{( 2_+ 3_- 4_+ 5_- l_{3+} )} (u)
    \frac{1}{q_{25}^2}
    \widehat{A}_{MHV(0)}^{( ( -l_3 )_{-} 6_+ 1_- )} (u) \, , ~~
    \label{6-5} \\
    &&
    \widehat{A}_{NMHV(1; {\rm Li}_2 )}^{( 1_- 2_+ 3_+ 4_- 5_- 6_+  )} (u)
    \nonumber \\
    &=&
    \!\!\!
    \widehat{A}_{MHV(1)}^{( 1_- 2_+  3_+ 4_- l_{3+} )} (u)
    \frac{1}{q_{14}^2}
    \widehat{A}_{MHV(0)}^{( ( -l_3 )_{-} 5_- 6_+ )} (u)
    \, + \,
    \widehat{A}_{MHV(1)}^{( 5_- 6_+ 1_- 2_+ l_{3+} )} (u)
    \frac{1}{q_{52}^2}
    \widehat{A}_{MHV(0)}^{( ( -l_3 )_{-} 3_+ 4_- )} (u)
    \nonumber \\
    &&
    \!\!\!
    + \,
    \widehat{A}_{MHV(1)}^{(  3_+ 4_- 5_- 6_+  l_{3+} )} (u)
    \frac{1}{q_{36}^2}
    \widehat{A}_{MHV(0)}^{( ( -l_3 )_{-} 1_- 2_+ )} (u)
    \, + \,
    \widehat{A}_{MHV(1)}^{( 2_+ 3_+ 4_- 5_- l_{3+} )} (u)
    \frac{1}{q_{52}^2}
    \widehat{A}_{MHV(0)}^{( ( -l_3 )_{-} 6_+ 1_- )} (u) \, . ~~
    \label{6-6}
\eeqar
As in (\ref{6-2})-(\ref{6-4}), each term can be calculated
explicitly in terms of the external momenta.
We do not list each term here. To illustrate our calculation, however,
we present an explicit form of the one-loop subamplitudes
$\widehat{A}_{MHV(1)}^{( 1_- 2_+  3_- 4_+ l_{3+} )} (u)$ in (\ref{6-5}):
\beqar
    &&
    \widehat{A}_{MHV(1)}^{( 1_- 2_+  3_- 4_+ l_{3+} )} (u)
    \nonumber \\
    &=&
    \frac{ig^2}{( 4 \pi )^2} \,
    \Biggl\{ \,
    \Tr ( t^{1} t^{2} t^{3} t^{4} t^{l_3}  )
    \left[
    {\rm Li}_2 \left( 1
    + \frac{( p_{4} \cdot p_{2} )}{( p_{1} \cdot p_{4} )}
    + \frac{( p_{4} \cdot p_{2} )}{( p_{1} \cdot p_{2} )}
    + \frac{( p_{4} \cdot p_{2} )^2}{( p_{1} \cdot p_{4} )( p_{1} \cdot p_{2} )}
    \right)
    \right.
    \nonumber  \\
    &&
    \left.
    - {\rm Li}_2 \left( 1 + \frac{( p_{4} \cdot p_{2} )}{( p_{1} \cdot p_{2} )} \right)
    - {\rm Li}_2 \left( 1 + \frac{( p_{4} \cdot p_{2} )}{( p_{1} \cdot p_{4} )} \right)
    + \frac{ \pi^2}{6} \,
    \right]
    \widehat{C}_{MHV(0)}^{( 1_- 2_+ 3_- 4_+ l_{3+} )} (u; {\bf 1})
    \, + \, \P(12|34)
    \,
    \Biggr\}
    \nonumber \\
    &&
    +
    \left(
    \begin{array}{c}
      1 \\
      2 \\
      3 \\
      4
    \end{array}
    \right)
    \leftrightarrow
    \left(
    \begin{array}{c}
      2 \\
      3 \\
      4 \\
      1
    \end{array}
    \right)
    \, .
    \label{6-7}
\eeqar

\noindent
\underline{Six-point type-II amplitudes}

We first consider the case of $( 1_- 2_- 3_- 4_+ 5_+ 6_+ )$.
In applying the formula (\ref{5-47}) in this helicity configuration,
we find that we can not have three $\widehat{C}_{MHV(0)}$'s, that is,
one of them can not have two negative-helicity states in any way; it has
only a single negative-helicity state.
Therefore the type-II part of the $( 1_- 2_- 3_- 4_+ 5_+ 6_+ )$ amplitudes vanish:
\beq
    \widehat{A}_{NMHV(1; {\rm Li}_3 )}^{( 1_{-} 2_{-} 3_{-} 4_{+} 5_{+} 6_{+} )} (u)
    \, = \, 0 \, .
    \label{6-8}
\eeq

For the other cases, the type-II amplitudes are nontrivial.
We can calculate them as follows.
\beqar
    &&
    \widehat{A}_{NMHV(1; {\rm Li}_3 )}^{( 1_{-} 2_{+} 3_{-} 4_{+} 5_{-} 6_{+} )} (u)
    \nonumber \\
    &=&
    \frac{i g^2}{(4 \pi )^3} \,
    \Biggl\{ \,
    \Tr ( t^{1} t^{2} t^{3} t^{4} t^{5} t^{6} )
    \left[
    {\rm Li}_3 \left( 1
    + \frac{( p_{2} \cdot p_{6} )}{( p_{5} \cdot p_{6} )}
    \right)
    -
    {\rm Li}_3 \left( 1
    + \frac{( p_{6} \cdot p_{4} )}{( p_{3} \cdot p_{6} )}
    \right)
    \right.
    \nonumber  \\
    &&
    \left.
    +
    {\rm Li}_3 \left( 1
    + \frac{( p_{3} \cdot p_{2} )}{( p_{1} \cdot p_{2} )}
    \right)
    -
    {\rm Li}_3 \left( 1
    + \frac{( p_{2} \cdot p_{6} )}{( p_{5} \cdot p_{2} )}
    \right)
    +
    {\rm Li}_3 \left( 1
    + \frac{( p_{6} \cdot p_{4} )}{( p_{3} \cdot p_{4} )}
    \right)
    -
    {\rm Li}_3 \left( 1
    + \frac{( p_{3} \cdot p_{2} )}{( p_{1} \cdot p_{3} )}
    \right)
    \right]
    \nonumber \\
    &&
    \sqrt{
    \widehat{C}_{MHV(0)}^{( 1_{-} 2_{+} 3_{-} 4_{+} )}
    (u; {\bf 1} ) \,
    \widehat{C}_{MHV(0)}^{( 3_{-} 4_{+} 5_{-} 6_{+} )}
    (u; {\bf 1} ) \,
    \widehat{C}_{MHV(0)}^{( 5_{-} 6_{+} 1_{-} 2_{+} )}
    (u; {\bf 1} )
    }
    \, + \, \P(12|34|56)
    \,
    \Biggr\}
    \nonumber \\
    &&
    +
    \left(
    \begin{array}{c}
      1 \\
      2 \\
      3 \\
      4 \\
      5 \\
      6
    \end{array}
    \right)
    \leftrightarrow
    \left(
    \begin{array}{c}
      2 \\
      3 \\
      4 \\
      5 \\
      6 \\
      1
    \end{array}
    \right),
    \label{6-9}
\eeqar
\beqar
    &&
    \widehat{A}_{NMHV(1; {\rm Li}_3 )}^{( 1_{-} 2_{+} 3_{+} 4_{-} 5_{-} 6_{+} )} (u)
    \nonumber \\
    &=&
    \frac{i g^2}{(4 \pi )^3} \,
    \Tr ( t^{1} t^{2} t^{3} t^{4} t^{5} t^{6} )
    \left[
    {\rm Li}_3 \left( 1
    + \frac{( p_{2} \cdot p_{6} )}{( p_{5} \cdot p_{6} )}
    \right)
    -
    {\rm Li}_3 \left( 1
    + \frac{( p_{6} \cdot p_{4} )}{( p_{3} \cdot p_{6} )}
    \right)
    \right.
    \nonumber  \\
    &&
    \left.
    +
    {\rm Li}_3 \left( 1
    + \frac{( p_{3} \cdot p_{2} )}{( p_{1} \cdot p_{2} )}
    \right)
    -
    {\rm Li}_3 \left( 1
    + \frac{( p_{2} \cdot p_{6} )}{( p_{5} \cdot p_{2} )}
    \right)
    +
    {\rm Li}_3 \left( 1
    + \frac{( p_{6} \cdot p_{4} )}{( p_{3} \cdot p_{4} )}
    \right)
    -
    {\rm Li}_3 \left( 1
    + \frac{( p_{3} \cdot p_{2} )}{( p_{1} \cdot p_{3} )}
    \right)
    \right]
    \nonumber \\
    &&
    \sqrt{
    \widehat{C}_{MHV(0)}^{( 1_{-} 2_{+} 3_{+} 4_{-} )}
    (u; {\bf 1} ) \,
    \widehat{C}_{MHV(0)}^{( 3_{+} 4_{-} 5_{-} 6_{+} )}
    (u; {\bf 1} ) \,
    \widehat{C}_{MHV(0)}^{( 5_{-} 6_{+} 1_{-} 2_{+} )}
    (u; {\bf 1} )
    }
    \, + \, \P(12|34|56) \, .
    \label{6-10}
\eeqar

From the above results, we can obtain all of the six-point one-loop NMHV
amplitudes.
These results are qualitatively different from the unitary-cut results
\cite{Bern:1994cg,Drummond:2008vq} in many ways.
First, the ${\rm Li}_3$-dependence
does not appear in the unitary-cut method.
Second, the non-holomorphic part of the one-loop amplitudes
are encoded in the $\frac{1}{q_{ij}}$ dependence of the type-I amplitudes
while, in the unitary-cut results,
the non-holomorphicity aries from rather involved
so-called spin factors.
Lastly and most significantly, our amplitudes are not necessarily
planar while the unitary-cut results are.
{\it
It is true that the amplitudes have the single-trace
structure but, as discussed in section 2,
this feature is a direct consequence of the CSW generalization
and is completely independent of the conventional large $N$ analysis.
In a large-$N$ independent nonperturbative analysis, the single-trace
color structure dose not necessarily mean the planarity of the amplitudes.
Thus, if necessary to tell the planarity, it is natural
to interpret our results as non-planar amplitudes.}

%%%%%%%%%%%%%%%%%%%%%%%%%%%%%%%%%%%%%%%%%%%%%%%%%%%%%%%%%%%
\section{One-loop N$^2$MHV amplitudes and beyond}
%\noindent
%\underline{One-loop N$^2$MHV amplitudes}

\begin{figure} [htbp]
\begin{center}
\includegraphics[width=135mm]{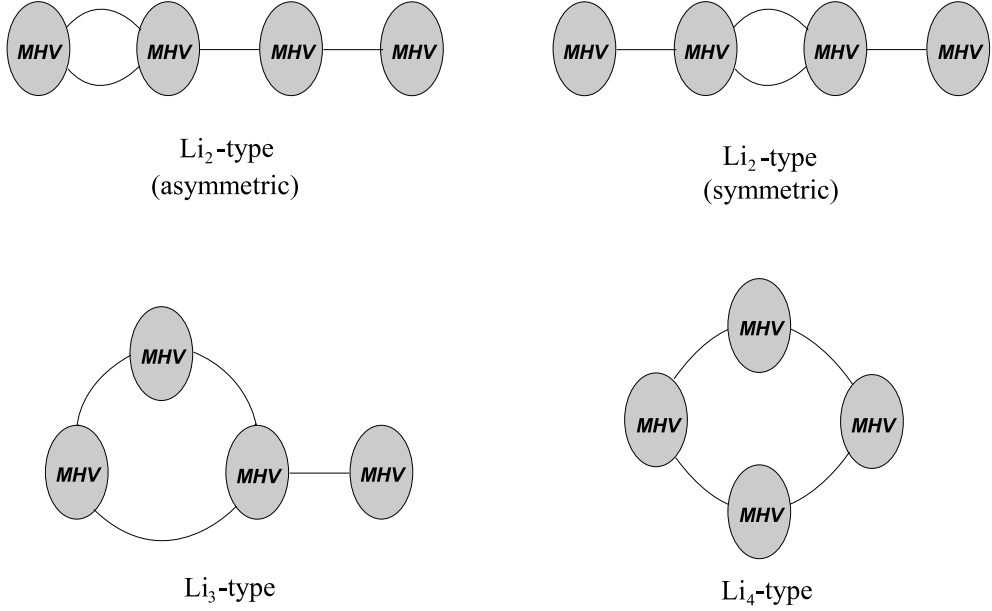}
\caption{MHV diagrams contributing to one-loop N$^{2}$MHV amplitudes}
\label{fighol0407}
\end{center}
\end{figure}

The MHV diagrams that contribute to one-loop N$^2$MHV amplitudes
(with four negative-helicity gluons and $(n-4)$ positive-helicity gluons)
are shown in Figure \ref{fighol0407}.

%%%%%%%%%%%%%%%%%%%%%%%%% figure %%%%%%%%%%%%%%%%%%%%%%%%%
%%%%%%%%%%%%%%%%%%%%%%%%% figure %%%%%%%%%%%%%%%%%%%%%%%%%

\noindent
\underline{${\rm Li}_2$- and ${\rm Li}_3$-type contributions}

\begin{figure} [htbp]
\begin{center}
\includegraphics[width=140mm]{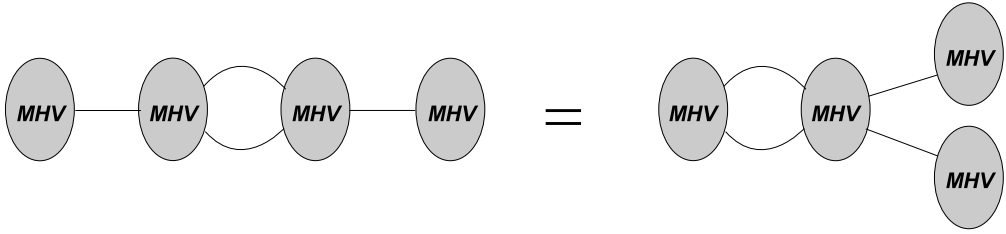}
\caption{Symmetric ${\rm Li}_2$-type MHV diagrams contributing to one-loop N$^{2}$MHV amplitudes}
\label{fighol0408}
\end{center}
\end{figure}

As in the NMHV case, the ${\rm Li}_2$- and ${\rm Li}_3$-type
amplitudes contributing the one-loop N$^2$MHV
amplitudes can be calculated by a direct application of the CSW rules.
In the momentum-space representation, the combined amplitudes can be calculated as
\beqar
    \widehat{A}_{N^2 MHV (1;{\rm Li}_{2,3})}^{( a_- b_- c_- d_- )} (u)
    \!\!
    & = &
    \!\!
    \left.
    \sum_{i=1}^{n} \sum_{r=1}^{n-3}
    \widehat{A}_{NMHV(1)}^{( i_+ \cdots a_- b_- c_- \cdots (i+r)_+ l_{4+} )} (u)
    \,
    \frac{1}{q_{i \, i+r}^{2}}
    \,
    \widehat{A}_{MHV(0)}^{( (-l_4)_- \cdots d_- \cdots (i-1)_+ )} (u)
    \right|_{l_4 = q_{i \, i+r} {\bar \eta} }
    \label{7-1}
\eeqar
where the one-loop information comes from
$\widehat{A}_{NMHV(1)}^{( i_+ \cdots a_- b_- c_- \cdots (i+r)_+ l_{4+} )} (u)$.
Using the results in section 5 and noticing that the
internal index $l_{4}$ is auxiliary, we find that this subamplitude can
be expressed as
\beq
    \widehat{A}_{NMHV(1)}^{( i_+ \cdots a_- b_- c_- \cdots (i+r)_+ l_{4+} )} (u)
    \, = \,
    \widehat{A}_{NMHV(1;{\rm Li}_2 )}^{( i_+ \cdots a_- b_- c_- \cdots (i+r)_+ l_{4+} )} (u)
    \,
    +
    \,
    \widehat{A}_{NMHV(1;{\rm Li}_3 )}^{( i_+ \cdots a_- b_- c_- \cdots (i+r)_+ l_{4+} )} (u)
    \label{7-2}
\eeq
where the ${\rm Li}_2$-type amplitudes are given by
\beqar
    &&
    \widehat{A}_{NMHV(1;{\rm Li}_2 )}^{( i_+ \cdots a_- b_- c_- \cdots (i+r)_+ l_{4+} )} (u)
    \nonumber \\
    &=&
    i g^2
    \sum_{j=i}^{i+r} \sum_{t_1 = 1}^{r-2} \sum_{k = j}^{j + t_1 }
    \sum_{t_2 = 1}^{\left\lfloor \frac{t_1 +1}{2} \right\rfloor -1 }
    \left( 1 - \frac{1}{2} \del_{ \frac{t_1 +1}{2} , t_2 +1} \right)
    \nonumber \\
    &&
    \sum_{\si^{(1)} \in \S_{t_2 + 1}}
    \sum_{\si^{(2)} \in \S_{t_1 - t_2}}
    \sum_{\si^{(3)} \in \S_{r - t_1}} \!\!\!
    \Tr ( t^{\si^{(1)}_{k}} \cdots t^{\si^{(1)}_{k+ t_2}} \, t^{\si^{(2)}_{k+t_2 + 1}} \cdots
    t^{\si^{(2)}_{k + t_1}} \, t^{\si^{(3)}_{j + t_1 + 1}} \cdots t^{\si^{(3)}_{j+r}} t^{l_4} )
    \,
    \L_{t_1 + 1: t_2 ; k}^{( l_1 , l_2 )}
    \nonumber \\
    &&
    \left[
    \widehat{C}_{MHV(0)}^{( k_+ \cdots a_- \cdots b_- (k+t_1)_+ l_{3+} )} (u; \si^{( 1 \otimes 2 )})
    \,
    \frac{1}{q_{k \, k+ t_1}^{2}}
    \,
    \widehat{C}_{MHV(0)}^{( (-l_3 )_{-} ( j+ t_1 + 1)_+ \cdots c_-
    \cdots (j+r)_+ l_{4+} )} (u; \si^{( 3 )})
    \right.
    \nonumber \\
    &&
    ~
    + \,
    \left( 1 - \frac{1}{2} \del_{ r - t_1 , n-r-1} \right)
    \widehat{C}_{MHV(0)}^{( l_{4+} k_+ \cdots a_- \cdots b_- (k+t_1)_+ l_{3+} )} (u; \si^{( 1 \otimes 2 )})
    \nonumber \\
    &&
    \left.
    ~ \times
    \,
    \frac{1}{q_{k \, k+ t_1}^{2}}
    \widehat{C}_{MHV(0)}^{( (-l_3 )_{-} ( j+ t_1 + 1)_+ \cdots c_-
    \cdots (j+r)_+  )} (u; \si^{( 3 )})
    \right]_{l_3 = q_{k \, k+t_1} \bar{\eta}} \, .
    \label{7-3}
\eeqar
As indicated in Figure \ref{fighol0407}, there are
asymmetric and symmetric MHV diagrams for the ${\rm Li}_2$-type amplitudes.
These are reflected in the two terms inside the square bracket of (\ref{7-3}).
In the second term, which corresponds to the symmetric MHV diagram,
we take care of the double counting due to the reflection symmetry.
Upon polylog regularization, we can treat the one-loop MHV diagram
as a unit of MHV diagrams. Thus the symmetric MHV diagram can also
be expressed as in Figure \ref{fighol0408}.

%%%%%%%%%%%%%%%%%%%%%%%%% figure %%%%%%%%%%%%%%%%%%%%%%%%%
%%%%%%%%%%%%%%%%%%%%%%%%% figure %%%%%%%%%%%%%%%%%%%%%%%%%

The ${\rm Li}_3$-type amplitudes in (\ref{7-2}) can also be
obtained from the tree-level CSW rules and the one-loop NMHV
amplitudes of ${\rm Li}_3$ type.
\beqar
    &&
    \widehat{A}_{NMHV(1;{\rm Li}_3 )}^{( i_+ \cdots a_- b_- c_- \cdots (i+r)_+ l_{4+} )} (u)
    \nonumber \\
    &=&
    i g^2
    \sum_{j=i}^{i+r}
    \sum_{r_1 = 1}^{\left\lfloor \frac{r +1}{3} \right\rfloor -1 }
    \sum_{r_2 = 1}^{\left\lfloor \frac{r +1}{3} \right\rfloor -1 }
    \left( 1 - \frac{2}{3} \del_{ \frac{r +1}{3} , r_1 +1} \del_{ \frac{r +1}{3} , r_2 +1} \right)
    \nonumber \\
    &&
    \sum_{\si^{(1)} \in \S_{r_1 + 1}}
    \sum_{\si^{(2)} \in \S_{r_2 + 1}}
    \sum_{\si^{(3)} \in \S_{r - r_1 - r_2 - 1}} \!\!\!\!\!
    \Tr ( t^{\si^{(1)}_{j}} \cdots t^{\si^{(1)}_{j+ r_1}} \, t^{\si^{(2)}_{j + r_1 + 1}} \cdots
    t^{\si^{(2)}_{j + r_1 + r_2 +1}} \, t^{\si^{(3)}_{j + r_1 + r_2 +2}} \cdots t^{\si^{(3)}_{j+r}} t^{l_4} )
    \,
    \nonumber \\
    &&
    \L_{r+1: r_1 , r_2 ; \si_j}^{( l_1 , l_2 , l_3 )}
    \,
    \left[
    \widehat{C}_{MHV(0)}^{( j_{+}  \cdots a_{-} \cdots b_{-} \cdots (j+r_1 + r_2 + 1)_{+} )}
    (u; \si^{(1 \otimes 2)}) \,
    \widehat{C}_{MHV(0)}^{( (j + r_1 + 1)_{1+}  \cdots b_{-} \cdots c_{-} \cdots (j+ r)_{+} l_4 )}
    (u; \si^{(2 \otimes 3)}) \,
    \right.
    \nonumber \\
    &&
    ~
    \left.
    \widehat{C}_{MHV(0)}^{( (j + r_1 + r_2 + 2 )_{+}  \cdots c_{-} \cdots (j+ r)_{+}
    l_{4+}  j_{+} \cdots a_{-} \cdots (j+ r_1 )_{+} )}
    (u; \si^{(3 \otimes 1)})
    \right]^{\frac{1}{2}} \, .
    \label{7-4}
\eeqar
As in the case of (\ref{5-11}), the position of $l_4$ can be chosen
arbitrarily.
In the above expression, we place the internal index $l_4$ adjacent to
$(j+r)$ and $j$.

\noindent
\underline{${\rm Li}_4$-type contributions}

In analogy to (\ref{5-47}), we can calculate the
${\rm Li}_4$-type amplitudes in Figure \ref{fighol0407} as follows.
\beqar
    &&
    \widehat{A}_{N^2 MHV(1;{\rm Li}_4 )}^{( a_- b_- c_- d_- )} (u)
    \nonumber \\
    &=&
    -ig^2
    \sum_{i=1}^{n}
    \sum_{r_1 = 1}^{\left\lfloor \frac{n}{4} \right\rfloor -1 }
    \sum_{r_2 = 1}^{\left\lfloor \frac{n}{4} \right\rfloor -1 }
    \sum_{r_3 = 1}^{\left\lfloor \frac{n}{4} \right\rfloor -1 }
    \left( 1 - \frac{3}{4} \del_{ \frac{n}{4} , r_1 +1} \del_{ \frac{n}{4} , r_2 +1}
    \del_{ \frac{n}{4} , r_3 +1} \right)
    \nonumber \\
    &&
    \sum_{ \si^{(1)} \in \S_{r_{1}+1}} \,
    \sum_{ \si^{(2)} \in \S_{r_{2}+1}} \,
    \sum_{ \si^{(3)} \in \S_{r_{3}+1}} \,
    \sum_{ \si^{(4)} \in \S_{r_{4}+1}} \,
    \nonumber \\
    &&
    \Tr ( t^{\si^{(1)}_{a_1}} \cdots t^{\si^{(1)}_{a_2}} \, t^{\si^{(2)}_{b_1}}
    \cdots t^{\si^{(2)}_{b_2}} \, t^{\si^{(3)}_{c_1}} \cdots t^{\si^{(3)}_{c_2}}
    \, t^{\si^{(4)}_{d_1}} \cdots t^{\si^{(4)}_{d_2}} )
    \nonumber \\
    &&
    \L_{n:r_1 , r_2 , r_3 ; \si_i}^{(l_1 , l_2 , l_3 , l_4 )}
    \,
    \left[
    \widehat{C}_{MHV(0)}^{( a_{1+}  \cdots a_{-} \cdots b_{-} \cdots b_{2+} )}
    (u; \si^{(1 \otimes 2)}) \,
    \widehat{C}_{MHV(0)}^{( b_{1+}  \cdots b_{-} \cdots c_{-} \cdots c_{2+} )}
    (u; \si^{(2 \otimes 3)})
    \right.
    \nonumber \\
    &&
    ~~
    \left.
    \widehat{C}_{MHV(0)}^{( c_{1+}  \cdots c_{-} \cdots d_{-} \cdots d_{2+} )}
    (u; \si^{(3 \otimes 4)}) \,
    \widehat{C}_{MHV(0)}^{( d_{1+}  \cdots d_{-} \cdots a_{-} \cdots a_{2+} )}
    (u; \si^{(4 \otimes 1)})
    \right]^{\frac{1}{2}}
    \label{7-5}
\eeqar
where the indices $a_1 , a_2 , \cdots , d_2$ are labeled by
\beqar
    &&
    a_1 = i \, , ~~ a_2 = i + r_1 \, ,
    \nonumber \\
    &&
    b_1 = i + r_1 + 1 \, , ~~ b_2 = i + r_1 + r_2 + 1 \, ,
    \nonumber \\
    &&
    c_1 = i + r_1 + r_2 + 2 \, , ~~ c_2 = i +  r_1 + r_2 + r_3 + 2 \, ,
    \nonumber \\
    &&
    d_1 = i + r_1 + r_2 + r_3 + 3 \, , ~~ d_2 = i-1 \, .
    \label{7-6}
\eeqar
The symbol
$\L_{n:r_1 , r_2 , r_3 ; \si_i}^{(l_1 , l_2 , l_3 , l_4 )}$ is
given by
\beq
    \L_{n:r_1 , r_2 , r_3 ; \si_i}^{(l_1 , l_2 , l_3 , l_4 )}
    \, = \,
    \frac{1}{(4 \pi )^4} \int
    \om_{1}^{(1)} \om_{2}^{(0)} \om_{3}^{(0)} \om_{4}^{(0)}
    \,
    \R_{n:r_1 , r_2 , r_3 ; \si_i}^{(l_1 , l_2 , l_3 , l_4 )}  \, ,
    \label{7-7}
\eeq
where $\R_{n:r_1 , r_2 , r_3 ; \si_i}^{(l_1 , l_2 , l_3 , l_4 )}$
can be obtained in the same way as
$\R_{n:r_1 , r_2 ; \si_i}^{(l_1 , l_2 , l_3  )}$  in (\ref{5-33})-(\ref{5-38}).
Namely, we can obtain it from
\beq
    \R_{n:r_1 , r_2 , r_3 ; i}^{(l_1 , l_2 , l_3 , l_4 )}
    \, = \,
    \left.
    \frac{
    ( l_1 ~ l_2 ) ( l_2 ~ l_3 ) ( l_3 ~ l_4 ) ( l_4 ~ l_1 )
    ( a_1 ~ a_2 )
    ( b_1 ~ b_2 )
    ( c_1 ~ c_2 )
    ( d_1 ~ d_2 )
    }{
    ( l_1 ~ a_1 )( l_2 ~ b_1 )( l_3 ~ c_1 )( l_4 ~ d_1 )
    ( l_1 ~ b_2 )( l_2 ~ c_2 )( l_3 ~ d_2 )( l_4 ~ a_2 )
    }
    \right|_{ a_1 \parallel b_1 \parallel c_1 \parallel d_1 , \, cycl.}
    \label{7-8}
\eeq
with replacements of
$a_1$ and $a_2$ by $\si^{(1)}_{a_1}$ and $\si^{(1)}_{a_2}$, and so on.

As in (\ref{3-36}), we can
make a tedious but straightforward calculation for the raw data:
\beqar
    &&
    \left.
    \frac{
    ( l_1 ~ l_2 ) ( l_2 ~ l_3 ) ( l_3 ~ l_4 ) ( l_4 ~ l_1 )
    ( a_1 ~ a_2 )
    ( b_1 ~ b_2 )
    ( c_1 ~ c_2 )
    ( d_1 ~ d_2 )
    }{
    ( l_1 ~ a_1 )( l_2 ~ b_1 )( l_3 ~ c_1 )( l_4 ~ d_1 )
    ( l_1 ~ b_2 )( l_2 ~ c_2 )( l_3 ~ d_2 )( l_4 ~ a_2 )
    }
    \right|_{ a_1 \parallel b_1 \parallel c_1 \parallel d_1 }
    \nonumber \\
    &=&
    \widehat{R}^{( l_1 , l_2 )}_{( b_2 , b_1 )}
    \widehat{R}^{( l_2 , l_3 )}_{( c_2 , c_1 )}
    \widehat{R}^{( l_3 , l_4 )}_{( d_2 , d_1 )}
    \widehat{R}^{( l_4 , l_1 )}_{( a_2 , a_1 )}
    \, - \,
    \widehat{R}^{( l_4 , l_1 )}_{( a_2 , a_1 )}
    \widehat{R}^{( l_1 , l_2 )}_{( b_2 , c_2 )}
    \widehat{R}^{( l_3 , l_1 )}_{( d_2 , c_2 )}
    \nonumber \\
    &&
    - \,
    \widehat{R}^{( l_1 , l_2 )}_{( b_2 , c_2 )}
    \widehat{R}^{( l_4 , l_1 )}_{( a_2 , c_2 )}
    \widehat{R}^{( l_3 , l_4 )}_{( d_2 , d_1 )}
    \, + \,
    \widehat{R}^{( l_3 , l_1 )}_{( d_2 , d_1 )}
    \widehat{R}^{( l_2 , l_4 )}_{( c_2 , b_2 )}
    \widehat{R}^{( l_1 , l_4 )}_{( b_2 , a_2 )}
     \nonumber \\
    &&
    - \,
    \widehat{R}^{( l_2 , l_3 )}_{( c_2 , c_1 )}
    \widehat{R}^{( l_4 , l_1 )}_{( a_2 , d_2 )}
    \widehat{R}^{( l_1 , l_3 )}_{( b_2 , d_2 )}
    \, + \,
    \widehat{R}^{( l_1 , l_3 )}_{( b_2 , b_1 )}
    \widehat{R}^{( l_2 , l_3 )}_{( c_2 , c_1 )}
    \widehat{R}^{( l_4 , l_1 )}_{( a_2 , a_1 )}
    \nonumber \\
    &&
    + \,
    \widehat{R}^{( l_2 , l_4 )}_{( c_2 , c_1 )}
    \widehat{R}^{( l_3 , l_4 )}_{( d_2 , a_2 )}
    \widehat{R}^{( l_1 , l_3 )}_{( b_2 , a_2 )}
    \, - \,
    \widehat{R}^{( l_2 , l_4 )}_{( c_2 , a_2 )}
    \widehat{R}^{( l_1 , l_2 )}_{( a_1 , a_2 )}
    \widehat{R}^{( l_1 , l_3 )}_{( b_2 , d_2 )}
    \nonumber \\
    &&
    - \,
    \widehat{R}^{( l_1 , l_2 )}_{( b_2 , b_1 )}
    \widehat{R}^{( l_2 , l_3 )}_{( c_2 , d_2 )}
    \widehat{R}^{( l_2 , l_4 )}_{( d_2 , a_2 )}
    \, + \,
    \widehat{R}^{( l_2 , l_4 )}_{( d_1 , a_2 )}
    \widehat{R}^{( l_1 , l_2 )}_{( b_2 , c_2 )}
    \widehat{R}^{( l_1 , l_3 )}_{( c_2 , d_2 )}
    \nonumber \\
    &&
    + \,
    \widehat{R}^{( l_2 , l_4 )}_{( c_2 , a_2 )}
    \widehat{R}^{( l_3 , l_4 )}_{( d_2 , d_1 )}
    \widehat{R}^{( l_1 , l_2 )}_{( b_2 , b_1 )}
    \, - \,
    \widehat{R}^{( l_2 , l_4 )}_{( c_2 , a_2 )}
    \widehat{R}^{( l_1 , l_2 )}_{( b_2 , b_1 )}
    \widehat{R}^{( l_1 , l_3 )}_{( d_1 , d_2 )}
    \nonumber \\
    &&
    + \,
    \widehat{R}^{( l_3 , l_4 )}_{( d_2 , a_2 )}
    \widehat{R}^{( l_2 , l_3 )}_{( c_2 , c_1 )}
    \widehat{R}^{( l_1 , l_3 )}_{( b_2 , b_1 )}
    \, - \,
    \widehat{R}^{( l_2 , l_4 )}_{( c_2 , a_2 )}
    \widehat{R}^{( l_3 , l_4 )}_{( c_1 , c_2 )}
    \widehat{R}^{( l_1 , l_3 )}_{( b_2 , d_2 )}
    \nonumber \\
    &&
    - \,
    \widehat{R}^{( l_3 , l_4 )}_{( d_2 , d_1 )}
    \widehat{R}^{( l_1 , l_3 )}_{( b_2 , b_1 )}
    \widehat{R}^{( l_2 , l_4 )}_{( c_2 , a_2 )}
    \, + \,
    \widehat{R}^{( l_1 , l_3 )}_{( b_2 , d_2 )}
    \widehat{R}^{( l_2 , l_4 )}_{( c_2 , a_2 )} \, .
    \label{7-9}
\eeqar
Taking account of the cyclic property of the
$\{ l_1 , l_2 , l_3 , l_4 \}$ indices, we can
carry out the polylog regularization in analogy to (\ref{5-40}) as
\beqar
    &&
    \frac{1}{(4 \pi )^4} \int
    \om_{1}^{(1)} \om_{2}^{(0)} \om_{3}^{(0)} \om_{4}^{(0)}
    \,
    \R_{n:r_1 , r_2 , r_3 ; i}^{(l_1 , l_2 , l_3 , l_4 )}  \, ,
    \nonumber \\
    & \longrightarrow &
    ~
    {\rm Li}_4 \left(
    1 - \Delta^{(12)}_{( b_2 , b_1 )}
    - \Delta^{(23)}_{( c_2 , c_1 )}
    - \Delta^{(34)}_{( d_2 , d_1 )}
    - \Delta^{(41)}_{( a_2 , a_1 )}
    \right)
    \, - \,
    {\rm Li}_4 \left(
    1 - \Delta^{(12)}_{( a_1 , a_2 )}
    - \Delta^{(23)}_{( c_2 , b_2 )}
    - \Delta^{(42)}_{( d_2 , c_2 )}
    \right)
    \nonumber \\
    &&
    \, - \,
    {\rm Li}_4 \left(
    1 - \Delta^{(23)}_{( c_1 , b_2 )}
    - \Delta^{(12)}_{( a_2 , c_2 )}
    - \Delta^{(41)}_{( d_1 , d_2 )}
    \right)
    \, + \,
    {\rm Li}_4 \left(
    1 - \Delta^{(42)}_{( d_2 , d_1 )}
    - \Delta^{(31)}_{( c_2 , b_2 )}
    - \Delta^{(12)}_{( b_2 , a_2 )}
    \right)
    \nonumber \\
    &&
    \, - \,
    {\rm Li}_4 \left(
    1 - \Delta^{(34)}_{( c_1 , c_2 )}
    - \Delta^{(12)}_{( d_2 , a_2 )}
    - \Delta^{(41)}_{( d_1 , d_2 )}
    \right)
    \, + \,
    {\rm Li}_4 \left(
    1 - \Delta^{(42)}_{( b_2 , b_1 )}
    - \Delta^{(34)}_{( c_2 , c_1 )}
    - \Delta^{(12)}_{( a_1 , a_2 )}
    \right)
    \nonumber \\
    &&
    \, + \,
    {\rm Li}_4 \left(
    1 - \Delta^{(31)}_{( c_2 , c_1 )}
    - \Delta^{(41)}_{( a_2 , d_2 )}
    - \Delta^{(42)}_{( b_2 , a_2 )}
    \right)
    \, - \,
    {\rm Li}_4 \left(
    1 - \Delta^{(31)}_{( c_2 , a_2 )}
    - \Delta^{(41)}_{( a_2 , a_1 )}
    - \Delta^{(42)}_{( b_2 , d_2 )}
    \right)
    \nonumber \\
    &&
    \, - \,
    {\rm Li}_4 \left(
    1 - \Delta^{(23)}_{( b_1 , b_2 )}
    - \Delta^{(34)}_{( d_2 , c_2 )}
    - \Delta^{(31)}_{( d_2 , a_2 )}
    \right)
    \, + \,
    {\rm Li}_4 \left(
    1 - \Delta^{(31)}_{( a_1 , a_2 )}
    - \Delta^{(23)}_{( c_2 , b_2 )}
    - \Delta^{(42)}_{( d_2 , c_2 )}
    \right)
    \nonumber \\
    &&
    \, + \,
    {\rm Li}_4 \left(
    1 - \Delta^{(31)}_{( c_2 , a_2 )}
    - \Delta^{(41)}_{( d_1 , d_2 )}
    - \Delta^{(23)}_{( b_1 , b_2 )}
    \right)
    \, - \,
    {\rm Li}_4 \left(
    1 - \Delta^{(31)}_{( c_2 , a_2 )}
    - \Delta^{(23)}_{( b_1 , b_2 )}
    - \Delta^{(42)}_{( d_2 , d_1 )}
    \right)
    \nonumber \\
    &&
    \, + \,
    {\rm Li}_4 \left(
    1 - \Delta^{(41)}_{( a_2 , d_2 )}
    - \Delta^{(34)}_{( a_2 , c_2 )}
    - \Delta^{(42)}_{( b_2 , b_1 )}
    \right)
    \, - \,
    {\rm Li}_4 \left(
    1 - \Delta^{(31)}_{( c_2 , a_2 )}
    - \Delta^{(23)}_{( c_2 , c_1 )}
    - \Delta^{(42)}_{( b_2 , d_2 )}
    \right)
    \nonumber \\
    &&
    \, - \,
    {\rm Li}_4 \left(
    1 - \Delta^{(41)}_{( d_1 , d_2 )}
    - \Delta^{(42)}_{( b_2 , b_1 )}
    - \Delta^{(31)}_{( c_2 , a_2 )}
    \right)
    \, + \,
    {\rm Li}_4 \left(
    1 - \Delta^{(31)}_{( c_2 , a_2 )}
    - \Delta^{(42)}_{( b_2 , d_2 )}
    \right)
    \label{7-10}
\eeqar
where $\Delta^{(12)}$'s, $\Delta^{(23)}$'s, $\Delta^{(34)}$'s and $\Delta^{(41)}$'s
are defined by
\beqar
    &&
    \Delta^{(12)}_{( b_2 , b_1 )} = c_{r_1 ; a_1}^{(1)} ( P^{(a)} + p_{a_2} )^2 \, ,
    ~~
    \Delta^{(12)}_{( a_1 , a_2 )} = c_{r_1 ; a_1}^{(1)} ( P^{(a)} + p_{b_2} )^2 \, ,
    ~~
    \Delta^{(12)}_{( b_2 , a_2 )} = c_{r_1 ; a_1}^{(1)}  P^{(a)2} \, ,
    \nonumber \\
    &&
    \Delta^{(23)}_{( c_2 , c_1 )} = c_{r_2 ; b_1}^{(2)} ( P^{(b)} + p_{b_2} )^2 \, ,
    ~~
    \Delta^{(23)}_{( b_1 , b_2 )} = c_{r_2 ; b_1}^{(2)} ( P^{(b)} + p_{c_2} )^2 \, ,
    ~~
    \Delta^{(23)}_{( c_2 , b_2 )} = c_{r_2 ; b_1}^{(2)}  P^{(b)2} \, ,
    \nonumber \\
    &&
    \Delta^{(34)}_{( d_2 , d_1 )} = c_{r_3 ; c_1}^{(3)} ( P^{(c)} + p_{c_2} )^2 \, ,
    ~~
    \Delta^{(34)}_{( c_1 , c_2 )} = c_{r_3 ; c_1}^{(3)} ( P^{(c)} + p_{d_2} )^2 \, ,
    ~~
    \Delta^{(34)}_{( d_2 , c_2 )} = c_{r_3 ; c_1}^{(3)}  P^{(c)2} \, ,
    \nonumber \\
    &&
    \Delta^{(41)}_{( a_2 , a_1 )} = c_{r_4 ; d_1}^{(4)} ( P^{(d)} + p_{d_2} )^2 \, ,
    ~~
    \Delta^{(41)}_{( d_1 , d_2 )} = c_{r_4 ; d_1}^{(4)} ( P^{(d)} + p_{a_2} )^2 \, ,
    ~~
    \Delta^{(41)}_{( a_2 , d_2 )} = c_{r_4 ; d_1}^{(4)}  P^{(d)2} \, ,
    \nonumber \\
    &&
    \Delta^{(12)}_{( d_2 , a_2 )} = c_{r_1 ; a_1}^{(1)}  P^{(a)2} \, ,
    ~~
    \Delta^{(34)}_{( a_2 , c_2 )} = c_{r_3 ; c_1}^{(3)}  P^{(c)2} \, .
    \label{7-11}
\eeqar
Explicit forms of these terms can be obtained as in the cases of
(\ref{5-41})-(\ref{5-43}).
{\it
Strictly speaking, however, the last two entries
$\Delta^{(12)}_{( a_2 , d_2 )}$ and $\Delta^{(34)}_{( c_2 , a_2 )}$
can not be defined directly from the previous cases. Namely, these
are expressions that do not appear in the BST-type polylog regularization
(\ref{4-16}). Thus, if we apply the BST method strictly, we can not
define these terms. }
We here simply regard the undefined indices (say, $d_2$ of
$\Delta^{(12)}_{( a_2 , d_2 )}$) as dummy indices to define these terms.
Similarly, we can {\it define} the rest of $\Delta$'s in
(\ref{7-10}) as below.
\beqar
    \begin{array}{rcl}
    \Delta^{(42)}_{( d_2 , c_2 )}
    & = & c_{r_{4} ; d_1}^{(4)} \,  P^{(d)2}
    \\
    \Delta^{(42)}_{( d_2 , d_1 )}
    & = & c_{r_{4} ; d_1}^{(4)} \, ( P^{(d)} + p_{b_2} )^2
    \\
    \Delta^{(42)}_{( b_2 , b_1 )}
    & = & c_{r_{2} ; b_1}^{(2)} \, ( P^{(b)} + p_{d_2} )^2
    \\
    \Delta^{(42)}_{( b_2 , a_2 )}
    & = & c_{r_{2} ; b_1}^{(2)} \,  P^{(b)  2}
    \\
    \Delta^{(42)}_{( b_2 , d_2 )}
    & = & c_{r_{2} ; b_1}^{(2)} \,  P^{(b)  2}
    \end{array}
    &&
    \begin{array}{rcl}
    \Delta^{(31)}_{( d_2 , a_2 )}
    & = & c_{r_{1} ; a_1}^{(1)} \,  P^{(a)2}
    \\
    \Delta^{(31)}_{( a_1 , a_2 )}
    & = & c_{r_{1} ; a_1}^{(1)} \, ( P^{(a)} + p_{c_2} )^2
    \\
    \Delta^{(31)}_{( c_2 , c_1 )}
    & = & c_{r_{3} ; c_1}^{(3)} \, ( P^{(c)} + p_{a_2} )^2
    \\
    \Delta^{(31)}_{( c_2 , b_2 )}
    & = & c_{r_{3} ; c_1}^{(3)} \,  P^{(c)  2}
    \\
    \Delta^{(31)}_{( c_2 , a_2 )}
    & = & c_{r_{3} ; c_1}^{(3)} \,  P^{(c)  2}
    \end{array}
    \label{7-12}
\eeqar
These may be considered as natural extension of the BST prescription
but it should be emphasized that these
are not obtained solely by the BST prescription;
there are ambiguities in these definitions.
For example, $\Delta^{(42)}_{( b_2 , d_2 )}$ can also be defined
as $c_{r_{4} ; d_1}^{(4)} \,  P^{(d)  2}$ if we treat the index $b_2$ as
a dummy index.
At present, we do not know how to fix such an ambiguity.
This indicates the limitation of our analysis.
We expect that some solution to this problem will be provided
by seeking a proof of the BST prescription in section 4.

\noindent
\underline{Generalization}
%\subsection{Generalization}

For the completion of our analysis, we now sketch how to obtain one-loop
N$^{m}$MHV amplitudes ($m = 1, 2, \cdots n-4$) in general.
The amplitudes can be obtained in a recursive manner.
Let  $\widehat{A}_{N^{m}MHV(1)}^{( a^{(1)}_{-} a^{(2)}_{-}  \cdots a^{(m+2)}_{-} )} (u)$
be the $n$-point one-loop N$^{m}$MHV amplitudes in the momentum-space representation.
Suppose that we know the form of
$\widehat{A}_{N^{m-1}MHV(1)}^{( a^{(1)}_{-} a^{(2)}_{-}  \cdots a^{(m+1)}_{-} )} (u)$.
Then the one-loop N$^{m}$MHV amplitudes can be decomposed as
\beq
    \widehat{A}_{N^{m}MHV(1)}^{( a^{(1)}_{-} a^{(2)}_{-}  \cdots a^{(m+2)}_{-} )} (u)
    \, = \,
    \widehat{A}_{N^{m}MHV(1;{\rm Li}_{2,3,\cdots, m+1})}^{( a^{(1)}_{-} a^{(2)}_{-}
    \cdots a^{(m+2)}_{-} )} (u)
    \, + \,
    \widehat{A}_{N^{m}MHV(1;{\rm Li}_{m+2})}^{( a^{(1)}_{-} a^{(2)}_{-}
    \cdots a^{(m+2)}_{-} )} (u) \, .
    \label{7-13}
\eeq
As in the N$^2$MHV case (\ref{7-1}), the first term
can be expressed by direct use of the CSW rules:
\beqar
    &&
    \widehat{A}_{N^{m} MHV (1;{\rm Li}_{2,3, \cdots , m+1})}^{( ( a^{(1)}_{-} a^{(2)}_{-}
    \cdots a^{(m+2)}_{-} )} (u)
    \nonumber \\
    & = &
    \left.
    \sum_{i=1}^{n} \sum_{r=1}^{n-3}
    \widehat{A}_{N^{m-1}MHV(1)}^{( i_+ \cdots ( a^{(1)}_{-} a^{(2)}_{-}
    \cdots a^{(m+1)}_{-} \cdots (i+r)_+ (l_{m+2})_{+} )} (u)
    \,
    \frac{1}{q_{i \, i+r}^{2}}
    \,
    \widehat{A}_{MHV(0)}^{( (-l_{m+2})_- \cdots a^{(m+2)}_{-} \cdots (i-1)_+ )} (u)
    \right|_{l_{m+2} = q_{i \, i+r} {\bar \eta} }
    \label{7-14}
\eeqar
where the one-loop N$^{m-1}$MHV subamplitudes are given by a sum of
${\rm Li}_k$-type contributions $(k = 2, 3, \cdots , m+1)$.
Explicitly, these are written as
\beq
    \widehat{A}_{N^{m-1}MHV(1)}^{( i_+ \cdots a^{(1)}_{-} a^{(2)}_{-}
    \cdots a^{(m+1)}_{-} \cdots (i+r)_+ (l_{m+2})_{+} )} (u)
    \, = \,
    \sum_{k=2}^{m+1}
    \widehat{A}_{N^{m-1}MHV(1; {\rm Li}_{k})}^{( i_+ \cdots a^{(1)}_{-} a^{(2)}_{-}
    \cdots a^{(m+1)}_{-} \cdots (i+r)_+ (l_{m+2})_{+} )} (u) \, .
    \label{7-15}
\eeq
As in the case of (\ref{7-3}), we need to take account of
symmetries of MHV diagrams in order to obtain explicit forms of
the amplitudes (\ref{7-14}).

The characteristic feature of the one-loop
N$^{m}$MHV amplitude (\ref{7-13}) arises from its second term.
In analogy to (\ref{7-5}) and (\ref{5-47}), we can compute
this term as follows.
\beqar
    &&
    \widehat{A}_{N^{m}MHV(1;{\rm Li}_{m+2})}^{( a^{(1)}_{-} a^{(2)}_{-}
    \cdots a^{(m+2)}_{-} )} (u)
    \nonumber \\
    &=&
    (-1)^{\left\lfloor \frac{m}{2} \right\rfloor + 2} \, i g^2
    \sum_{i=1}^{n}
    \sum_{r_1 = 1}^{\left\lfloor \frac{n}{m+2} \right\rfloor -1 }
    \sum_{r_2 = 1}^{\left\lfloor \frac{n}{m+2} \right\rfloor -1 }
    \!\! \cdots \!\!
    \sum_{r_{m+1} = 1}^{\left\lfloor \frac{n}{m+2} \right\rfloor -1 }
    \!\!
    \left( 1 - \frac{m+1}{m+2} \del_{ \frac{n}{m+2} , r_1 +1} \del_{ \frac{n}{m+2} , r_2 +1}
    \cdots \del_{ \frac{n}{m+2} , r_{m+1} +1} \right)
    \nonumber \\
    &&
    \sum_{ \si^{(1)} \in \S_{r_{1}+1}} \,
    \sum_{ \si^{(2)} \in \S_{r_{2}+1}} \,
    \cdots
    \sum_{ \si^{(m+2)} \in \S_{r_{m+2}+1}}
    \nonumber \\
    &&
    \Tr \left( t^{\si^{(1)}_{a^{(1)}_{1}}} \cdots t^{\si^{(1)}_{a^{(1)}_{2}}} \,
    t^{\si^{(2)}_{a^{(2)}_{1}}} \cdots t^{\si^{(2)}_{a^{(2)}_{2}}} \,
    \cdots
    \, t^{\si^{(m+2)}_{a^{(m+2)}_{1}}} \cdots t^{\si^{(m+2)}_{a^{(m+2)}_{2}}} \right)
    \nonumber \\
    &&
    \L_{n:r_1 , r_2 , \cdots , r_{m+1} ; \si_i}^{(l_1 , l_2 , \cdots , l_{m+2} )}
    \,
    \left[
    \widehat{C}_{MHV(0)}^{( a^{(1)}_{1+}  \cdots a^{(1)}_{-} \cdots a^{(2)}_{-} \cdots a^{(2)}_{2+} )}
    (u; \si^{(1 \otimes 2)}) \,
    \widehat{C}_{MHV(0)}^{( a^{(2)}_{1+}  \cdots a^{(2)}_{-} \cdots a^{(3)}_{-} \cdots a^{(3)}_{2+} )}
    (u; \si^{(2 \otimes 3)}) \times \cdots
    \right.
    \nonumber \\
    &&
    \left.
    \cdots \times
    \widehat{C}_{MHV(0)}^{( a^{(m+1)}_{1+}  \cdots a^{(m+1)}_{-} \cdots a^{(m+2)}_{-} \cdots a^{(m+2)}_{2+} )}
    (u; \si^{(m+1 \, \otimes \, m+2)}) \,
    \widehat{C}_{MHV(0)}^{( a^{(m+2)}_{1+}  \cdots a^{(m+2)}_{-} \cdots a^{(1)}_{-} \cdots a^{(1)}_{2+} )}
    (u; \si^{( m+2 \, \otimes \, 1)})
    \right]^{\frac{1}{2}}
    \nonumber \\
    \label{7-16}
\eeqar
where the indices $a^{(1)}_{1} , a^{(2)}_{2} , \cdots , a^{(m+2)}_{1} , a^{(m+2)}_{2}$ are labeled by
\beqar
    &&
    a^{(1)}_{1} = i \, , ~~ a^{(1)}_{2} = i + r_1 \, ,
    \nonumber \\
    &&
    a^{(2)}_{1} = i + r_1 + 1 \, , ~~ a^{(2)}_{2} = i + r_1 + r_2 + 1 \, ,
    \nonumber \\
    &&
    \vdots
    \nonumber \\
    &&
    a^{(m+1)}_{1} = i + r_1 + r_2 + \cdots + r_{m} + m \, , ~~
    a^{(m+1)}_{2} = i +  r_1 + r_2 + \cdots + r_{m+1} + m \, ,
    \nonumber \\
    &&
    a^{(m+2)}_{1} = i + r_1 + r_2 + \cdots + r_{m+1} + m+1 \, , ~~ a^{(m+2)}_{2} = i-1 \, .
    \label{7-17}
\eeqar
The auxiliary index $r_{m+2}$ is given by $r_{m+2} = n-(r_1 + r_2 + \cdots + r_{m+1} + m + 2)$.
In terms of the indices (\ref{7-17}), the set of permutations are explicitly denoted as
\beq
    \si^{(1)} \, = \,
    \left(
      \begin{array}{c}
        a^{(1)}_{1} ~~ \cdots ~~ a^{(1)}_{2} \\
        \si^{(1)}_{a^{(1)}_{1}} ~ \cdots ~ \si^{(1)}_{a^{(1)}_{2}} \\
      \end{array}
    \right) , \,
    \si^{(2)} \, = \,
    \left(
      \begin{array}{c}
        a^{(2)}_{1} ~~ \cdots ~~ a^{(2)}_{2} \\
        \si^{(2)}_{a^{(2)}_{1}} ~ \cdots ~ \si^{(2)}_{a^{(2)}_{2}} \\
      \end{array}
    \right) ,  \cdots , \,
    \si^{(m+2)} \, = \,
    \left(
      \begin{array}{c}
        a^{(m+2)}_{1} ~~ \cdots ~~ a^{(m+2)}_{2} \\
        \si^{(m+2)}_{a^{(m+2)}_{1}} ~ \cdots ~ \si^{(m+2)}_{a^{(m+2)}_{2}} \\
      \end{array}
    \right) .
    \label{7-18}
\eeq

The polylog regularization is to be realized in the symbol
$\L_{n:r_1 , r_2 , \cdots , r_{m+1} ; \si_i}^{(l_1 , l_2 , \cdots , l_{m+2} )}$
which can be expressed as
\beq
    \L_{n:r_1 , r_2 , \cdots , r_{m+1} ; \si_i}^{(l_1 , l_2 , \cdots , l_{m+2} )}
    \, = \,
    \frac{1}{(4 \pi )^{m+2}} \int
    \om_{1}^{(1)} \om_{2}^{(0)} \om_{3}^{(0)} \cdots \om_{m+2}^{(0)}
    \,
    \R_{n:r_1 , r_2 , \cdots , r_{m+1} ; \si_i}^{(l_1 , l_2 , \cdots , l_{m+2} )}
    \label{7-19}
\eeq
where as in (\ref{7-8}) the integrand can be written as
\beqar
    &&
    \R_{n:r_1 , r_2 , \cdots , r_{m+1} ; \si_i}^{(l_1 , l_2 , \cdots , l_{m+2} )}
    \nonumber \\
    &=&
    \frac{
    ( l_1 ~ l_2 ) ( l_2 ~ l_3 ) \cdots ( l_{m+2} ~ l_1 )
    }{
    \left( l_1 ~ \si^{(1)}_{a^{(1)}_{1}} \right)
    \left( l_2 ~ \si^{(2)}_{a^{(2)}_{1}} \right) \cdots
    \left( l_{m+2} ~ \si^{(m+2)}_{a^{(m+2)}_{1}} \right)
    }
    \nonumber \\
    &&
    \times
    \left.
    \frac{
    \left( \si^{(1)}_{a^{(1)}_{1}} ~ \si^{(1)}_{a^{(1)}_{2}} \right)
    \left( \si^{(2)}_{a^{(2)}_{1}} ~ \si^{(2)}_{a^{(2)}_{2}} \right)
    \cdots
    \left( \si^{(m+2)}_{a^{(m+2)}_{1}} ~ \si^{(m+2)}_{a^{(m+2)}_{2}} \right)
    }{
    \left( l_1 ~ \si^{(2)}_{a^{(2)}_{2}} \right)
    \left( l_2 ~ \si^{(3)}_{a^{(3)}_{2}} \right) \cdots
    \left( l_{m+2} ~ \si^{(1)}_{a^{(1)}_{2}} \right)
    }
    \right|_{ \si^{(1)}_{a^{(1)}_{1}} \parallel
    \si^{(2)}_{a^{(2)}_{1}} \parallel \cdots \parallel
    \si^{(m+2)}_{a^{(m+2)}_{1}} , \, cycl.}
    \label{7-20}
\eeqar
Using the Schouten identities, we can in principle factorize the above
quantity into a sum of the products of $\widehat{R}$'s as shown in (\ref{7-9}).
Its final form is not clear at the present state although we find that
it involves at most $2^{m+2}$ such terms.
The fact that such a factorization is possible, however,
guarantees that we can utilize the polylog regularization in simplifying the expression (\ref{7-19}).
To be more concrete, by use of the polylog regularization,
we can naturally express
$\L_{n:r_1 , r_2 , \cdots , r_{m+1} ; \si_i}^{(l_1 , l_2 , \cdots , l_{m+2} )}$
in terms of a set of ${\rm Li}_{m+2}$ contributions.
For $m=0$, this is analytically confirmed by the BST method.
In the previous sections, we have developed the BST method in the holonomy formalism
so as to show that this picture can also be applicable to the cases of $m=1, 2$
in a systematic way.
It is therefore natural to consider that we can generalize
this picture to arbitrary $m$. The above expressions (\ref{7-19}), (\ref{7-20})
firmly suggest the possibility of such a generalization in a concrete fashion.

%%%%%%%%%%%%%%%%%%%%%%%%%%%%%%%%%%%%%%%%%%%%%%%%%%%%%%
\section{Concluding remarks}

In the present paper, we carry out a first serious analysis
on quantum aspects of the holonomy formalism introduced
in the recent paper \cite{Abe:2009kn}.
As mentioned in the concluding section of \cite{Abe:2009kn},
we can extend the holonomy formalism to obtain one-loop
amplitudes without any modifications.
In this paper, we basically confirm this proposition
by (a) considering the off-shell continuation of the Nair measure
which is necessary to incorporate the loop integrals into the
formalism and (b) considering the one-loop amplitudes
in the coordinate-space representation, rather than the conventional
momentum-space representation.

To be more specific in part (b), we define
the $x$-space contraction operator (\ref{3-20}).
It is this operator that embodies the CSW rules in the
functional derivation of gluon amplitudes in terms of
the $x$-space S-matrix functional (\ref{3-14}).
The basic result of this paper is
that we can systematically obtain one-loop gluon amplitudes
from this S-matrix functional by a conventional functional method,
presenting what we consider the most natural generalization of
the CSW rules or the BST method to one-loop amplitudes
in the framework of the holonomy formalism.

As mentioned at the end of section 2,
the CSW-based calculation of one-loop amplitudes
are qualitatively different from the previously known unitary-cut method
(or the open-string calculation).
The former leads to single-trace color structure while the latter
has multi-trace color decomposition in describing the one-loop amplitudes.
Notice that the the single-trace property does not necessarily mean
the planarity of the amplitudes in the present case since this property
is a direct consequence of the CSW generalization to one-loop amplitudes
and is completely independent of the conventional large $N$ analysis.
In the literature, as far as the author notices, there exist no
analytic expressions for the CSW-based one-loop non-MHV amplitudes.
In this sense, our new analytic results for one-loop
NMHV and N$^2$MHV amplitudes of $\N = 4$ super Yang-Mills theory
are conjectural in nature.

Lastly, we comment on the polylog regularization introduced in section 4.
The regularization can be executed by use of what we call the BST prescription (\ref{4-16}).
In this paper we could not provide a proof of this prescription.
Thus the polylog regularization itself is to some extent speculative.
It does, however, lead to the correct BST representation of
one-loop MHV amplitudes in (\ref{3-41})-(\ref{3-49}).
Furthermore, with the knowledge of the iterated-integral representation of polylogarithm
functions (\ref{4-12}), we can naturally generalize
the polylog regularization scheme to one-loop non-MHV amplitudes.
It is widely recognized in the literature that the CSW rules should
apply not only to one-loop MHV amplitudes but also to any type of loop amplitudes, and
besides, the recent developments indicate that reminder functions of higher-loop
amplitudes can be described systematically in terms of polylogarithms.
Given these facts, partly in search of a new perspective to one-loop amplitudes,
we find it of some significance to pursue the application
of the polylog regularization to one-loop non-MHV amplitudes despite of its conjectural nature.
The latter half of this paper has been developed along these lines of reasonings.

%%%%%%%%%%%%%%%%%%%%%%%%%%%%%%%%%%%%%%%%%%%%%%%%%%%%%%%%%%%%%%%%

%%%%%%%%%%%%%%%%%%%%%%%%%%%%%%%%%%%%%%%%%%%%%%%%%%%%%%%%%%%%%%%%


\begin{thebibliography}{99}
 \small
 \setlength\itemsep{-0.25pt}
 \setlength\baselineskip{11pt}

%%%%%%  CSW  %%%%%%
%\cite{Cachazo:2004kj}
\bibitem{Cachazo:2004kj}
  F.~Cachazo, P.~Svrcek and E.~Witten,
  %``MHV vertices and tree amplitudes in gauge theory,''
  JHEP {\bf 0409}, 006 (2004)
  [arXiv:hep-th/0403047].
  %%CITATION = JHEPA,0409,006;%%

%%%%%%  CSW-related earlier developments %%%%%%
%\cite{Georgiou:2004wu}
\bibitem{Georgiou:2004wu}
  G.~Georgiou and V.~V.~Khoze,
  %``Tree amplitudes in gauge theory as scalar MHV diagrams,''
  JHEP {\bf 0405}, 070 (2004)
  [arXiv:hep-th/0404072].
  %%CITATION = JHEPA,0405,070;%%
%\cite{Georgiou:2004by}
\bibitem{Georgiou:2004by}
  G.~Georgiou, E.~W.~N.~Glover and V.~V.~Khoze,
  %``Non-MHV Tree Amplitudes in Gauge Theory,''
  JHEP {\bf 0407}, 048 (2004)
  [arXiv:hep-th/0407027].
  %%CITATION = JHEPA,0407,048;%%
%\cite{Wu:2004fba}
\bibitem{Wu:2004fba}
  J.~B.~Wu and C.~J.~Zhu,
  %``MHV vertices and scattering amplitudes in gauge theory,''
  JHEP {\bf 0407}, 032 (2004)
  [arXiv:hep-th/0406085].
  %%CITATION = JHEPA,0407,032;%%
%\cite{Abe:2004ep}
\bibitem{Abe:2004ep}
  Y.~Abe, V.~P.~Nair and M.~I.~Park,
  %``Multigluon amplitudes, N = 4 constraints and the WZW model,''
  Phys.\ Rev.\  D {\bf 71}, 025002 (2005)
  [arXiv:hep-th/0408191].
  %%CITATION = PHRVA,D71,025002;%%

%%%%%  loop amplitudes in twistor space  %%%%%
%\cite{Cachazo:2004zb}
\bibitem{Cachazo:2004zb}
  F.~Cachazo, P.~Svrcek and E.~Witten,
  %``Twistor space structure of one-loop amplitudes in gauge theory,''
  JHEP {\bf 0410}, 074 (2004)
  [arXiv:hep-th/0406177].
  %%CITATION = JHEPA,0410,074;%%
%\cite{Cachazo:2004by}
\bibitem{Cachazo:2004by}
  F.~Cachazo, P.~Svrcek and E.~Witten,
  %``Gauge theory amplitudes in twistor space and holomorphic anomaly,''
  JHEP {\bf 0410}, 077 (2004)
  [arXiv:hep-th/0409245].
  %%CITATION = JHEPA,0410,077;%%
%\cite{Cachazo:2004dr}
\bibitem{Cachazo:2004dr}
  F.~Cachazo,
  %``Holomorphic anomaly of unitarity cuts and one-loop gauge theory
  %amplitudes,''
  arXiv:hep-th/0410077.
  %%CITATION = HEP-TH/0410077;%%

%%%%%%%  BST method  for MHV one-loop amplitudes  %%%%%%%%%%%
%\cite{Brandhuber:2004yw}
\bibitem{Brandhuber:2004yw}
  A.~Brandhuber, B.~J.~Spence and G.~Travaglini,
  %``One-loop gauge theory amplitudes in N = 4 super Yang-Mills from MHV
  %vertices,''
  Nucl.\ Phys.\  B {\bf 706}, 150 (2005)
  [arXiv:hep-th/0407214].
  %%CITATION = NUPHA,B706,150;%%
%\cite{Bedford:2004py}
\bibitem{Bedford:2004py}
  J.~Bedford, A.~Brandhuber, B.~J.~Spence and G.~Travaglini,
  %``A twistor approach to one-loop amplitudes in N = 1 supersymmetric
  %Yang-Mills theory,''
  Nucl.\ Phys.\  B {\bf 706}, 100 (2005)
  [arXiv:hep-th/0410280].
  %%CITATION = NUPHA,B706,100;%%
%\cite{Bedford:2004nh}
\bibitem{Bedford:2004nh}
  J.~Bedford, A.~Brandhuber, B.~J.~Spence and G.~Travaglini,
  %``Non-supersymmetric loop amplitudes and MHV vertices,''
  Nucl.\ Phys.\  B {\bf 712}, 59 (2005)
  [arXiv:hep-th/0412108].
  %%CITATION = NUPHA,B712,59;%%
%\cite{Brandhuber:2005kd}
\bibitem{Brandhuber:2005kd}
  A.~Brandhuber, B.~Spence and G.~Travaglini,
  %``From trees to loops and back,''
  JHEP {\bf 0601}, 142 (2006)
  [arXiv:hep-th/0510253].
  %%CITATION = JHEPA,0601,142;%%

%%%%  BST related works  %%%%
%\cite{Luo:2004ss}
\bibitem{Luo:2004ss}
  M.~x.~Luo and C.~k.~Wen,
  %``One-loop maximal helicity violating amplitudes in N = 4 super  Yang-Mills
  %theories,''
  JHEP {\bf 0411}, 004 (2004)
  [arXiv:hep-th/0410045].
  %%CITATION = JHEPA,0411,004;%%
%\cite{Quigley:2004pw}
\bibitem{Quigley:2004pw}
  C.~Quigley and M.~Rozali,
  %``One-loop MHV amplitudes in supersymmetric gauge theories,''
  JHEP {\bf 0501}, 053 (2005)
  [arXiv:hep-th/0410278].
  %%CITATION = JHEPA,0501,053;%%
%%%%  BST review  %%%%
%\cite{Brandhuber:2011ke}
\bibitem{Brandhuber:2011ke}
  A.~Brandhuber, B.~Spence and G.~Travaglini,
  %``Tree-Level Formalism,''
  arXiv:1103.3477 [hep-th].
  %%CITATION = ARXIV:1103.3477;%%

%%%%  Nair measure  %%%%
%\cite{Nair:1988bq}
\bibitem{Nair:1988bq}
  V.~P.~Nair,
  %``A CURRENT ALGEBRA FOR SOME GAUGE THEORY AMPLITUDES,''
  Phys.\ Lett.\  B {\bf 214}, 215 (1988).
  %%CITATION = PHLTA,B214,215;%%

%%% one-loop MHV amplitudes %%%
%\cite{Bern:1994zx}
\bibitem{Bern:1994zx}
  Z.~Bern, L.~J.~Dixon, D.~C.~Dunbar and D.~A.~Kosower,
  %``One-Loop n-Point Gauge Theory Amplitudes, Unitarity and Collinear Limits,''
  Nucl.\ Phys.\  B {\bf 425}, 217 (1994)
  [arXiv:hep-ph/9403226].
  %%CITATION = NUPHA,B425,217;%%
%%% 6-point one-loop NMHV amplitudes %%%
%\cite{Bern:1994cg}
\bibitem{Bern:1994cg}
  Z.~Bern, L.~J.~Dixon, D.~C.~Dunbar and D.~A.~Kosower,
  %``Fusing gauge theory tree amplitudes into loop amplitudes,''
  Nucl.\ Phys.\  B {\bf 435}, 59 (1995)
  [arXiv:hep-ph/9409265].
  %%CITATION = NUPHA,B435,59;%%

%%%% developments of unitary cut methods %%%
%\cite{Britto:2004nc}
\bibitem{Britto:2004nc}
  R.~Britto, F.~Cachazo and B.~Feng,
  %``Generalized unitarity and one-loop amplitudes in N = 4  super-Yang-Mills,''
  Nucl.\ Phys.\  B {\bf 725}, 275 (2005)
  [arXiv:hep-th/0412103].
  %%CITATION = NUPHA,B725,275;%%
%\cite{Britto:2005ha}
\bibitem{Britto:2005ha}
  R.~Britto, E.~Buchbinder, F.~Cachazo and B.~Feng,
  %``One-loop amplitudes of gluons in SQCD,''
  Phys.\ Rev.\  D {\bf 72}, 065012 (2005)
  [arXiv:hep-ph/0503132].
  %%CITATION = PHRVA,D72,065012;%%

%\cite{Bena:2004xu}
\bibitem{Bena:2004xu}
  I.~Bena, Z.~Bern, D.~A.~Kosower and R.~Roiban,
  %``Loops in twistor space,''
  Phys.\ Rev.\  D {\bf 71}, 106010 (2005)
  [arXiv:hep-th/0410054].
  %%CITATION = PHRVA,D71,106010;%%
%\cite{Bern:2004ky}
\bibitem{Bern:2004ky}
  Z.~Bern, V.~Del Duca, L.~J.~Dixon and D.~A.~Kosower,
  %``All non-maximally-helicity-violating one-loop seven-gluon amplitudes in  N
  %= 4 super-Yang-Mills theory,''
  Phys.\ Rev.\  D {\bf 71}, 045006 (2005)
  [arXiv:hep-th/0410224].
  %%CITATION = PHRVA,D71,045006;%%
%\cite{Bern:2004bt}
\bibitem{Bern:2004bt}
  Z.~Bern, L.~J.~Dixon and D.~A.~Kosower,
  %``All next-to-maximally helicity-violating one-loop gluon amplitudes in N  =
  %4 super-Yang-Mills theory,''
  Phys.\ Rev.\  D {\bf 72}, 045014 (2005)
  [arXiv:hep-th/0412210].
  %%CITATION = PHRVA,D72,045014;%%

%%%%  planar amplitudes: ABDK, BDS, AM, DHKS, et al., etc  %%%%%
%\cite{Anastasiou:2003kj}
\bibitem{Anastasiou:2003kj}
  C.~Anastasiou, Z.~Bern, L.~J.~Dixon and D.~A.~Kosower,
  %``Planar amplitudes in maximally supersymmetric Yang-Mills theory,''
  Phys.\ Rev.\ Lett.\  {\bf 91}, 251602 (2003)
  [arXiv:hep-th/0309040].
  %%CITATION = PRLTA,91,251602;%%
%\cite{Bern:2005iz}
\bibitem{Bern:2005iz}
  Z.~Bern, L.~J.~Dixon, V.~A.~Smirnov,
  %``Iteration of planar amplitudes in maximally supersymmetric
  % Yang-Mills theory at three loops and beyond,''
  Phys.\ Rev.\  {\bf D72}, 085001 (2005).
  [hep-th/0505205].

%%%%  reminder function at two loop MHV %%%%
%\cite{Bern:2008ap}
\bibitem{Bern:2008ap}
  Z.~Bern, L.~J.~Dixon, D.~A.~Kosower, R.~Roiban, M.~Spradlin, C.~Vergu and A.~Volovich,
  %``The Two-Loop Six-Gluon MHV Amplitude in Maximally Supersymmetric Yang-Mills
  %Theory,''
  Phys.\ Rev.\  D {\bf 78}, 045007 (2008)
  [arXiv:0803.1465 [hep-th]].
  %%CITATION = PHRVA,D78,045007;%%
%\cite{Drummond:2008aq}
\bibitem{Drummond:2008aq}
  J.~M.~Drummond, J.~Henn, G.~P.~Korchemsky and E.~Sokatchev,
  %``Hexagon Wilson loop = six-gluon MHV amplitude,''
  Nucl.\ Phys.\  B {\bf 815}, 142 (2009)
  [arXiv:0803.1466 [hep-th]].
  %%CITATION = NUPHA,B815,142;%%
%\cite{DelDuca:2010zg}
\bibitem{DelDuca:2010zg}
  V.~Del Duca, C.~Duhr and V.~A.~Smirnov,
  %``The Two-Loop Hexagon Wilson Loop in N = 4 SYM,''
  JHEP {\bf 1005}, 084 (2010)
  [arXiv:1003.1702 [hep-th]].
  %%CITATION = JHEPA,1005,084;%%
%\cite{Zhang:2010tr}
\bibitem{Zhang:2010tr}
  J.~H.~Zhang,
  %``On the two-loop hexagon Wilson loop remainder function in N=4 SYM,''
  arXiv:1004.1606 [hep-th].
  %%CITATION = ARXIV:1004.1606;%%
%\cite{Brandhuber:2010bj}
\bibitem{Brandhuber:2010bj}
  A.~Brandhuber, P.~Heslop, P.~Katsaroumpas, D.~Nguyen, B.~Spence, M.~Spradlin and G.~Travaglini,
  %``A Surprise in the Amplitude/Wilson Loop Duality,''
  JHEP {\bf 1007}, 080 (2010)
  [arXiv:1004.2855 [hep-th]].
  %%CITATION = JHEPA,1007,080;%%
%\cite{Goncharov:2010jf}
\bibitem{Goncharov:2010jf}
  A.~B.~Goncharov, M.~Spradlin, C.~Vergu and A.~Volovich,
  %``Classical Polylogarithms for Amplitudes and Wilson Loops,''
  Phys.\ Rev.\ Lett.\  {\bf 105}, 151605 (2010)
  [arXiv:1006.5703 [hep-th]].
  %%CITATION = PRLTA,105,151605;%%
%\cite{Lipatov:2010ad}
\bibitem{Lipatov:2010ad}
  L.~N.~Lipatov and A.~Prygarin,
  %``BFKL approach and six-particle MHV amplitude in N=4 super Yang-Mills,''
  arXiv:1011.2673 [hep-th].
  %%CITATION = ARXIV:1011.2673;%%
%\cite{Gaiotto:2011dt}
\bibitem{Gaiotto:2011dt}
  D.~Gaiotto, J.~Maldacena, A.~Sever and P.~Vieira,
  %``Pulling the straps of polygons,''
  arXiv:1102.0062 [hep-th].
  %%CITATION = ARXIV:1102.0062;%%

%%%%  Wilson loop and gluon amplitudes  %%%%
%\cite{Alday:2007hr}
\bibitem{Alday:2007hr}
  L.~F.~Alday and J.~M.~Maldacena,
  %``Gluon scattering amplitudes at strong coupling,''
  JHEP {\bf 0706}, 064 (2007)
  [arXiv:0705.0303 [hep-th]].
  %%CITATION = JHEPA,0706,064;%%
%\cite{Alday:2008yw}
\bibitem{Alday:2008yw}
  L.~F.~Alday and R.~Roiban,
  %``Scattering Amplitudes, Wilson Loops and the String/Gauge Theory
  %Correspondence,''
  Phys.\ Rept.\  {\bf 468}, 153 (2008)
  [arXiv:0807.1889 [hep-th]].
  %%CITATION = PRPLC,468,153;%%
%\cite{Henn:2009bd}
\bibitem{Henn:2009bd}
  J.~M.~Henn,
  %``Duality between Wilson loops and gluon amplitudes,''
  arXiv:0903.0522 [hep-th].
  %%CITATION = ARXIV:0903.0522;%%

%%%%%% dual superconformal symmetry %%%%%%%%
%\cite{Drummond:2008vq}
\bibitem{Drummond:2008vq}
  J.~M.~Drummond, J.~Henn, G.~P.~Korchemsky and E.~Sokatchev,
  %``Dual superconformal symmetry of scattering amplitudes in N=4
  %super-Yang-Mills theory,''
  Nucl.\ Phys.\  B {\bf 828}, 317 (2010)
  [arXiv:0807.1095 [hep-th]].
  %%CITATION = NUPHA,B828,317;%%

%%%%%% in relation to the dual superconformal symmetry %%%%%%%%
%\cite{Elvang:2009ya}
\bibitem{Elvang:2009ya}
  H.~Elvang, D.~Z.~Freedman and M.~Kiermaier,
  %``Dual conformal symmetry of 1-loop NMHV amplitudes in N=4 SYM theory,''
  JHEP {\bf 1003}, 075 (2010)
  [arXiv:0905.4379 [hep-th]].
  %%CITATION = JHEPA,1003,075;%%
%\cite{Brandhuber:2009xz}
\bibitem{Brandhuber:2009xz}
  A.~Brandhuber, P.~Heslop and G.~Travaglini,
  %``One-Loop Amplitudes in N=4 Super Yang-Mills and Anomalous Dual Conformal
  %Symmetry,''
  JHEP {\bf 0908}, 095 (2009)
  [arXiv:0905.4377 [hep-th]].
  %%CITATION = JHEPA,0908,095;%%
%\cite{Korchemsky:2009hm}
\bibitem{Korchemsky:2009hm}
  G.~P.~Korchemsky and E.~Sokatchev,
  %``Symmetries and analytic properties of scattering amplitudes in N=4 SYM
  %theory,''
  Nucl.\ Phys.\  B {\bf 832}, 1 (2010)
  [arXiv:0906.1737 [hep-th]].
  %%CITATION = NUPHA,B832,1;%%
%\cite{Brandhuber:2009kh}
\bibitem{Brandhuber:2009kh}
  A.~Brandhuber, P.~Heslop and G.~Travaglini,
  %``Proof of the Dual Conformal Anomaly of One-Loop Amplitudes in N=4 SYM,''
  JHEP {\bf 0910}, 063 (2009)
  [arXiv:0906.3552 [hep-th]].
  %%CITATION = JHEPA,0910,063;%%
%%%%  recent review  %%%%%
%\cite{Henn:2011xk}
\bibitem{Henn:2011xk}
  J.~M.~Henn,
  %``Dual conformal symmetry at loop level: massive regularization,''
  arXiv:1103.1016 [hep-th].
  %%CITATION = ARXIV:1103.1016;%%

%%%%  Yangian symmetry in relation to dual superconformal symmetry  %%%%
%\cite{Drummond:2009fd}
\bibitem{Drummond:2009fd}
  J.~M.~Drummond, J.~M.~Henn and J.~Plefka,
  %``Yangian symmetry of scattering amplitudes in N=4 super Yang-Mills theory,''
  JHEP {\bf 0905}, 046 (2009)
  [arXiv:0902.2987 [hep-th]].
  %%CITATION = JHEPA,0905,046;%%
%\cite{Beisert:2010gn}
\bibitem{Beisert:2010gn}
  N.~Beisert, J.~Henn, T.~McLoughlin and J.~Plefka,
  %``One-Loop Superconformal and Yangian Symmetries of Scattering Amplitudes in
  %N=4 Super Yang-Mills,''
  JHEP {\bf 1004}, 085 (2010)
  [arXiv:1002.1733 [hep-th]].
  %%CITATION = JHEPA,1004,085;%%
%\cite{Drummond:2010uq}
\bibitem{Drummond:2010uq}
  J.~M.~Drummond and L.~Ferro,
  %``The Yangian origin of the Grassmannian integral,''
  JHEP {\bf 1012}, 010 (2010)
  [arXiv:1002.4622 [hep-th]].
  %%CITATION = JHEPA,1012,010;%%
%%%%%%%  review: Yangian symmetry  %%%%%%%%%
%\cite{Bargheer:2011mm}
\bibitem{Bargheer:2011mm}
  T.~Bargheer, N.~Beisert and F.~Loebbert,
  %``Exact Superconformal and Yangian Symmetry of Scattering Amplitudes,''
  arXiv:1104.0700 [hep-th].
  %%CITATION = ARXIV:1104.0700;%%

%%%%  Arkani-Hamed: momentum twistors  %%%%
%\cite{ArkaniHamed:2009si}
\bibitem{ArkaniHamed:2009si}
  N.~Arkani-Hamed, F.~Cachazo, C.~Cheung and J.~Kaplan,
  %``The S-Matrix in Twistor Space,''
  JHEP {\bf 1003}, 110 (2010)
  [arXiv:0903.2110 [hep-th]].
  %%CITATION = JHEPA,1003,110;%%
%\cite{ArkaniHamed:2009dn}
\bibitem{ArkaniHamed:2009dn}
  N.~Arkani-Hamed, F.~Cachazo, C.~Cheung and J.~Kaplan,
  %``A Duality For The S Matrix,''
  JHEP {\bf 1003}, 020 (2010)
  [arXiv:0907.5418 [hep-th]].
  %%CITATION = JHEPA,1003,020;%%
%\cite{ArkaniHamed:2010kv}
\bibitem{ArkaniHamed:2010kv}
  N.~Arkani-Hamed, J.~L.~Bourjaily, F.~Cachazo, S.~Caron-Huot and J.~Trnka,
  %``The All-Loop Integrand For Scattering Amplitudes in Planar N=4 SYM,''
  arXiv:1008.2958 [hep-th].
  %%CITATION = ARXIV:1008.2958;%%

%%%%%  momentum twistor space method  %%%%
%\cite{Drummond:2010mb}
\bibitem{Drummond:2010mb}
  J.~M.~Drummond and J.~M.~Henn,
  %``Simple loop integrals and amplitudes in N=4 SYM,''
  arXiv:1008.2965 [hep-th].
  %%CITATION = ARXIV:1008.2965;%%
%\cite{Bullimore:2010pj}
\bibitem{Bullimore:2010pj}
  M.~Bullimore, L.~Mason and D.~Skinner,
  %``MHV Diagrams in Momentum Twistor Space,''
  arXiv:1009.1854 [hep-th].
  %%CITATION = ARXIV:1009.1854;%%
%\cite{Mason:2010yk}
\bibitem{Mason:2010yk}
  L.~Mason and D.~Skinner,
  %``The Complete Planar S-matrix of N=4 SYM as a Wilson Loop in Twistor
  %Space,''
  arXiv:1009.2225 [hep-th].
  %%CITATION = ARXIV:1009.2225;%%
%\cite{Brandhuber:2010mi}
\bibitem{Brandhuber:2010mi}
  A.~Brandhuber, B.~Spence, G.~Travaglini and G.~Yang,
  %``A Note on Dual MHV Diagrams in N=4 SYM,''
  arXiv:1010.1498 [hep-th].
  %%CITATION = ARXIV:1010.1498;%%
%\cite{He:2010ju}
\bibitem{He:2010ju}
  S.~He and T.~McLoughlin,
  %``On All-loop Integrands of Scattering Amplitudes in Planar N=4 SYM,''
  arXiv:1010.6256 [hep-th].
  %%CITATION = ARXIV:1010.6256;%%
%\cite{ArkaniHamed:2010gg}
\bibitem{ArkaniHamed:2010gg}
  N.~Arkani-Hamed, J.~L.~Bourjaily, F.~Cachazo, A.~Hodges and J.~Trnka,
  %``A Note on Polytopes for Scattering Amplitudes,''
  arXiv:1012.6030 [hep-th].
  %%CITATION = ARXIV:1012.6030;%%
%\cite{ArkaniHamed:2010gh}
\bibitem{ArkaniHamed:2010gh}
  N.~Arkani-Hamed, J.~L.~Bourjaily, F.~Cachazo and J.~Trnka,
  %``Local Integrals for Planar Scattering Amplitudes,''
  arXiv:1012.6032 [hep-th].
  %%CITATION = ARXIV:1012.6032;%%
%\cite{Bullimore:2011ni}
\bibitem{Bullimore:2011ni}
  M.~Bullimore and D.~Skinner,
  %``Holomorphic Linking, Loop Equations and Scattering Amplitudes in Twistor
  %Space,''
  arXiv:1101.1329 [hep-th].
  %%CITATION = ARXIV:1101.1329;%%
%\cite{Adamo:2011cb}
\bibitem{Adamo:2011cb}
  T.~Adamo and L.~Mason,
  %``MHV diagrams in twistor space and the twistor action,''
  arXiv:1103.1352 [hep-th].
  %%CITATION = ARXIV:1103.1352;%%
%\cite{Eden:2011ku}
\bibitem{Eden:2011ku}
  B.~Eden, P.~Heslop, G.~P.~Korchemsky and E.~Sokatchev,
  %``The super-correlator/super-amplitude duality: Part II,''
  arXiv:1103.4353 [hep-th].
  %%CITATION = ARXIV:1103.4353;%%

%%%%  recent unitary cut developments  %%%%
%\cite{Bern:2009xq}
\bibitem{Bern:2009xq}
  Z.~Bern, J.~J.~M.~Carrasco, H.~Ita, H.~Johansson and R.~Roiban,
  %``On the Structure of Supersymmetric Sums in Multi-Loop Unitarity Cuts,''
  Phys.\ Rev.\  D {\bf 80}, 065029 (2009)
  [arXiv:0903.5348 [hep-th]].
  %%CITATION = PHRVA,D80,065029;%%
%\cite{Bern:2010tq}
\bibitem{Bern:2010tq}
  Z.~Bern, J.~J.~M.~Carrasco, L.~J.~Dixon, H.~Johansson and R.~Roiban,
  %``The Complete Four-Loop Four-Point Amplitude in N=4 Super-Yang-Mills
  %Theory,''
  arXiv:1008.3327 [hep-th].
  %%CITATION = ARXIV:1008.3327;%%
%\cite{Kosower:2010yk}
\bibitem{Kosower:2010yk}
  D.~A.~Kosower, R.~Roiban and C.~Vergu,
  %``The Six-Point NMHV amplitude in Maximally Supersymmetric Yang-Mills
  %Theory,''
  arXiv:1009.1376 [hep-th].
  %%CITATION = ARXIV:1009.1376;%%
%\cite{Schabinger:2011wh}
\bibitem{Schabinger:2011wh}
  R.~M.~Schabinger,
  %``One-Loop N = 4 Super Yang-Mills Scattering Amplitudes to All Orders in the
  %Dimensional Regularization Parameter,''
  arXiv:1103.2769 [hep-th].
  %%CITATION = ARXIV:1103.2769;%%

%%%%%%%%%%  review of unitary-cut method  %%%%%%%%
%\cite{Britto:2010xq}
\bibitem{Britto:2010xq}
  R.~Britto,
  %``Loop amplitudes in gauge theories: modern analytic approaches,''
  arXiv:1012.4493 [hep-th].
  %%CITATION = ARXIV:1012.4493;%%
%\cite{Bern:2011qt}
\bibitem{Bern:2011qt}
  Z.~Bern and Y.~t.~Huang,
  %``Basics of Generalized Unitarity,''
  arXiv:1103.1869 [hep-th].
  %%CITATION = ARXIV:1103.1869;%%
%\cite{Carrasco:2011hw}
\bibitem{Carrasco:2011hw}
  J.~J.~M.~Carrasco and H.~Johansson,
  %``Generic multiloop methods and application to N=4 super-Yang-Mills,''
  arXiv:1103.3298 [hep-th].
  %%CITATION = ARXIV:1103.3298;%%
%\cite{Schabinger:2011kb}
\bibitem{Schabinger:2011kb}
  R.~M.~Schabinger,
  %``One-loop N = 4 Super Yang-Mills Scattering Amplitudes Calculated to All
  %Orders in the Dimensional Regularization Parameter,''
  arXiv:1104.3873 [hep-th].
  %%CITATION = ARXIV:1104.3873;%%

%%%%   holonomy formalism   %%%%%
%\cite{Abe:2009kn}
\bibitem{Abe:2009kn}
  Y.~Abe,
  %``Holonomies of gauge fields in twistor space 1: bialgebra, supersymmetry,
  %and gluon amplitudes,''
  Nucl.\ Phys.\  B {\bf 825}, 242 (2010)
  [arXiv:0906.2524 [hep-th]].
  %%CITATION = NUPHA,B825,242;%%
%\cite{Abe:2009kq}
\bibitem{Abe:2009kq}
  Y.~Abe,
  %``Holonomies of gauge fields in twistor space 2: Hecke algebra,
  %diffeomorphism, and graviton amplitudes,''
  Nucl.\ Phys.\  B {\bf 825}, 268 (2010)
  [arXiv:0906.2526 [hep-th]].
  %%CITATION = NUPHA,B825,268;%%

%%%%%  CSW generalization to loop amplitudes  %%%%
%\cite{Sever:2009aa}
\bibitem{Sever:2009aa}
  A.~Sever and P.~Vieira,
  %``Symmetries of the N=4 SYM S-matrix,''
  arXiv:0908.2437v2 [hep-th].
  %%CITATION = ARXIV:0908.2437;%%

%%%%%%%%  color decomposition: early works  %%%%%%%
%\cite{Bern:1990ux}
\bibitem{Bern:1990ux}
  Z.~Bern, D.~A.~Kosower,
  %``Color decomposition of one loop amplitudes in gauge theories,''
  Nucl.\ Phys.\  {\bf B362}, 389-448 (1991).
%\cite{Bern:1991aq}
\bibitem{Bern:1991aq}
  Z.~Bern, D.~A.~Kosower,
  %``The Computation of loop amplitudes in gauge theories,''
  Nucl.\ Phys.\  {\bf B379}, 451-561 (1992).


%%%%  Kohno's book: on iterated representation  %%%
%\cite{Kohno:2002bk)
\bibitem{Kohno:2002bk}
  T.~Kohno, {\it Conformal Field Theory and Topology}, Translations of
  Mathematical Monographs, Volume 210, American Mathematical Society (2002).

%%%%  Ueno's book: on Iwahori-Hecke algebra  %%%
%\cite{Ueno:2008bk)
\bibitem{Ueno:2008bk}
  K.~Ueno, {\it Conformal Field Theory with Gauge Symmetry}, Fields Institute
  Monographs, Volume 24, American Mathematical Society (2008).

%%%%  Kohno-Drinfel'd monodoromy theorem  %%%
%\cite{Chari:1994pz}
\bibitem{Chari:1994pz}
  V.~Chari and A.~Pressley,
  {\it A Guide To Quantum Groups},
  %\href{/spires/find/hep/www?irn=3125289}{SPIRES entry}
  Cambridge University Press (1995).

%%%%   abelian holonomy formalism   %%%%%
%\cite{Abe:2010sa}
\bibitem{Abe:2010sa}
  Y.~Abe,
  %``Application of abelian holonomy formalism to the elementary theory of
  %numbers,''
  arXiv:1005.4299 [hep-th].
  %%CITATION = ARXIV:1005.4299;%%


\end{thebibliography}
\end{document}